\documentclass[3p,times]{elsarticle}

\usepackage{amssymb}
\usepackage{amsmath, bbm}
\usepackage[figuresright]{rotating}
\usepackage{capt-of}

\newcommand{\La}{{\Lambda}}
\newcommand{\Si}{{\Sigma}}
\newcommand{\Kb}{{\overline{K}}}

\newcommand{\be}{\begin{eqnarray}}
\newcommand{\ee}{\end{eqnarray}}

\newcommand{\countzero}{\setcounter{equation}{0}%
         \setcounter{figure}{0}%
         \setcounter{table}{0}}
\newlength{\feynwidth} 
\newlength{\feynwidthbig} 
\DeclareMathOperator{\tr}{tr}
\newcommand{\trace}[1]{\ensuremath{\tr\left(#1\right)}} 
\newcommand{\MMBB}{
\!\begin{picture}(24,5)(0,0)
\put(6,1){\footnotesize $M_1M_2$}
\put(0,4){\line(3,-2){25}}
\put(0,-14){\footnotesize $B_{il}B_{ir}$}
\end{picture}
}
\newcommand{\MMB}{
\!\begin{picture}(24,5)(0,0)
\put(6,0){\footnotesize $M_1M_2$}
\put(0,3){\line(3,-2){25}}
\put(2,-13){\footnotesize $B_{i}$}
\end{picture}
}
\usepackage{xcolor} 

\begin{document}

\begin{frontmatter}


\title{Hyperon-nucleon interaction at next-to-leading order\\ in chiral effective field theory}
\author[J]{J. Haidenbauer}
\author[M]{S. Petschauer}
\author[M]{N. Kaiser}
\author[B,J]{U.-G. Mei{\ss}ner}
\author[J]{A. Nogga}
\author[M,E]{W. Weise}
\address[J]{Institute for Advanced Simulation, Institut f{\"u}r Kernphysik and
J\"ulich Center for Hadron Physics, Forschungszentrum J{\"u}lich, D-52425 J{\"u}lich, Germany}
\address[M]{Physik Department, Technische Universit\"at M\"unchen, D-85747
  Garching, Germany}
\address[B]{Helmholtz Institut f\"ur Strahlen- und Kernphysik and Bethe Center
 for Theoretical Physics, Universit\"at Bonn, D-53115 Bonn, Germany}
\address[E]{ECT*, Villa Tambosi, I-38123 Villazzano (Trento), Italy }
\begin{abstract}
Results for the $\Lambda N$ and $\Sigma N$ interactions obtained
at next-to-leading order in chiral effective field theory are reported.
At the order considered there are contributions from one- and two-pseudoscalar-meson 
exchange diagrams and from four-baryon contact terms without and with two 
derivatives. SU(3) flavor symmetry is imposed for constructing the hyperon-nucleon 
interaction while the explicit SU(3) symmetry breaking by the physical masses of
the pseudoscalar mesons ($\pi$, $K$, $\eta$) is taken into account. 
An excellent description of the hyperon-nucleon system can be achieved at next-to-leading order.
It is on the same level of quality as the one obtained by the most advanced 
phenomenological hyperon-nucleon interaction models.
\end{abstract}
\begin{keyword}
Hyperon-nucleon interaction \sep 
Effective field theory 
\PACS{13.75.Ev \sep 12.39.Fe \sep 14.20.Pt}
\end{keyword}
\end{frontmatter}

\section{Introduction}
\label{sec:1}

While there is a steady interest in physics involving baryons with 
strangeness and a corresponding increase of empirical information 
ever since the discovery of the $\Lambda$-hyperon many decades ago, the present
times seem to be particularly rewarding. 
First, at new experimental facilities like J-PARC in Japan or FAIR in 
Germany a significant amount of beam time will be devoted to strangeness 
physics research. The proposed experiments encompass accurate measurements 
of level spectra and decay properties of strangeness $S=-1$ and 
$S=-2$ hypernuclei \cite{JPARC,Panda} but also of elementary cross 
sections for $\Sigma^+ p$ scattering \cite{Hiruma}. 
Information on $\La p$ scattering, specifically on the scattering lengths, 
might emerge from ongoing studies of the final-state interaction in
production reactions like $pp\to K^+\La p$ \cite{COSY}
and $\gamma d\to K^+\La n$ \cite{Hicks}. 

Parallel to this development, techniques for dealing with few- and 
many-body systems have reached a high degree of sophistication
\cite{Pieper,Kievsky,NCSM,Roth:2011vt,Borasoy:2006qn,Lee,Hagen:2007ew,Hagen:2007hi}.
Some of these allow one to consider nuclei with much more than four nucleons,
the limit for standard few-body calculations with the Faddeev-Yakubovsky
theory \cite{Kievsky}. Of particular interest in this context are nuclear
lattice simulations as they offer a new many-body technique directly tailored
to the effective field theory description of baryon-baryon interactions, 
as high-lighted recently by the first ever {\it ab initio} calculation of the
Hoyle state in the spectrum of $^{12}$C~\cite{Epelbaum:2011md}.
Thus, it seems to be feasible to perform similar calculations of 
hypernuclei too, with comparable accuracy as those for ordinary nuclei,
which would open a completely new testing ground for the hyperon-nucleon ($YN$) interaction.
Though few-body calculations of hypernuclei can be already found in the 
literature \cite{Nemura,Hiyama}, for the latter
aspect it would be desirable to employ techniques that allow one to
use directly the elementary $YN$ interaction (i.e.\ without any approximation)
and, in particular, to include the important $\La$-$\Si$ conversion. Only 
then one can connect the properties of the hypernuclei unambiguously with 
those of the underlying $\La N$ (and $\Si N$) interaction.

Finally, and on a different frontier, 
lattice QCD calculations have matured to a certain degree,
as documented in recent review articles~\cite{Beane11,Aoki12}, and 
are coming closer to a level where they can provide additional 
constraints on the baryon-baryon interactions
in the strangeness sector \cite{Beane12}.

To keep up with these developments we present here a study of the
$YN$ interaction performed at next-to-leading order (NLO) in 
chiral effective field theory (EFT). It builds upon and extends
a previous investigation by the Bonn-J\"ulich group carried out at
leading order (LO) \cite{Polinder:2006zh}. 
Using chiral EFT for the $YN$ interaction is prompted by the great
success that this scheme has met in the application to the nucleon-nucleon ($NN$)
interaction. Indeed, proposed by Weinberg \cite{Wei90,Wei91} more than two 
decades ago, chiral EFT has turned out to be a rather powerful tool for the
derivation of nuclear forces. Its most salient feature is that
there is an underlying power counting which allows one to improve calculations
systematically by going to higher orders in a perturbative expansion.
In addition, it is possible to derive two- and three-nucleon 
forces as well as external current operators in a consistent way.
We are now at a stage that the latter aspect will also be important 
for realistic studies of hypernuclear interactions. In the past
there have been already discussions on the role of a three-body $YN$ 
interaction \cite{Bhaduri,Gal71} in hypernuclei and specifically for the 
properties of neutron star matter \cite{Schaffner,Vidana}. 
 
As the most recent applications demonstrate, the nucleon-nucleon interaction 
can be described accurately within the chiral EFT approach \cite{Entem:2003ft,Epe05}. 
In line with the original suggestion of Weinberg, the power counting is 
applied to the $NN$ potential rather than to the reaction amplitude. The latter is 
obtained from solving a regularized Lippmann-Schwinger equation for the derived 
interaction potential. The chiral $NN$ potential contains pion-exchanges and a series of 
contact interactions with an increasing number of derivatives. The latter 
represent the short-range part of the $NN$ force and are parametrized by low-energy constants (LECs), that need to be 
fixed by a fit to data. For reviews we refer the reader
to the recent Refs.~\cite{Epelbaum:2008ga,Machleidt:2011zz}.

In our study of the $YN$ interaction, we follow the scheme that has been 
applied by Epelbaum et al.~\cite{Epe05,Epe98,Epe00} to the $NN$ interaction. 
For investigations of the $YN$ interaction based on other schemes 
see \cite{Kor01,Beane:2003yx}.
Still, there are some essential differences between the $\La N$, $\Si N$ systems
and the $NN$ case that have an influence on how one proceeds in the application of
chiral EFT in practice. First and foremost, there is no phase-shift analysis for the
$S=-1$ sector and, therefore, we have to fix the LECs by a direct fit to data 
rather than by a fit to individual partial waves as it is done in the $NN$ case. 
Secondly, the amount of $YN$ data is rather limited. Indeed, there are basically only 
integrated cross sections, often with large uncertainties. 
Thus, we follow here the practice of previous investigations of the $YN$ interaction, 
notably those performed in the meson-exchange picture \cite{Hol89,Hai05,Rij99,Rij10}, 
and impose constraints from SU(3) flavor symmetry in order to reduce the number of
free parameters. In particular, all the baryon-baryon-meson coupling constants
are fixed from SU(3) symmetry and the symmetry is also exploited to derive relations
between the various LECs. In the actual calculation the SU(3) symmetry is broken,
however, by the mass differences between the Goldstone bosons ($\pi$, $K$, $\eta$) 
and between the baryons. For these masses we use the known physical values. 
In any case, we want to stress that we consider the imposed SU(3) symmetry 
primarily as a working hypothesis and not as a rigorous constraint. Future data with 
higher precision will possibly demand to depart from SU(3) symmetry in some way. 
In that sense our present investigation certainly has primarily an exploratory 
character. 
At the moment we are able to describe the available $\La N$ and $\Si N$ data 
consistently without any explicit SU(3) breaking in the contact interactions 
as will be demonstrated below. 
A simultaneous description of the $NN$ interaction with contact 
terms that fulfil SU(3) symmetry turned out, however, to be not possible. 
 
As mentioned above, in order to obtain the reaction amplitude from the
interaction potential derived within chiral EFT, one has to 
solve a regularized Lippmann-Schwinger equation. The question how this
regularization should be performed is an open issue and is still 
controversially discussed in the literature, see, e.g.\ 
\cite{Nog05,Pav06,Mach12,Phillips13}.
In the present work, we refrain from touching this 
certainly very important question. Rather we follow closely
the procedure adopted by Epelbaum et al.~\cite{Epe05,Epe00} and 
others \cite{Entem:2003ft}, in their study of the $NN$ interaction 
and introduce a momentum-dependent exponential regulator function into the
scattering equation. 

The present paper is structured as follows: 
In Sect.~2, a review of the chiral EFT approach is given with special emphasis
on the imposed SU(3) symmetry. In particular, the structure of the contact 
interactions at LO and NLO is specified and the expression for the one-meson
exchange contributions are reproduced. A detailed description of the
two-boson exchange potential that arises at NLO is presented in the appendix. 
The strategy followed in the fit to the data is outlined in Sect.~3. 
In Sect.~4 results for the $\Lambda N$ and $\Sigma N$ interactions 
obtained at NLO are discussed and compared to available experimental 
information. Results of our LO calculation and of the J\"ulich '04
$YN$ interaction \cite{Hai05}, a conventional meson-exchange model, 
are presented, too. The paper ends with a short summary. 
In the appendix, we provide expressions for the two-boson exchange potential.
Furthermore we summarize SU(3) breaking effects which arise at NLO from quark mass 
insertions in the interaction Lagrangian. 

\section{Chiral potential at next-to-leading order}
\label{sec:2}

\begin{figure}[t]
 \centering
 \includegraphics[width=\feynwidth]{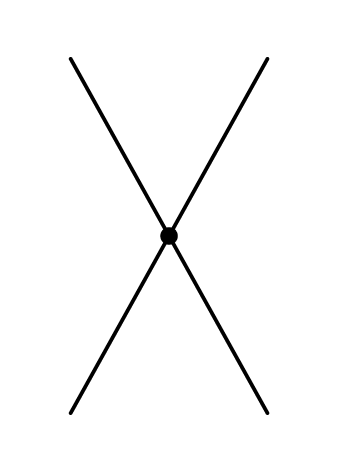}
 \includegraphics[width=\feynwidth]{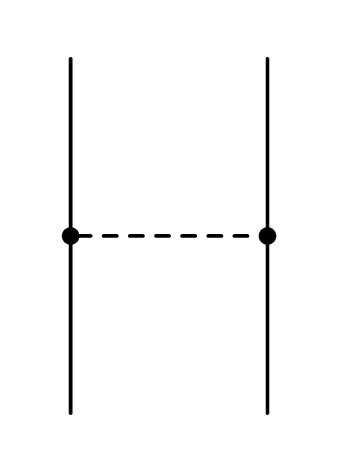}
 \includegraphics[width=\feynwidth]{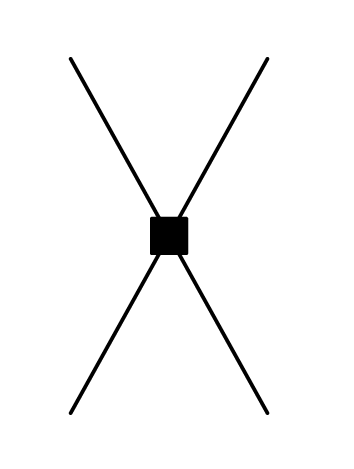}
 \includegraphics[width=\feynwidth]{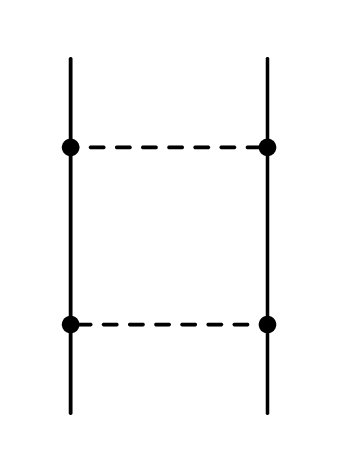}
 \includegraphics[width=\feynwidth]{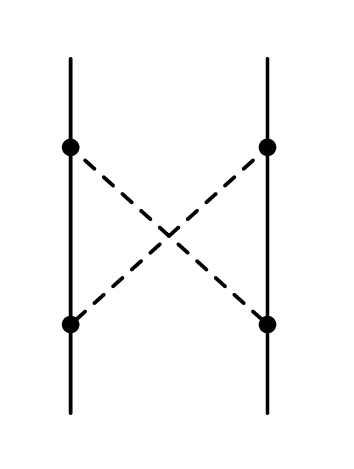}
 \includegraphics[width=\feynwidth]{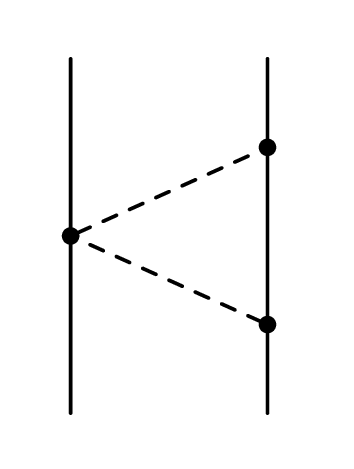}
 \includegraphics[width=\feynwidth]{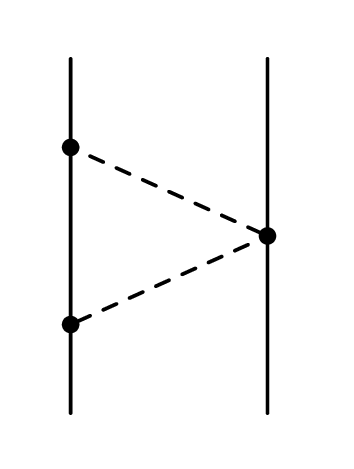}
 \includegraphics[width=\feynwidth]{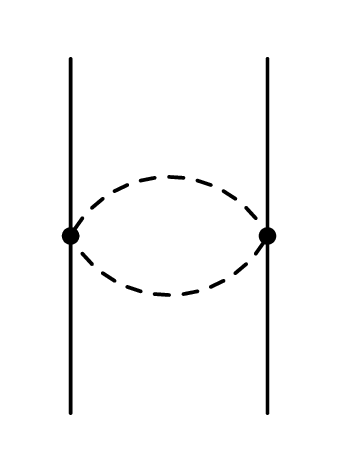}
 \caption{Relevant Feynman diagrams up-to-and-including next-to-leading order. Solid and dashed lines denote octet baryons and pseudoscalar mesons, respectively. The square symbolizes a contact vertex with two derivatives. 
From left to right: LO contact term, one-meson exchange, NLO contact term, planar box, crossed box, left triangle, right triangle, football diagram.} \label{fig:feynman}
\end{figure}

The derivation of the chiral baryon-baryon potentials for the strangeness sector 
at LO using the Weinberg power counting has been outlined in Refs.~\cite{Polinder:2006zh,Hai10a,Haidenbauer:2007ra}. 
The NLO contributions
for the $NN$ case are described in detail in Ref.~\cite{Epe00}, while the 
extension to the baryon-baryon case has been worked out in Ref.~\cite{Pet11}. 
The LO potential consists of four-baryon contact terms without derivatives and of 
one-pseudoscalar-meson exchanges. At NLO contact terms with two derivatives 
arise, together with loop contributions from (irreducible) two-pseudoscalar-meson exchanges.
The corresponding Feynman diagrams are shown in Fig.~\ref{fig:feynman}.

\subsection{Contact terms}
\label{sec:2CT}

The spin dependence of the potentials due to leading order contact terms is given by \cite{Epe00}
\begin{eqnarray}
V^{(0)}_{BB\to BB} &=& C_{S} + C_{T}\,
\mbox{\boldmath $\sigma$}_1\cdot\mbox{\boldmath $\sigma$}_2\,,
\end{eqnarray}
where the parameters $C_{S}$ and $C_{T}$ are low-energy 
constants (LECs) that need to be determined in a fit to data. 
At next-to-leading order the spin- and momentum-dependence of the contact terms reads 
\begin{eqnarray}
V^{(2)}_{BB\to BB} &=& C_1 {\bf q}^{\,2}+ C_2 {\bf k}^{\,2} + (C_3 {\bf q}^{\,2}+ C_4 {\bf k}^{\,2})
\,\mbox{\boldmath $\sigma$}_1\cdot\mbox{\boldmath $\sigma$}_2 
+ \frac{i}{2} C_5 (\mbox{\boldmath $\sigma$}_1+\mbox{\boldmath $\sigma$}_2)\cdot ({\bf q} \times {\bf k}) \nonumber \\
&+& C_6 ({\bf q} \cdot \mbox{\boldmath $\sigma$}_1) ({\bf q} \cdot \mbox{\boldmath $\sigma$}_2)
+ C_7 ({\bf k} \cdot \mbox{\boldmath $\sigma$}_1) ({\bf k} \cdot \mbox{\boldmath $\sigma$}_2)
+ \frac{i}{2} C_8 (\mbox{\boldmath $\sigma$}_1-\mbox{\boldmath $\sigma$}_2)\cdot ({\bf q} \times {\bf k}) \ . 
\end{eqnarray}
The transferred 
and average momentum, ${\bf q}$ and ${\bf k}$, are defined in terms of the final and initial 
center-of-mass momenta of the baryons, ${\bf p}'$ and ${\bf p}$, as 
${\bf q}={\bf p}'-{\bf p}$ and ${\bf k}=({\bf p}'+{\bf p})/2$. 
The $C_i$ ($i=1,\dots,8$) are additional LECs depending on the considered baryon-baryon channel. 
When performing a partial wave projection, these terms contribute to the two $S$--wave
($^1S_0$, $^3S_1$) potentials, the four $P$--wave
($^1P_1$, $^3P_0$, $^3P_1$, $^3P_2$) potentials, and the $^3S_1$-$^3D_1$ and $^1P_1$-$^3P_1$ 
transition potentials in the following way \cite{Epe05}:
\begin{eqnarray}
\label{VC0}
V(^1S_0) &=& {4\pi} \, (C_S-3C_T) + \pi \, ( 4C_1 + C_2 -12C_3
-3C_4 -4C_6 -C_7) ({p}^2+{p}'^2)~, \nonumber \\
&=& \tilde{C}_{^1S_0} + {C}_{^1S_0} ({p}^2+{p}'^2)~, \label{C1S0}\\
V(^3S_1) &=& {4\pi} \, (C_S+C_T) + \frac{\pi}{3} \, ( 12C_1 + 3C_2 +12C_3
+3C_4 +4C_6 +C_7) ({p}^2+{p}'^2)~, \nonumber \\
&=& \tilde{C}_{^3S_1} + {C}_{^3S_1} ({p}^2+{p}'^2)~, \label{C3S1} \\
V(^1P_1) &=& \frac{2\pi}{3} \, ( -4C_1 + C_2 +12C_3
-3C_4 +4C_6 -C_7) \, {p}\, {p}'
= {C}_{^1P_1}\, {p}\, {p}'~,\\
V(^3P_1) &=& \frac{2\pi}{3} \, ( -4C_1 + C_2 - 4C_3
+C_4 + 2C_5 -8C_6 +2C_7) \, {p}\, {p}'
= {C}_{^3P_1}\, {p}\, {p}'~, \\ 
\label{VC1}
V(^3P_1 - {^1P_1}) &=& -\frac{4\sqrt{2}\pi}{3} \, C_8\,
{p}\, {p}' = {C}_{^3P_1 - ^1P_1}\, {p}\, {p}'~,
\label{VC2}
\end{eqnarray}
\begin{eqnarray}
V(^1P_1 - {^3P_1}) &=& -\frac{4\sqrt{2}\pi}{3} \, C_8\,
{p}\, {p}' = {C}_{^1P_1 - ^3P_1}\, {p}\, {p}'~,\\
V(^3P_0) &=& \frac{2\pi}{3} \, ( -4C_1 + C_2 - 4C_3
+C_4 + 4C_5 +12C_6 - 3C_7) \, {p}\, {p}'
= {C}_{^3P_0}\, {p}\, {p}'~,\\
V(^3P_2) &=& \frac{2\pi}{3} \, ( -4C_1 + C_2 - 4C_3
+C_4 - 2C_5 ) \, {p}\, {p}'
= {C}_{^3P_2}\,  {p}\, {p}'~,\\
V(^3D_1 -\, ^3S_1) &=& \frac{2\sqrt{2}\pi}{3} \, ( 4C_6 + C_7)\,
{p'}^2 = {C}_{^3S_1 -\, ^3D_1}\, {p'}^2~,\\
V(^3S_1 -\, ^3D_1) &=& \frac{2\sqrt{2}\pi}{3} \, ( 4C_6 + C_7)\,
{p}^2 = {C}_{^3S_1 -\, ^3D_1}\, {p}^2~, 
\label{VC}
\end{eqnarray}
with $p = |{\bf p}\,|$ and ${p}' = |{\bf p}\,'|$.
Note that the term proportional to $C_8$ in Eqs.~(\ref{VC1}) and (\ref{VC2}) represents
an antisymmetric spin-orbit force and gives rise to spin singlet-triplet 
transitions (i.e.\ $^1P_1 - {^3P_1}$). Such transitions cannot occur in the $NN$
interaction, unless isospin symmetry breaking is included, and, therefore, this term 
is absent in the equations given in Ref.~\cite{Epe00}. However, in general, 
this antisymmetric spin-orbit term is allowed. Specifically, it does not break 
SU(3) symmetry.

Assuming only isospin symmetry, the LECs for each 
spin-isospin channel of the various $BB\to BB$ interaction potentials are
independent. When imposing ${\rm SU(3)}$ flavor symmetry one obtains 
relations between the LECs and, thereby, the number of terms
that need to be fitted to data gets reduced. 
The relevant ${\rm SU(3)}$ structure for the scattering of two octet baryons follows
from the tensor product decomposition
$8$ $\otimes$ $8$ = $1$ $\oplus$ $8_a$ $\oplus$ $8_s$ $\oplus$ $10^*$ $\oplus$ $10$ $\oplus$ $27$ 
(for details see Refs.~\cite{Swa63,Dover1991}). With that one can express all the 
$C_{^1S_0,{\nu}}$, $C_{^3S_1,{\nu }},\ldots$, in Eqs.~(\ref{VC0}) -- (\ref{VC})
($\nu$= $NN\to NN$, $\Lambda N\to \Lambda N$, $\Lambda N\to \Sigma N$, $\Sigma N\to \Sigma N$)
by coefficients corresponding to the {$\mathrm{SU}(3)$} irreducible representations: 
$C^1$, $C^{8_a}$, $C^{8_s}$, $C^{10^*}$, $C^{10}$, $C^{27}$.
The particular combinations of LECs in the various $BB\to BB$ channels and for 
the various partial waves are summarized in Tab.~\ref{tab:SU3}.
For example, for the potential in the $^1S_0$ partial wave of the 
$\La N \to \La N$ channel we get
\begin{eqnarray}
V_{\La N \to \La N}(^1S_0) = \frac1{10}\left[9\tilde C^{27}_{^1S_0} + \tilde C^{8_s}_{^1S_0}
+ (9C^{27}_{^1S_0} + C^{8_s}_{^1S_0})(p^2+p'^2)\right]  \ . 
\end{eqnarray}
Note that Tab.~\ref{tab:SU3} gives the weight factors of the various baryon-baryon channels 
with respect to the irreducible SU(3) representations. In addition, it
reflects the constraints from the generalized Pauli principle. The interaction in partial waves like 
the $^3S_1$, $^3D_1$, and $^1P_1$, which are symmetric with regard to their 
spin-space component, is given by linear combinations of coefficients corresponding
to antisymmetric SU(3) representations ($C^{8_a}$, $C^{10^*}$, $C^{10}$),
whereas those with antisymmetric spin-space part ($^1S_0$, $^3P_0$, $^3P_1$, $^3P_2$) receive 
only contributions from symmetric representations ($C^{8_s}$, $C^{27}$). 
The $C_8$-term induces transitions between singlet and triplet states in the 
octet-representation $8_a$ and $8_s$, respectively \cite{Dover1991}.
For a detailed derivation of the ${\rm SU(3)}$ constraints on the
LECs see Ref.~\cite{Polinder:2006zh} or \cite{Pet13}. 

Due to the imposed ${\rm SU(3)}$ constraints at LO there are only 
five independent LECs for the $NN$ and the $YN$ sectors together,
as outlined in Ref.~\cite{Polinder:2006zh}. Note that without ${\rm
  SU(3)}$ symmetry, there would be twice as many.
At NLO ${\rm SU(3)}$ symmetry implies that in case of the $NN$ and 
$YN$ interactions there are eight new LECs entering the $S$-waves and $S$-$D$ 
transitions, respectively, and ten coefficients in the $P$-waves. 
Note that the sixth leading-order LEC corresponding to the singlet representation 
($C^1$) is present in the strangeness $S=-2$ channels with 
isospin $I=0$~\cite{Polinder:2007mp} and there are four more LECs that contribute 
to the $S=-2$ sector at NLO.

\begin{table}[ht]
\caption{SU(3) relations for the various contact potentials in the isospin basis.
$C^{27}_{\xi}$ etc. refers to the corresponding irreducible SU(3) representation 
for a particular partial wave ${\xi}$. The actual potential still needs to be 
multiplied by pertinent powers of the momenta $p$ and $p'$. 
}
\label{tab:SU3}
\vskip 0.1cm
\renewcommand{\arraystretch}{1.2}
\centering
\begin{tabular}{|l|c|c|l|l|l|}
\hline
&Channel &I &\multicolumn{3}{|c|}{$V({\xi})$} \\ 
\hline
&        &  &$\xi= \, ^1S_0, \, ^3P_0, \, ^3P_1, \, ^3P_2 $ 
& $\xi = \, ^3S_1, \, ^3S_1$-$^3D_1, \, ^1P_1$ & $\xi = \, ^1P_1$-$^3P_1$ \\
\hline
${S=\phantom{-}0}$&$NN\rightarrow NN$ &$0$ & \ \ -- & $C^{10^*}_{\xi}$ & \ \ -- \\
                       &$NN\rightarrow NN$ &$1$ & $C^{27}_{\xi}$ & \ \ -- & \ \ -- \\
\hline
${S=-1}$&$\La N \rightarrow \La N$ &$\frac{1}{2}$ &$\frac{1}{10}\left(9C^{27}_{\xi}+C^{8_s}_{\xi}\right)$
& $\frac{1}{2}\left(C^{8_a}_{\xi}+C^{10^*}_{\xi}\right)$ & $\frac{-1}{\sqrt{20}} C^{8_s8_a}_{\xi}$\\
&$\La N \rightarrow \Si N$ &$\frac{1}{2}$        &$\frac{3}{10}\left(-C^{27}_{\xi}+C^{8_s}_{\xi}\right)$
& $\frac{1}{2}\left(-C^{8_a}_{\xi}+C^{10^*}_{\xi}\right)$ & $\frac{3}{\sqrt{20}} C^{8_s8_a}_{\xi}$\\
&$\Si N \rightarrow \La N$ & & & & $\frac{-1}{\sqrt{20}} C^{8_s8_a}_{\xi}$\\
&$\Si N \rightarrow \Si N$  &$\frac{1}{2}$        &$\frac{1}{10}\left(C^{27}_{\xi}+9C^{8_s}_{\xi}\right)$
& $\frac{1}{2}\left(C^{8_a}_{\xi}+C^{10^*}_{\xi}\right)$ & $\frac{3}{\sqrt{20}} C^{8_s8_a}_{\xi} $\\
&$\Si N \rightarrow \Si N$  &$\frac{3}{2}$        &$C^{27}_{\xi}$
& $C^{10}_{\xi}$ & \ \ -- \\
\hline
\end{tabular}
\renewcommand{\arraystretch}{1.0}
\end{table}

\subsection{Goldstone boson exchange}
\label{sec:2MEX}
 
The one- and two-pseudoscalar-meson-exchange potentials follow from 
the \(\mathrm{SU(3)}\)-invariant meson-baryon interaction Lagrangian
\begin{equation}
 {\mathcal L}_\mathrm{MB}=\tr\left(\bar B \left(\mathrm i \gamma^\mu D_\mu -M_0\right) B\right) - \frac D 2 \tr\left(\bar B \gamma^\mu \gamma_5 \lbrace u_\mu,B\rbrace\right) - \frac F 2 \tr\left(\bar B \gamma^\mu \gamma_5 [u_\mu,B]\right)\,,
\end{equation}
with
\( D_\mu B = \partial_\mu B + [\Gamma_\mu,B]\),
\( \Gamma_\mu = \frac 1 2 ( u^\dagger \partial_\mu u + u \partial_\mu u^\dagger )\) and
\(u_\mu = \mathrm i ( u^\dagger \partial_\mu u - u \partial_\mu u^\dagger )\),
and where the trace is taken in flavor space. 
%
The constant \(M_0\) denotes the baryon mass in the three-flavor chiral limit.
The coupling constants $F$ and $D$ satisfy the relation $F+D=g_A\simeq 1.26$, where
$g_A$ is the axial-vector strength measured in neutron $\beta$--decay.
For the pseudoscalar mesons and octet baryons, collected in traceless \(3\times3\) matrices, 
\begin{equation}
 P =
 \begin{pmatrix}
  \frac{\pi^0}{\sqrt 2} + \frac{\eta}{\sqrt 6} & \pi^+ & K^+ \\
  \pi^- & -\frac{\pi^0}{\sqrt 2} + \frac{\eta}{\sqrt 6} & K^0 \\
   K^- & \Kb^0 & -\frac{2\eta}{\sqrt 6}
 \end{pmatrix}\,,\qquad
  B=
 \begin{pmatrix}
  \frac{\Sigma^0}{\sqrt 2} + \frac{\Lambda}{\sqrt 6} & \Sigma^+ & p \\
  \Sigma^- & -\frac{\Sigma^0}{\sqrt 2} + \frac{\Lambda}{\sqrt 6} & n \\
  -\Xi^- & \Xi^0 & -\frac{2\Lambda}{\sqrt 6}
 \end{pmatrix}\,,
\end{equation}
we use the usual non-linear realization of chiral symmetry with 
\(U(x) = u^2(x)=\exp\left(\mathrm i\sqrt 2 P(x)/ f_0\right)\), and
$f_0$ is the Goldstone boson decay constant in the chiral limit.
These fields transform under the chiral  group 
\(\mathrm{SU}(3)_\mathrm L \times \mathrm{SU}(3)_\mathrm R\) as
\( U \rightarrow RUL^\dagger\) and
\(B \rightarrow K B K^\dagger\)
with \(L\in\mathrm{SU}(3)_\mathrm L\,, R\in\mathrm{SU}(3)_\mathrm R\) and the SU(3) 
valued compensator field \(K=K(L,R,U)\), cf.\ Ref.~\cite{Bernard1995}.
After an expansion of the interaction Lagrangian in powers of \(P\) one obtains from the terms proportional to \(D\) and \(F\) the pseudovector coupling term
\begin{equation} \label{eq:pseudovector}
  \mathcal{L}_1 = -\frac{\sqrt2}{2f_0}\tr\left(D \bar B \gamma^\mu\gamma_5\left\{\partial_\mu P,B\right\} + F \bar B \gamma^\mu\gamma_5\left[\partial_\mu P,B\right]\right) \,,\\
 \end{equation}
which leads to a vertex between two baryons and one meson.
In the same way, the term involving the chiral connection \(\Gamma_\mu\) gives
\begin{equation}
  \mathcal{L}_2 = \frac1{4f_0^2}\tr\left(\mathrm i \bar B \gamma^\mu \left[\left[P,\partial_\mu P\right],B\right]\right)\,,\\
\end{equation}
which describes a (Weinberg-Tomozawa) vertex between two baryons and two mesons.

When writing the pseudovector interaction Lagrangian \(\mathcal L _1\) explicitly in the isospin basis, 
one gets
\begin{eqnarray}
{\mathcal L_1}&=&-f_{NN\pi}\bar{N}\gamma^\mu\gamma_5\mbox{\boldmath $\tau$}N\cdot\partial_\mu\mbox{\boldmath $\pi$} +if_{\Sigma\Sigma\pi
}\bar{\mbox{\boldmath $ \Sigma$}}\gamma^\mu\gamma_5\times{\mbox{\boldmath $ \Sigma$}}\cdot\partial_\mu\mbox{\boldmath $\pi$} \nonumber \\
&&-f_{\Lambda\Sigma\pi}\left[\bar{\Lambda}\gamma^\mu\gamma_5{\mbox{\boldmath $ \Sigma$}}+\bar{\mbox{\boldmath $\Sigma$}}\gamma^\mu
\gamma_5\Lambda\right]\cdot\partial_\mu\mbox{\boldmath $\pi$}-f_{\Xi\Xi\pi}\bar{\Xi}\gamma^\mu\gamma_5\mbox{\boldmath $\tau$}\Xi\cdot
\partial_\mu\mbox{\boldmath $\pi$} \nonumber \\
&&-f_{\Lambda NK}\left[\bar{N}\gamma^\mu\gamma_5\Lambda\partial_\mu K+ {\rm h.c.}\right]
-f_{\Xi\Lambda K}\left[\bar{\Xi}\gamma^\mu\gamma_5\Lambda\partial_\mu \Kb + {\rm h.c.} \right]
\nonumber \\&&
-f_{\Sigma NK}\left[\bar{N}\gamma^\mu\gamma_5\mbox{\boldmath $\tau$}\partial_\mu K\cdot{\mbox{\boldmath $ \Sigma$}}
+{\rm h.c.}\right]
-f_{\Sigma \Xi K}\left[\bar{ \Xi}\gamma^\mu\gamma_5\mbox{\boldmath $\tau$}\partial_\mu \Kb\cdot{\mbox{\boldmath $ \Sigma$}}
+{\rm h.c.}\right]
\nonumber \\&&
-f_{NN\eta_8}\bar{N}\gamma^\mu\gamma_5N\partial_\mu\eta
-f_{\Lambda\Lambda\eta_8}\bar{\Lambda}\gamma^\mu\gamma_5\Lambda\partial_\mu\eta
\nonumber \\&&
-f_{\Sigma\Sigma\eta_8}\bar{\mbox{\boldmath $ \Sigma$}}
\cdot\gamma^\mu\gamma_5{\mbox{\boldmath $ \Sigma$}}\partial_\mu\eta
-f_{\Xi\Xi\eta_8}\bar{\Xi}\gamma^\mu\gamma_5\Xi\partial_\mu\eta \ .
\label{eq:3.7}
\end{eqnarray}
Here, we have introduced the isospin doublets
\begin{equation}
N=\left(\begin{array}{r}p\\n\end{array}\right)\ ,\ \ \Xi=\left(\begin{array}{r}\Xi^0\\\Xi^-\end{array}\right)\ ,\ \ 
K=\left(\begin{array}{r}K^+\\K^0\end{array}\right)\ ,\ \  \Kb=\left(\begin{array}{r}\Kb^0\\-K^-\end{array}\right)\ .
\label{eq:3.8}
\end{equation}
The signs have
been chosen according to the conventions of Ref.~\cite{Swa63}, 
such that the inner product of the isovector $\mbox{\boldmath $\Sigma$}$ 
(or $\mbox{\boldmath $\pi$}$) defined in spherical components reads
\begin{equation}
\mbox{\boldmath $\Sigma$}\cdot\mbox{\boldmath $\Sigma$}=\sum_m (-1)^m\Sigma_m\Sigma_{-m}=
\Sigma^+\Sigma^-+\Sigma^0\Sigma^0+\Sigma^-\Sigma^+\ .
\label{eq:A2.8}
\end{equation}

Since the original interaction Lagrangian in Eq.~(\ref{eq:pseudovector}) is SU(3)-invariant, the various coupling constants are related to each other by \cite{Swa63} 
\begin{equation}
\begin{array}{rlrlrl}
f_{NN\pi}  = & f, & f_{NN\eta_8}  = & \frac{1}{\sqrt{3}}(4\alpha -1)f, & f_{\Lambda NK} = & -\frac{1}{\sqrt{3}}(1+2\alpha)f, \\
f_{\Xi\Xi\pi}  = & -(1-2\alpha)f, &  f_{\Xi\Xi\eta_8}  = & -\frac{1}{\sqrt{3}}(1+2\alpha )f, & f_{\Xi\Lambda K} = & \frac{1}{\sqrt{3}}(4\alpha-1)f, \\
f_{\Lambda\Sigma\pi}  = & \frac{2}{\sqrt{3}}(1-\alpha)f, & f_{\Sigma\Sigma\eta_8}  = & \frac{2}{\sqrt{3}}(1-\alpha )f, & f_{\Sigma NK} = & (1-2\alpha)f, \\
f_{\Sigma\Sigma\pi}  = & 2\alpha f, &  f_{\Lambda\Lambda\eta_8}  = & -\frac{2}{\sqrt{3}}(1-\alpha )f, & f_{\Xi\Sigma K} = & -f.
\end{array}
\label{su3}
\end{equation}
Evidently, all coupling constants are given in terms of $f\equiv g_A/2f_0$ and the ratio $\alpha=F/(F+D)$.
 
The expression for the one--pseudoscalar-meson exchange potential is similar to the 
standard one-pion-exchange potential \cite{Epe00}  
\begin{eqnarray}
V^{OBE}_{B_1B_2\to B_3B_4} 
&=&-f_{B_1B_3P}f_{B_2B_4P}\frac{\left(\mbox{\boldmath $\sigma$}_1\cdot{\bf q}\right)
\left(\mbox{\boldmath $\sigma$}_2\cdot{\bf q}\right)}{{\bf q}^2+m^2_P}\,{\mathcal I}_{B_1B_2\to B_3B_4}\ .
\label{OBE}
\end{eqnarray}
Here, $m_P$ is the mass of the exchanged pseudoscalar meson. 
In the present calculation we use the physical masses $m_\pi,m_K,m_\eta$ in Eq.~(\ref{OBE}). Thus, the explicit ${\rm SU(3)}$ breaking reflected in the mass splitting 
between the pseudoscalar mesons is taken into account. The $\eta$ meson is identified with the 
octet-state $\eta_8$.
The isospin factors ${\mathcal I}_{B_1B_2\to B_3B_4}$ are
given in Tab.~\ref{tab:3.1}.

\begin{table}[t]
\caption{Isospin factors ${\mathcal I}$ for the various one--pseudoscalar-meson exchanges.}
\label{tab:3.1}
\vskip 0.1cm
\renewcommand{\arraystretch}{1.2}
\centering
\begin{tabular}{|c|c|c|r|r|r|}
\hline
&Channel &Isospin &$\pi$ &$K$ &$\eta$\\
\hline
$S=0$ &$NN\rightarrow NN$ &$0$ &$-3$ &$0$ &$1$ \\
&                  &$1$ &$1$  &$0$ &$1$ \\
\hline
$S=-1$ &$\Lambda N\rightarrow \Lambda N$ &$\frac{1}{2}$ &$0$ &$1$ &$1$ \\
&$\Lambda N\rightarrow \Sigma N$ &$\frac{1}{2}$ &$-\sqrt{3}$ &$-\sqrt{3}$ &$0$ \\
&$\Sigma N\rightarrow \Sigma N$ &$\frac{1}{2}$ &$-2$ &$-1$ &$1$ \\
&                              &$\frac{3}{2}$ &$1$ &$2$ &$1$ \\
\hline
\end{tabular}
\renewcommand{\arraystretch}{1.0}
\end{table}

The two--pseudoscalar-meson exchange potential, built up by a set of one-loop diagrams, is described in detail in 
Appendix A.
Relativistic corrections to the one-meson exchange potential that arise at NLO  
due to differences of the baryon masses are discussed in Appendix~B. 

\subsection{Scattering equation}

In the actual calculation a partial-wave projection of the interaction 
potentials is performed, as described in detail in Ref.~\cite{Polinder:2006zh}. 
The reaction amplitudes are obtained from the solution of a coupled-channel 
Lippmann-Schwinger (LS) equation: 
\begin{eqnarray}
&&T^{\rho''\rho',J}_{\nu''\nu'}(p'',p';\sqrt{s})=V^{\rho''\rho',J}_{\nu''\nu'}(p'',p')+
\sum_{\rho,\nu}\int_0^\infty \frac{dpp^2}{(2\pi)^3} \, V^{\rho''\rho\, ,J}_{\nu''\nu}(p'',p)
\frac{2\mu_{\nu}}{q_{\nu}^2-p^2+i\eta}T^{\rho\rho',J}_{\nu\nu'}(p,p';\sqrt{s})\ .
\label{LS} 
\end{eqnarray}
Here,
the label $\nu$ indicates the particle channels and the label $\rho$ the partial wave. $\mu_\nu$ 
is the pertinent reduced baryon mass. The on-shell momentum $q_{\nu}$ in the intermediate state, is 
determined by $\sqrt{s}=\sqrt{M^2_{B_{1,\nu}}+q_{\nu}^2}+\sqrt{M^2_{B_{2,\nu}}+q_{\nu}^2}$. 
Relativistic kinematics is used for relating the laboratory momentum $p_{{\rm lab}}$ of the hyperons 
to the center-of-mass momentum. 
 
We solve the LS equation in the particle basis, in order to incorporate the correct physical
thresholds. Depending on the total charge, up to three baryon-baryon channels can couple. 
The Coulomb interaction is taken into account appropriately via the Vincent-Phatak 
method \cite{VP}.
The potentials in the LS 
equation are cut off with a regulator function, $f_R(\Lambda) =
\exp\left[-\left(p'^4+p^4\right)/\Lambda^4\right]$, 
in order to remove high-energy components \cite{Epe05}.
We consider cutoff values in the range $\Lambda=450$ -- $700\,$MeV, similar to what 
was used for chiral $NN$ potentials \cite{Epe05}, but anticipate here already 
that the best results are achieved for cutoffs located in the interval 
$500$ -- $650$ MeV.

\section{Fitting procedure}
\label{sec:3}

For the fitting procedure we consider the same ``standard'' set of 36 $YN$ data points 
that have been used in our previous works \cite{Polinder:2006zh,Hai05} as also done by the Nijmegen group in their investigations \cite{Rij99}.  
This data set consists of low-energy total cross sections for the reactions:
$\Lambda p \to \Lambda p$ 
from Ref.~\cite{Sec68} (6 data points) and Ref.~\cite{Ale68} (6 data points),
$\Sigma^- p \to \Lambda n$ 
\cite{Eng66} (6 data points), 
$\Sigma^- p \to \Sigma^0 n$ 
\cite{Eng66} (6 data points), 
$\Sigma^- p \to \Sigma^- p$ 
\cite{Eis71} (7 data points), 
$\Sigma^+ p \to \Sigma^+ p$ 
\cite{Eis71} (4 data points), 
and the inelastic capture ratio at rest \cite{Hep68,Ste70}.
Besides these $YN$ data the empirical binding energy of the hypertriton $^3_\La \rm H$
is used as a further constraint. 
Otherwise it would not be possible to fix the relative strength of the
spin-singlet and spin-triplet $S$-wave contributions to the $\Lambda p$ interaction. 

We recall that there are in total five independent 
LECs at LO, that describe the $NN$ and $YN$ interactions, see Tab.~\ref{tab:SU3}. 
In Ref.~\cite{Polinder:2006zh} a fit to the $YN$ scattering data at LO
was presented utilizing these five contact terms. It turned out that 
already in that scenario a fairly reasonable description 
of the 36 low-energy $YN$ scattering data could be achieved for cutoffs
$\Lambda=550 - 700$ MeV and for natural values of the LECs. 
At NLO there are eight new contact terms contributing to the $S$-waves and
the $^3S_1-{^3D_1}$ transition, and ten in the $P$-waves. 
As described in Sect.~\ref{sec:2CT}, we impose ${\rm SU(3)}$ flavor symmetry in 
order to reduce the number of LECs that need to be determined. Without implementing
this constraint there would be 20 independent contact terms for the 
$\Lambda N$ and $\Sigma N$ systems in the $S$-waves (and the $S$-$D$ transitions) 
alone, and, given the low number of data points together with their large error bars, 
it is simply impossible to fix all those LECs by a fit to the available empirical information. 

In the actual fitting procedure, first, we have considered only the 13 LECs in the
$S$-waves and the $S$-$D$ transitions. After all, the available $YN$ data consist 
practically only of total cross sections at low energies and these are predominantly 
determined by the $S$-wave amplitudes. The fits are performed for fixed values 
of the cutoff scale where we started with $\Lambda = 600$ MeV. The subsequent fits 
for other cutoffs were done under the constraint that the results should
stay as close as possible to those obtained with $\Lambda = 600$ MeV, for the
singlet and triplet cross sections separately.
This procedure is demanded by our requirement to reproduce the hypertriton as mentioned
above. 

Note that for the $\Sigma^+ p\to \Sigma^+ p$ and $\Sigma^- p\rightarrow \Sigma^- p$
channels the experimental total cross sections were obtained by incomplete angular coverage \cite{Eis71}:
\begin{eqnarray}
\label{eq:sigtot}
\sigma&=&\frac{2}{\cos \theta_{{\rm max}}-\cos \theta_{{\rm min}}}
\int_{\cos \theta_{{\rm min}}}^{\cos \theta_{{\rm max}}}\frac{d\sigma(\theta)}{d\cos \theta}d\cos \theta \ .
\end{eqnarray}
Following Ref.~\cite{Rij99}, 
we use $\cos \theta_{{\rm min}}=-0.5$ and $\cos \theta_{{\rm max}}=0.5$ in our calculations 
for the $\Sigma^+ p\rightarrow \Sigma^+ p$ and $\Sigma^- p\rightarrow \Sigma^- p$ cross sections, 
in order to stay as close as possible to the experimental procedure.
The total cross sections for the other channels are evaluated by simply
integrating the differential cross sections over the whole angular region.

For the capture ratio at rest, $r_R$, we follow the definition of Ref.~\cite{Swa62}:
\begin{equation}
   r_R=\frac{1}{4}\,\frac{\sigma_s(\Sigma^-p\rightarrow\Sigma^0n)}
                    {\sigma_s(\Sigma^-p\rightarrow\Lambda n)
                    +\sigma_s(\Sigma^-p\rightarrow\Sigma^0n)}
      +\frac{3}{4}\,\frac{\sigma_t(\Sigma^-p\rightarrow\Sigma^0n)}
                    {\sigma_t(\Sigma^-p\rightarrow\Lambda n)
                    +\sigma_t(\Sigma^-p\rightarrow\Sigma^0n)}\,,
\label{rR}
\end{equation}
where $\sigma_s$ is the total reaction cross section in the singlet
$^1S_0$ partial wave, and $\sigma_t$ the total reaction cross section
in the triplet-coupled $^3S_1$-$^3D_1$ partial waves. The cross sections 
are the ones at zero momentum, but following common practice \cite{Rij99} we 
evaluate the cross sections at a small non-zero momentum, 
namely $p_{\rm lab}=10\,$MeV/$c$. 

While the $\chi^2$ fit to the 36 data points allowed us to fix the majority of 
the $S$-wave LECs it turned out that concerning the $^3S_1$ partial wave in the $I=3/2$ $\Sigma N$ channel,
solutions with either a positive (attractive) or a negative (repulsive)
phase shift are possible. This can be understood from the SU(3) structure as given
in Table~\ref{tab:SU3} which shows that this partial wave is controlled by the 
``isolated'' $10$ representation such that the corresponding LECs do not 
enter in any of the other $YN$ channels. 
We adopt here the repulsive solution in accordance with evidence from 
recently measured $(\pi^-,K^+$) inclusive spectra related to $\Sigma^-$-formation 
in heavy nuclei \cite{SIG1,SIG2,SIG3},
which suggest a repulsive $\Sigma$-nucleus single-particle potential 
\cite{Kohno06,Dab08}. 

We should also mention that we observe some correlations between the values of 
the $S$-wave LECs at LO and NLO, i.e.\ $\tilde C$ and $C$ in 
Eqs.~(\ref{C1S0}) and (\ref{C3S1}). This is a consequence of the fact that the
fitted $\Sigma N$ cross sections lie all within a rather narrow energy 
interval near threshold so that there is only a fairly weak sensitivity to the momentum-dependent
$(p^2 + p'^2)$ terms. 
The $\Lambda p$ cross sections alone, which
are known over a larger energy range, are not sufficient for separating the strength of $\tilde C$ and $C$.

\begin{table}
\caption{The $YN$ contact terms for the $^1S_0$ and $^3S_1$-$^3D_1$
partial waves for various cut--offs. The values of the $\tilde C$'s are in 
$10^4$ ${\rm GeV}^{-2}$ the ones of the $C$'s in $10^4$ ${\rm GeV}^{-4}$; 
the values of $\Lambda$ in MeV.
}
\renewcommand{\arraystretch}{1.2}
\label{tab:F1}
\vspace{0.2cm}
\centering
\begin{tabular}{|c|c|rrrrrr|}
\hline
\multicolumn{2}{|c|}{$\Lambda$} & $450$ & $500$ & $550$& $600$& $650$& $700$  \\
\hline
$^1S_0$ 
&$\tilde C^{27}_{^1S_0}$   &$-0.0893$ &$-0.0672$  &$0.00648$ &$0.1876$ &$0.6140$ &$1.145$\\
&$\tilde C^{8_s}_{^1S_0}$  &$0.2000$ &$0.1970$  &$0.1930$ &$0.1742$ &$0.1670$ &$0.1730$\\
&$C^{27}_{^1S_0}$          &$1.500$ &$1.800$  &$2.010$ &$2.200$ &$2.400$ &$2.410$\\
&$C^{8_s}_{^1S_0}$         &$-0.200$ &$-0.200$  &$-0.206$ &$-0.0816$ &$-0.0597$ &$0.1000$\\
\hline
$^3S_1$-$^3D_1$ 
&$\tilde C^{10}_{^3S_1}$   &$0.104$ &$0.541$  &$0.149$  &$0.344$  &$0.499$  &$0.560$\\
&$\tilde C^{10^*}_{^3S_1}$ &$0.171$ &$0.209$  &$0.635$  &$1.420$  &$2.200$  &$2.960$\\
&$\tilde C^{8_a}_{^3S_1}$  &$0.0218$ &$0.00715$ &$-0.0143$ &$-0.0276$  &$-0.0269$  &$0.00173$\\
&$C^{10}_{^3S_1}$          &$2.240$ &$2.310$  &$2.450$ &$2.740$  &$2.530$  &$2.030$\\
&$C^{10^*}_{^3S_1}$        &$0.310$ &$0.143$  &$0.741$ &$1.090$  &$1.150$  &$1.120$\\
&$C^{8_a}_{^3S_1}$         &$0.373$ &$0.469$  &$0.627$ &$0.775$  &$0.854$  &$0.964$\\
&$C^{10}_{^3S_1-\,^3D_1}$  &$-0.360$ &$-0.429$  &$-0.428$  &$-0.191$  &$-0.191$  &$-0.122$\\
&$C^{10^*}_{^3S_1-\,^3D_1}$&$-0.300$ &$-0.300$  &$-0.356$ &$-0.380$  &$-0.380$  &$-0.228$\\
&$C^{8_a}_{^3S_1-\,^3D_1}$ &$0.0356$ &$0.0475$  &$0.0453$ &$-0.00621$  &$-0.00621$  &$-0.0497$\\
\hline
\end{tabular}
\renewcommand{\arraystretch}{1.0}
\end{table}
\begin{table}
\caption{The $YN$ contact terms for the $P$-waves for various cut--offs. 
The values of the LECs are in 
$10^4$ ${\rm GeV}^{-4}$; the values of $\Lambda$ in MeV.
}
\renewcommand{\arraystretch}{1.2}
\label{tab:F2}
\vspace{0.2cm}
\centering
\begin{tabular}{|c|c|rrrrrr|}
\hline
\multicolumn{2}{|c|}{$\Lambda$} & $450$ & $500$ & $550$& $600$& $650$& $700$  \\
\hline
$^3P_0$   
&$C^{27}_{^3P_0}$  &$1.47$ &$1.49$  &$1.51$ &$1.55$ &$1.60$ &$1.71$\\
&$C^{8_s}_{^3P_0}$ &$2.50$ &$2.50$  &$2.50$ &$2.50$ &$2.50$ &$2.50$\\
$^3P_1$  
&$C^{27}_{^3P_1}$  &$-0.43$ &$-0.43$  &$-0.43$ &$-0.43$ &$-0.43$ &$-0.43$\\
&$C^{8_s}_{^3P_1}$ &$0.65$ &$0.65$  &$0.65$ &$0.65$ &$0.65$ &$0.65$\\
$^3P_2$   
&$C^{27}_{^3P_2}$  &$-0.096$ &$-0.063$  &$-0.041$ &$-0.025$ &$-0.012$ &$0.000$\\
&$C^{8_s}_{^3P_2}$ &$1.00$ &$1.00$  &$1.00$ &$1.00$ &$1.00$ &$1.00$\\
\hline
$^1P_1$   
&$C^{10}_{^1P_1}$  &$0.49$ &$0.49$  &$0.49$ &$0.49$  &$0.49$  &$0.49$\\
&$C^{10^*}_{^1P_1}$&$-0.14$ &$-0.14$  &$-0.14$ &$-0.14$  &$-0.14$  &$-0.14$\\
&$C^{8_a}_{^1P_1}$ &$-0.65$ &$-0.60$  &$-0.58$ &$-0.56$  &$-0.54$  &$-0.52$\\
\hline
$^1P_1$-$^3P_1$   
&$C^{8_s8_a}_{^1P_1-\,^3P_1}$ &$0$ &$ 0$  &$ 0$ &$ 0$  &$0$  &$0$\\
\hline
\end{tabular}
\renewcommand{\arraystretch}{1.0}
\end{table}

The limited number (and quality) of differential cross sections and the complete 
lack of polarization observables makes a determination of the contact terms in
the $P$-waves from $YN$ data practically impossible. 
Therefore, in this case and in line with the power counting we assume 
strict SU(3) symmetry for the contact terms and use the $NN$ $P$-wave phase 
shifts as a further constraint.
In particular, we fix the LECs $C^{27}$ and $C^{10^*}$
from fitting to empirical $^1P_1$, $^3P_0$, and $^3P_1$ $NN$-phase shifts 
\cite{Stoks} in the region of $25 \leq T_{\rm lab} \leq 50$ MeV \cite{Epe00}. 
The $^3P_2$ partial wave is special. Here a NLO calculation with the
pertinent LEC determined from the low-energy region yields results at
higher energies that strongly overestimate the empirical phase shifts,
see, e.g.\ Ref.~\cite{Epe05}. Such a LEC would likewise lead to a considerable
overestimation of the $\La p$ cross section around and above the $\Si N$ 
threshold. In order to avoid this we fix this specific LEC from the 
$NN$-phase shifts in an energy region corresponding to the $\Si N$
threshold, namely $T_{\rm lab} \approx 150$ MeV. 

Utilizing the $NN$-phase shifts reduces the number of $P$-wave contact terms that 
need to be determined in the $YN$ sector by roughly a factor two. 
Here the most important constraint is provided by the 
$\Lambda p$ cross section above the $\Sigma N$ threshold, i.e.\ at around
$p_{\rm lab}\approx 800$ MeV/c, which is roughly 10~$mb$ according to 
experiments \cite{Kad71,Hau77}. Agreement with these data 
can be only achieved if the contributions from each of the $P$-waves
($^1P_1$, $^3P_0$, $^3P_1$, $^3P_2$) is kept small, which means in turn
that the corresponding phase shifts have to be small. The differential
cross section for $\Sigma^- p \to \Lambda n$ has been measured at two
energies near the $\Sigma N$ threshold and it is sensitive to the 
$P$-waves, too. 
We used this empirical information to fix the 
remaining six $P$-wave contact terms. But it should be said that
this information is not sufficient to pin them down reliably.
Note that we set the LEC corresponding to the $^1P_1$-$^3P_1$ transition 
to zero. 

Besides of contact terms with LECs that need to be determined
in a fit to data the potential includes also 
contributions from one-meson and two-meson exchanges.
The latter do not involve any free parameters. The coupling constant 
$f\equiv g_A/2f_0$ is fixed by utilizing the values $g_A= 1.26$ and 
$f_0 \approx f_\pi = 93$~MeV. For the $F/(F+D)$-ratio we adopt the 
${\rm SU(6)}$ value $\alpha=0.4$. It is close to the empirical value of 
$\alpha \approx 0.36 - 0.37$, as determined recently in analyses of
hyperon semi-leptonic decay data \cite{Ratcliffe,Yamanishi}. 

\begin{figure}[t]
\begin{center}
\includegraphics[height=64mm]{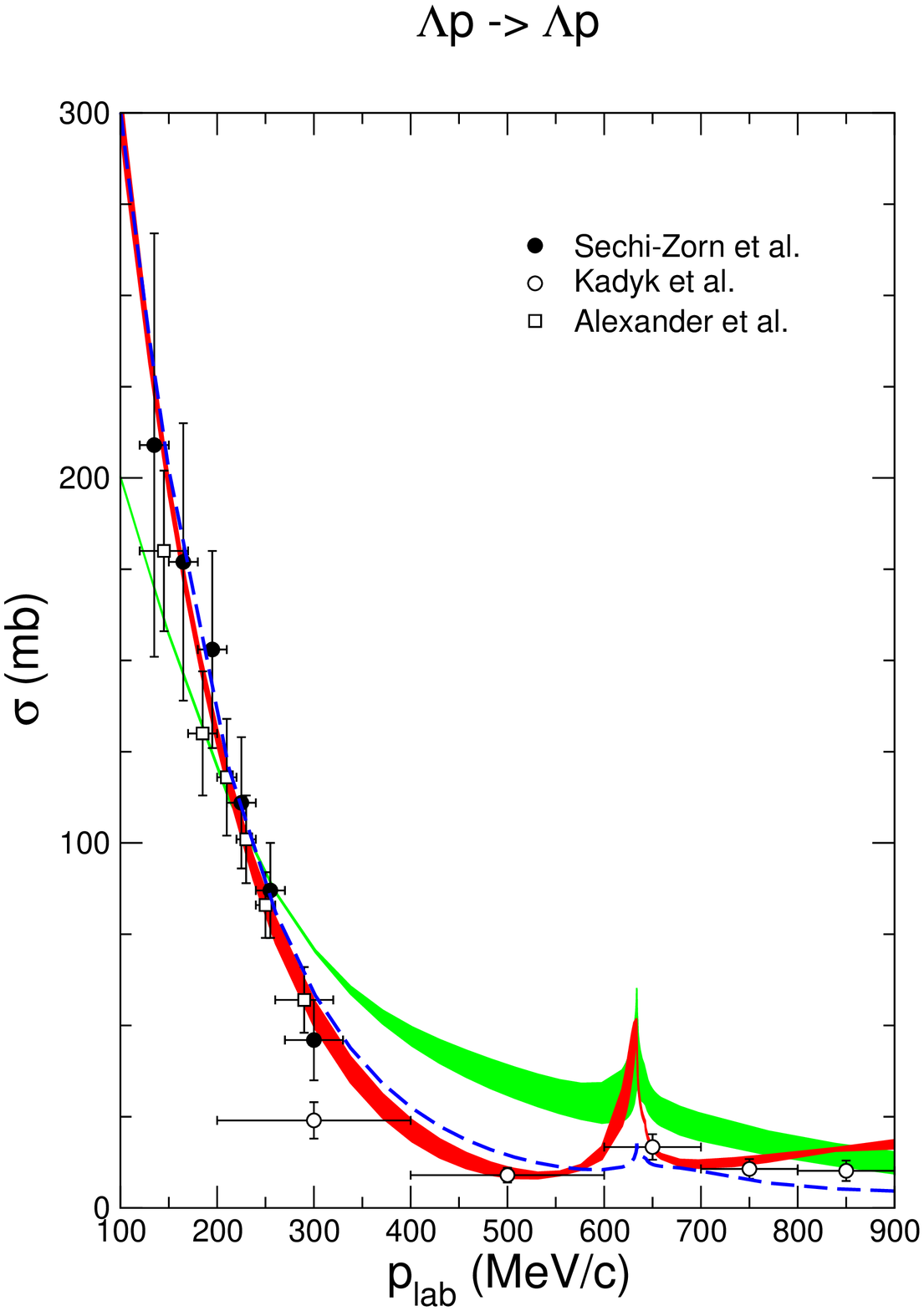}
\includegraphics[height=64mm]{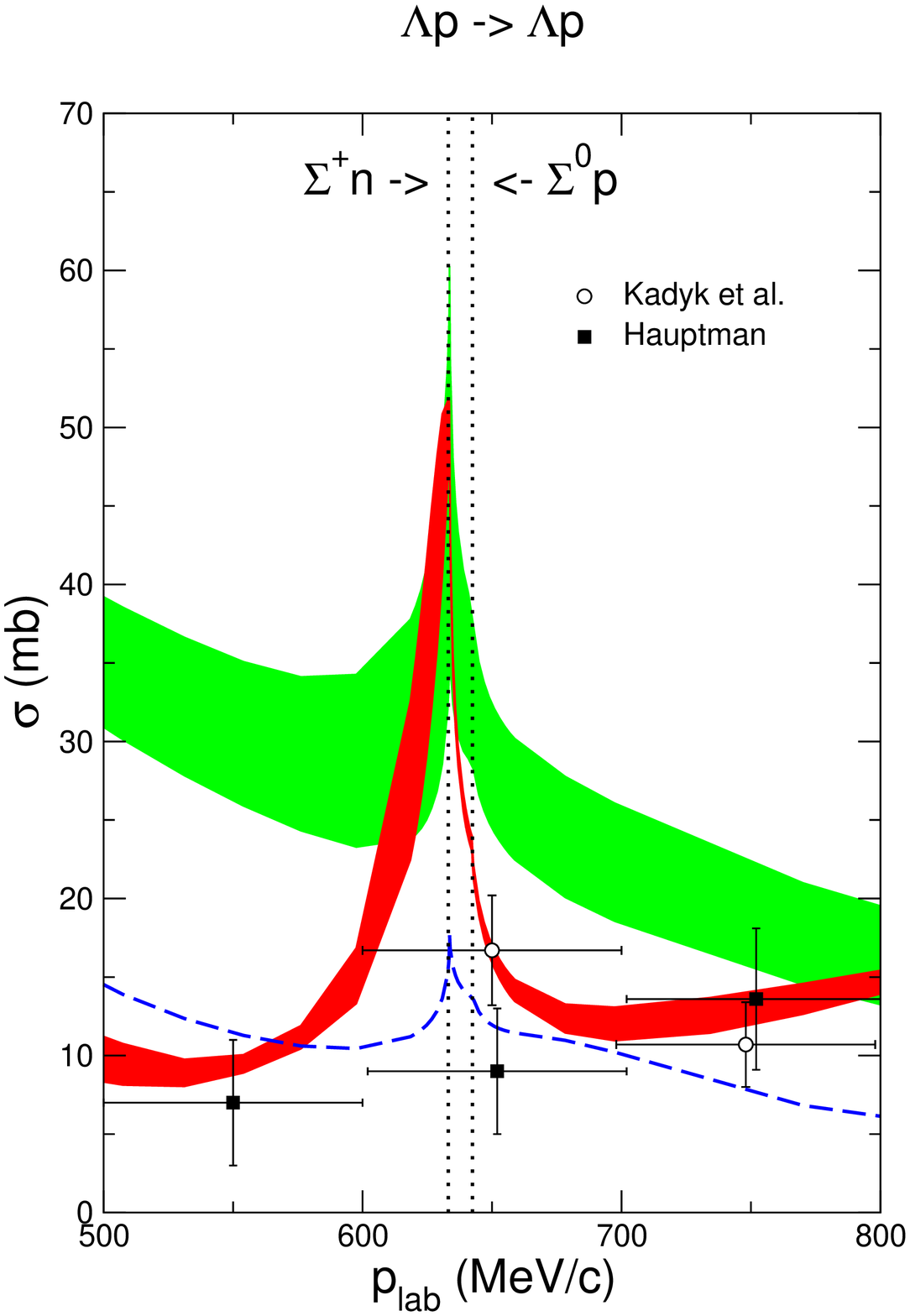}
\includegraphics[height=64mm]{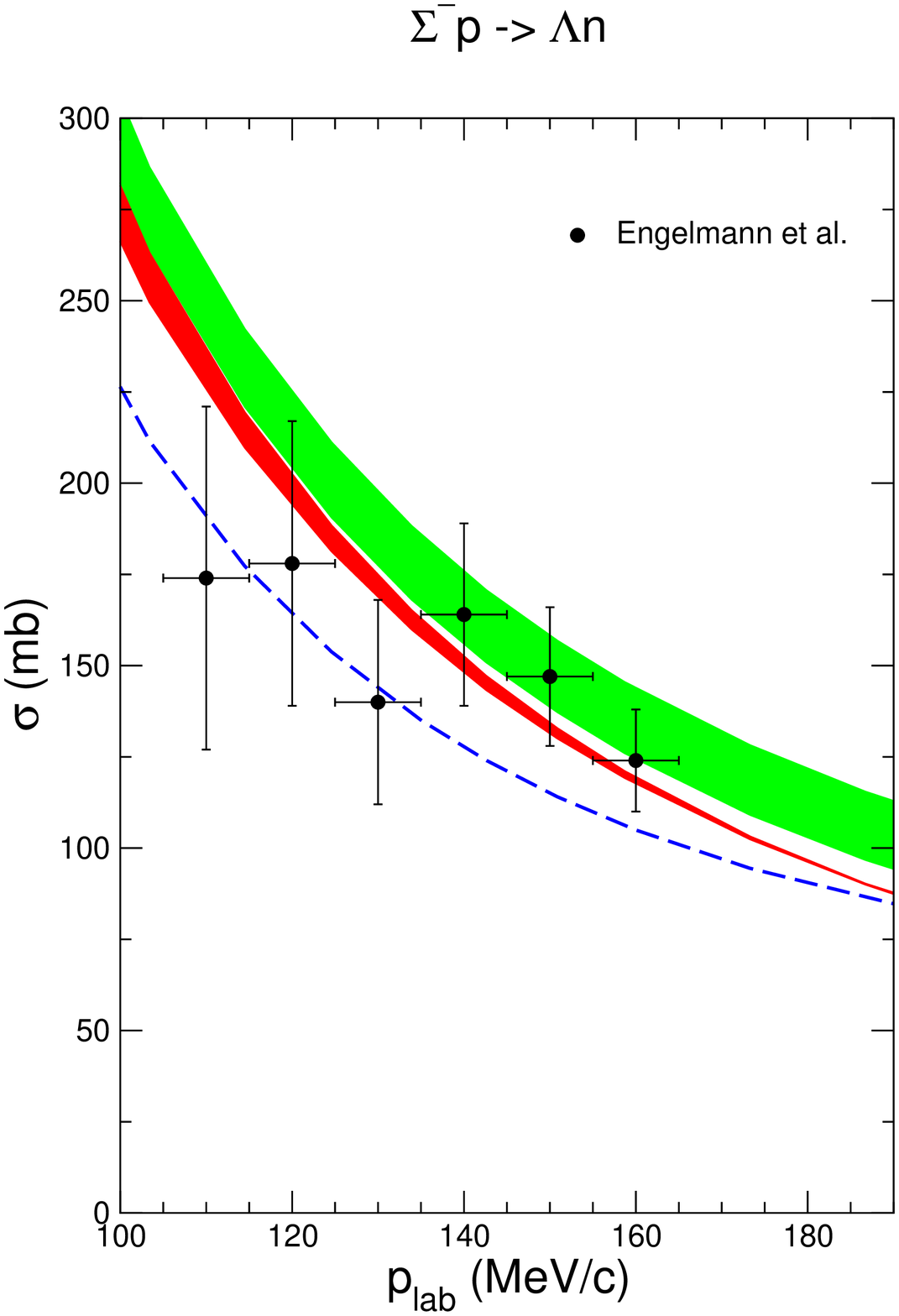}

\includegraphics[height=64mm]{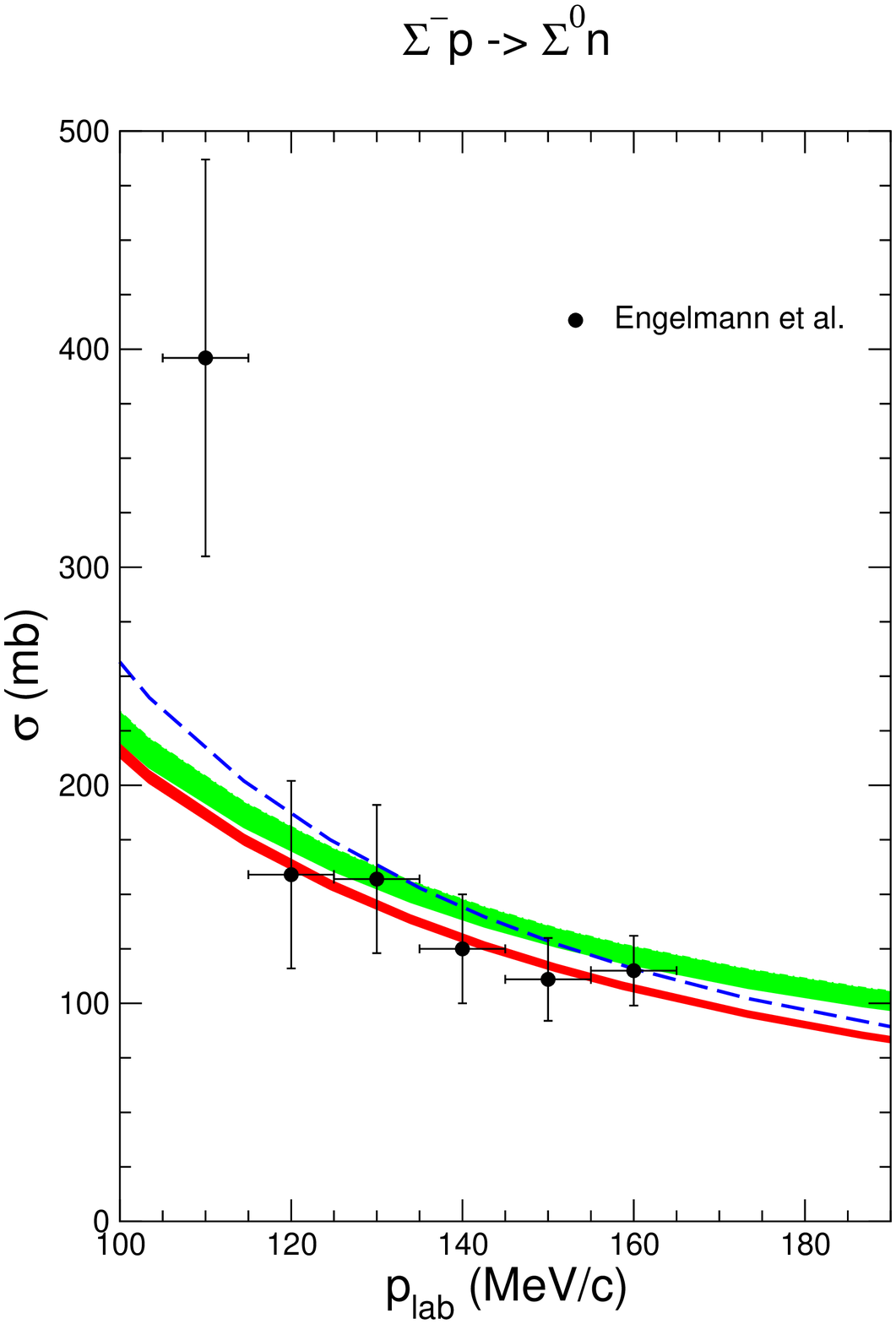}
\includegraphics[height=64mm]{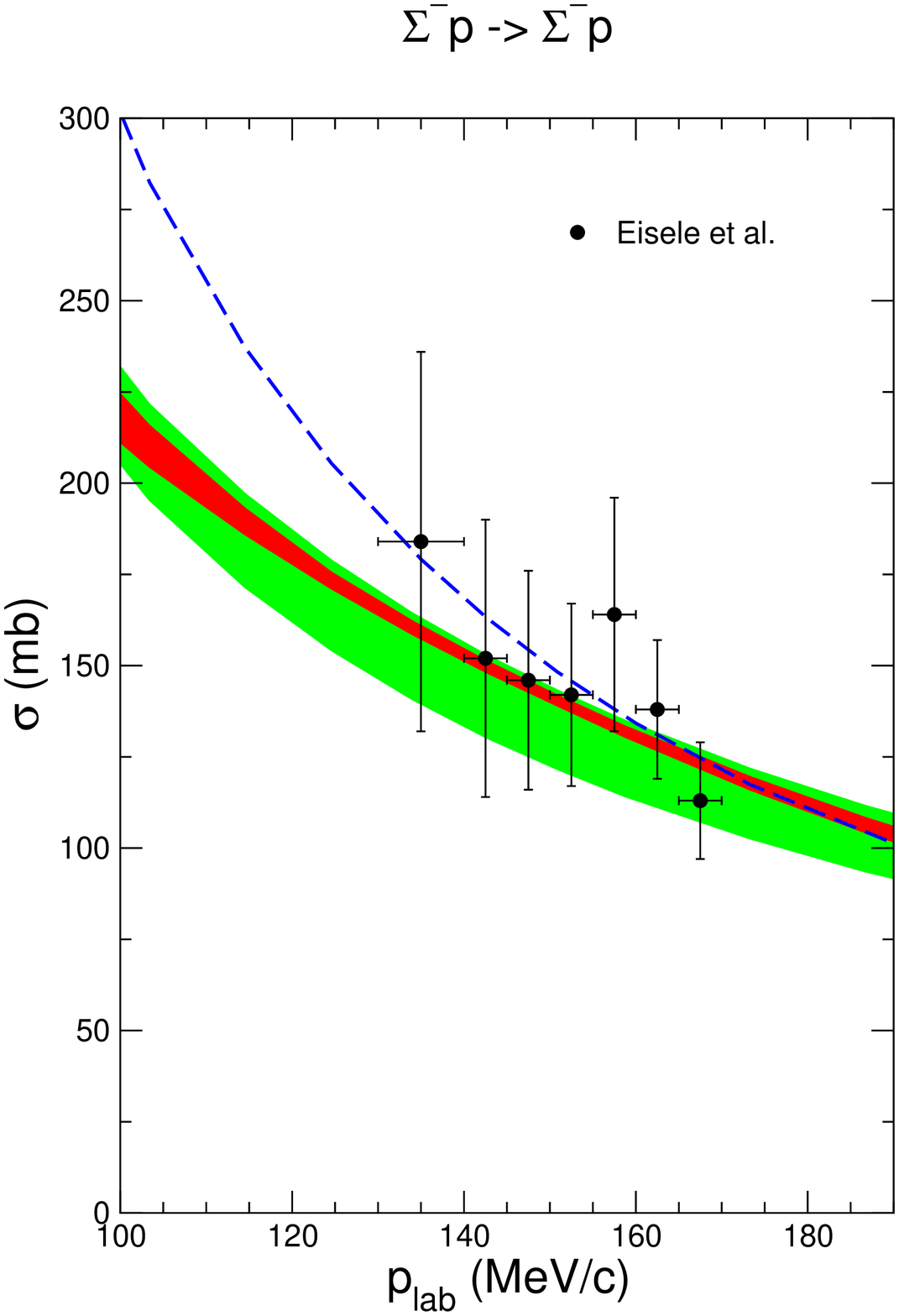}
\includegraphics[height=64mm]{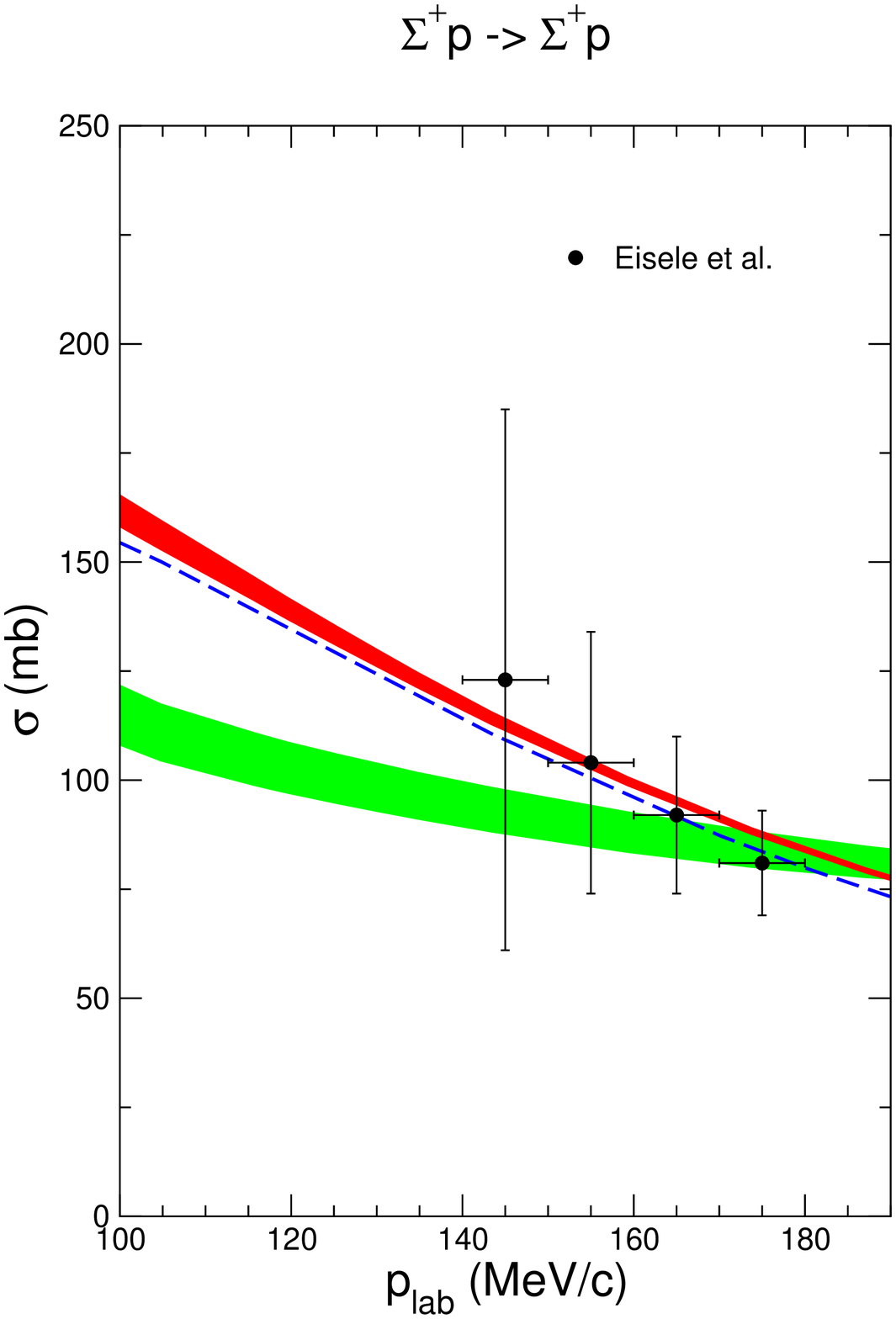}
\caption{
''Total'' cross section $\sigma$ (as defined in Eq.~(\ref{eq:sigtot}))
as a function of $p_{{\rm lab}}$.
The experimental cross sections are taken from Refs.~\cite{Sec68} (filled circles),
\cite{Ale68} (open squares), \cite{Kad71} (open circles), and \cite{Hau77} (filled squares)
($\Lambda p \to \Lambda p$), 
from \cite{Eng66} ($\Sigma^-p \to \Lambda n$, $\Sigma^-p \to \Sigma^0 n$)
and from \cite{Eis71} ($\Sigma^-p \to \Sigma^- p$, $\Sigma^+p \to \Sigma^+ p$). 
The red/dark band shows the chiral EFT results to NLO for variations of the cutoff
in the range $\Lambda =$ 500,$\ldots$,650~MeV, 
while the green/light band are results to LO for $\Lambda =$ 550,$\ldots$,700~MeV. 
The dashed curve is the result of 
the J{\"u}lich '04 meson-exchange potential \cite{Hai05}.
}
\label{fig:R1}
\end{center}
\end{figure}

In the spirit of the imposed SU(3) symmetry we keep all contributions
from one- and two-meson exchanges, i.e.\ also those from $\pi \eta$, $\eta K$, $K K$ exchange. The large mass splitting between the Goldstone bosons 
induces a sizable SU(3) breaking in the actual $YN$ potential that is
taken into account in our calculation. 
There is also an explicit SU(3) symmetry breaking in the coupling
constants as reflected in the empirical values of the decay 
constants $f_\pi$, $f_K$ and $f_\eta$ \cite{PDG}
which, in principle, should be taken into account in the NLO calculation. 
However, we have ignored such effects in the results reported in the present paper.
As a matter of fact, we have explored various scenarios in the course
of our investigation.
In particular, we performed fits based on the empirical
decay constants. We explored also the situation when only $\pi\pi$ exchange
diagrams are kept and all two-meson exchanges involving the heavier $K,\eta$ mesons 
are omitted. Furthermore, in Ref.~\cite{Hai10} we had presented results based 
on an incomplete NLO calculation, i.e.\ where only the NLO contact terms were 
taken into account but no two-meson--exchange contributions. 
In all these cases a comparable description of the $YN$ data could
be achieved, i.e.\ with a $\chi^2$ within 10-15\% of the best values achieved.
Seemingly, all two-meson exchange effects could be absorbed into the LECs 
and, moreover, one could still maintain SU(3) symmetry for these contact terms. 
A further uncertainty in our calculation is the value of the $\eta$-baryon coupling, 
since we identified the physical $\eta$ with the octet $\eta_8$. In our earlier
investigation \cite{Polinder:2006zh} we varied the $\eta$ coupling between zero and its 
octet value and we found very little influence on the description of the data.
Thus, we refrain from introducing a singlet coupling in the present study. 
It is possible though that future calculations of hypernuclei based on these chiral 
EFT interactions could indicate a preference for one or the other scenarios 
and/or yield evidence for the need of an explicit SU(3) breaking in the
contact terms. 
 
Finally, let us mention that we did consider also the $S$-waves for $NN$ scattering. 
In particular, we fixed the pertinent five LECs from a fit to the $np$ $^1S_0$ 
and $^3S_1$-$^3D_1$ phase shifts and mixing parameter in the energy 
range $T_{lab} \le 50$ MeV, independently of the $YN$ interaction.
Thereby it turned out that the LECs determined from the
$NN$ phase shifts are incompatible with those needed for the 
description of the $YN$ data with regard to the SU(3) symmetry. 
The most obvious case is the $^1S_0$ partial wave, where SU(3) symmetry 
implies that $V_{NN} \equiv V_{\Sigma N}$ (I=3/2), 
see Tab.~\ref{tab:SU3}, so that the (hadronic part of the) interaction in the 
$\Sigma^- n$ and $\Sigma^+ p$ channels is unambiguously fixed once the LECs 
are determined from the $np$ phases. However, with LECs determined from the 
latter channel a near-threshold bound state is obtained in the $\Sigma^+ p$ 
system and, as a consequence, the
empirical $\Sigma^+ p\to \Sigma^+ p$ cross section is grossly overestimated. 
This happens despite of the SU(3) breaking in the interaction that arises from
the contributions due to one-meson and two-meson exchanges and despite of the 
additional Coulomb repulsion in the $\Sigma^+ p$ system. 
Therefore, we must conclude that within the scheme followed here 
a combined quantitative description of the $NN$ and $YN$ sectors based 
on strictly SU(3) symmetric (LO and NLO) contact terms is not possible. 
Since at NLO an explicit SU(3) breaking in the LO contact terms arises anyway,
see Appendix B, we considered also a scenario where we took over the NLO 
contact terms from the fit to the $NN$ phase shifts and we fitted the 
remaining (LO and NLO) $S$-wave contact terms to the $YN$ data. In this
case a description of the $\Lambda N$ and $\Sigma N$ data is possible, but 
with a noticeably increased $\chi^2$. Moreover, we observe the questionable 
tendency of the $\La p$ amplitude in the $^3S_1$ partial wave to become 
rather large for momenta above the $\Si N$ threshold.
Thus, we decided to determine all contact terms in the $S$-waves and the
$S$-$D$ transition from a fit to the $YN$ sector alone where it turns out 
that SU(3) symmetry for the LECs can be preserved. 

The values of the contact terms obtained in the fitting procedure for the various
cutoffs are listed in Tables~\ref{tab:F1} and \ref{tab:F2}. 

\section{Results and discussion}
\label{sec:4}

The results obtained at NLO are presented in Fig.~\ref{fig:R1} (red/dark bands), 
together with those at LO (green/light bands). The bands represent the variation 
of the cross sections based on chiral EFT within the cutoff region of 
$\Lambda = 500 - 650$~MeV. Note that in the LO case variations of 
$\Lambda = 550 - 700$~MeV were considered \cite{Polinder:2006zh}.
For comparison also results for the 
J{\"u}lich '04 \cite{Hai05} meson-exchange model are shown (dashed lines),

Obviously, and as expected, the energy dependence exhibited by the data can be significantly 
better reproduced within our NLO calculation. This concerns in particular the $\Sigma^+p$
channel. But also for $\Lambda p$ the NLO results are now well in line with the data even 
up to the $\Sigma N$ threshold. Furthermore, one can see that the dependence on the cutoff
mass is strongly reduced in the NLO case. 

A quantitative comparison with the experiments is provided in Tab.~\ref{tab:R1}. 
There we list the obtained overall $\chi^2$ but also separate values for 
each data set that was included in the fitting procedure. Obviously the best
results are achieved in the range $\Lambda = 500 - 650$~MeV. Here, in addition, 
the $\chi^2$ exhibits also a fairly weak cutoff dependence so that one can really
speak of a plateau region. For larger cutoff values the $\chi^2$ increases smoothly while
it grows dramatically when going to lower values. Therefore, in Fig.~\ref{fig:R1} and
in the figures below we show only results based on variations of the cutoff within
this plateau region. 

\begin{figure}[t]
\begin{center}
\includegraphics[height=64mm]{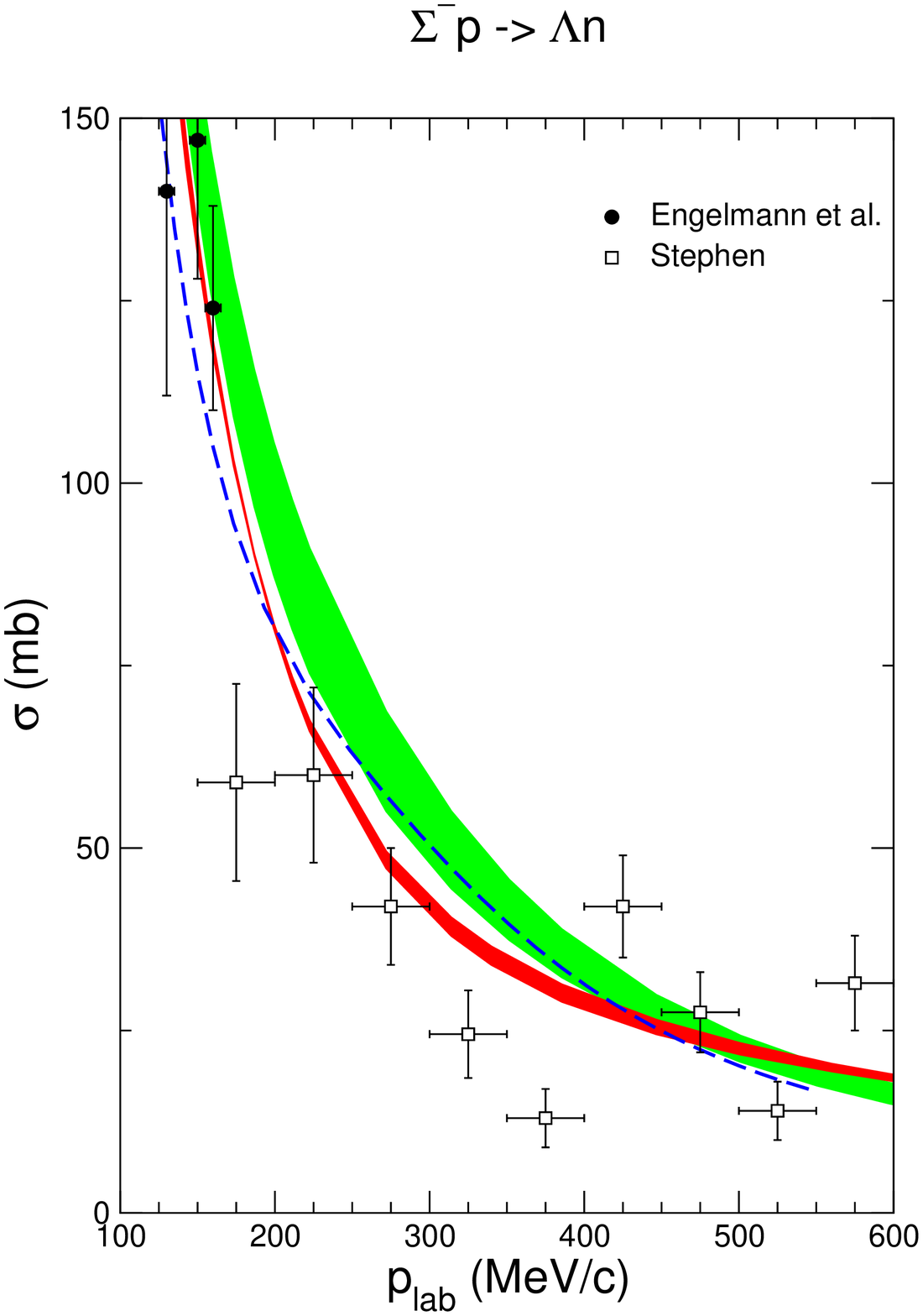}
\includegraphics[height=64mm]{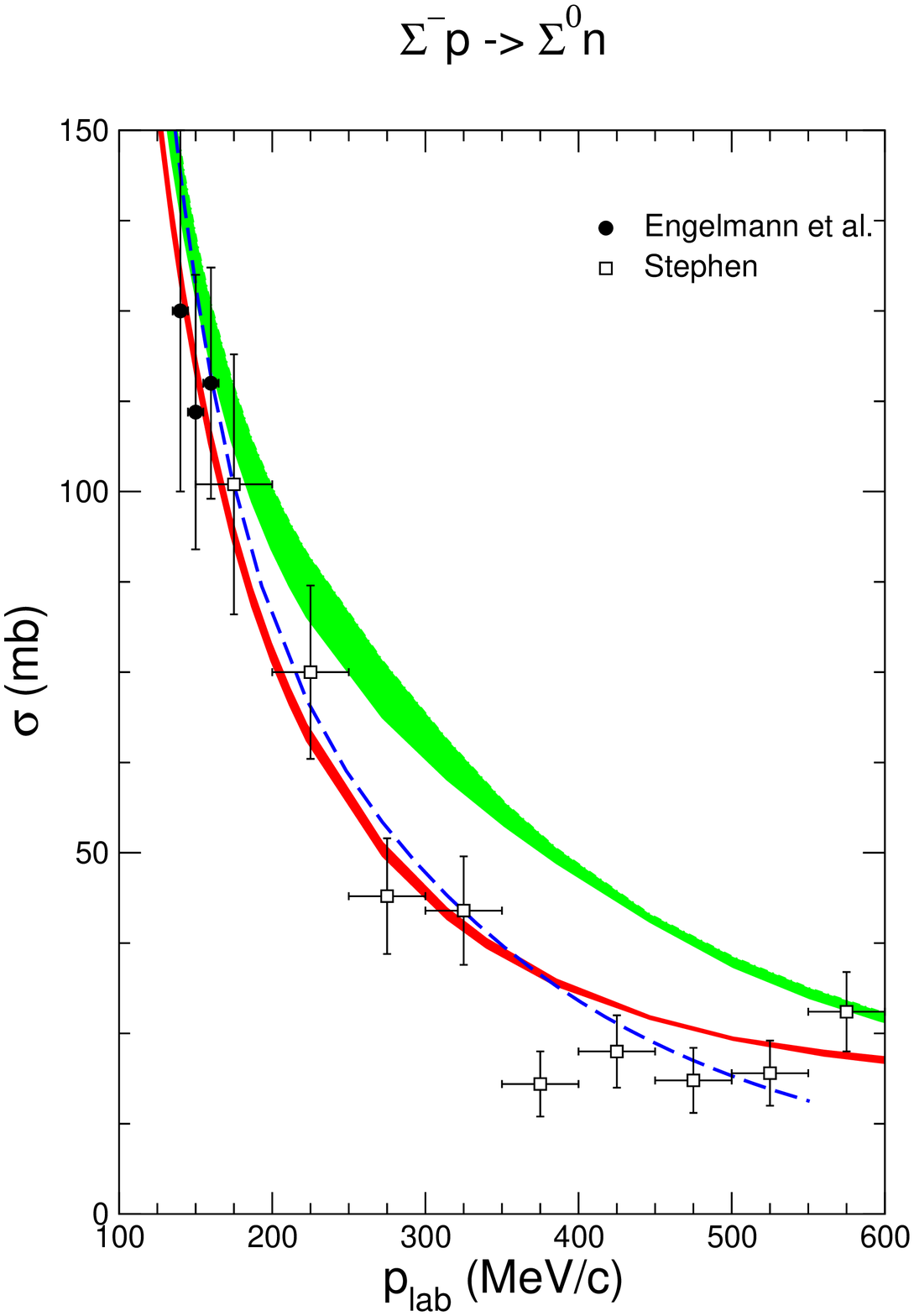}

\includegraphics[height=64mm]{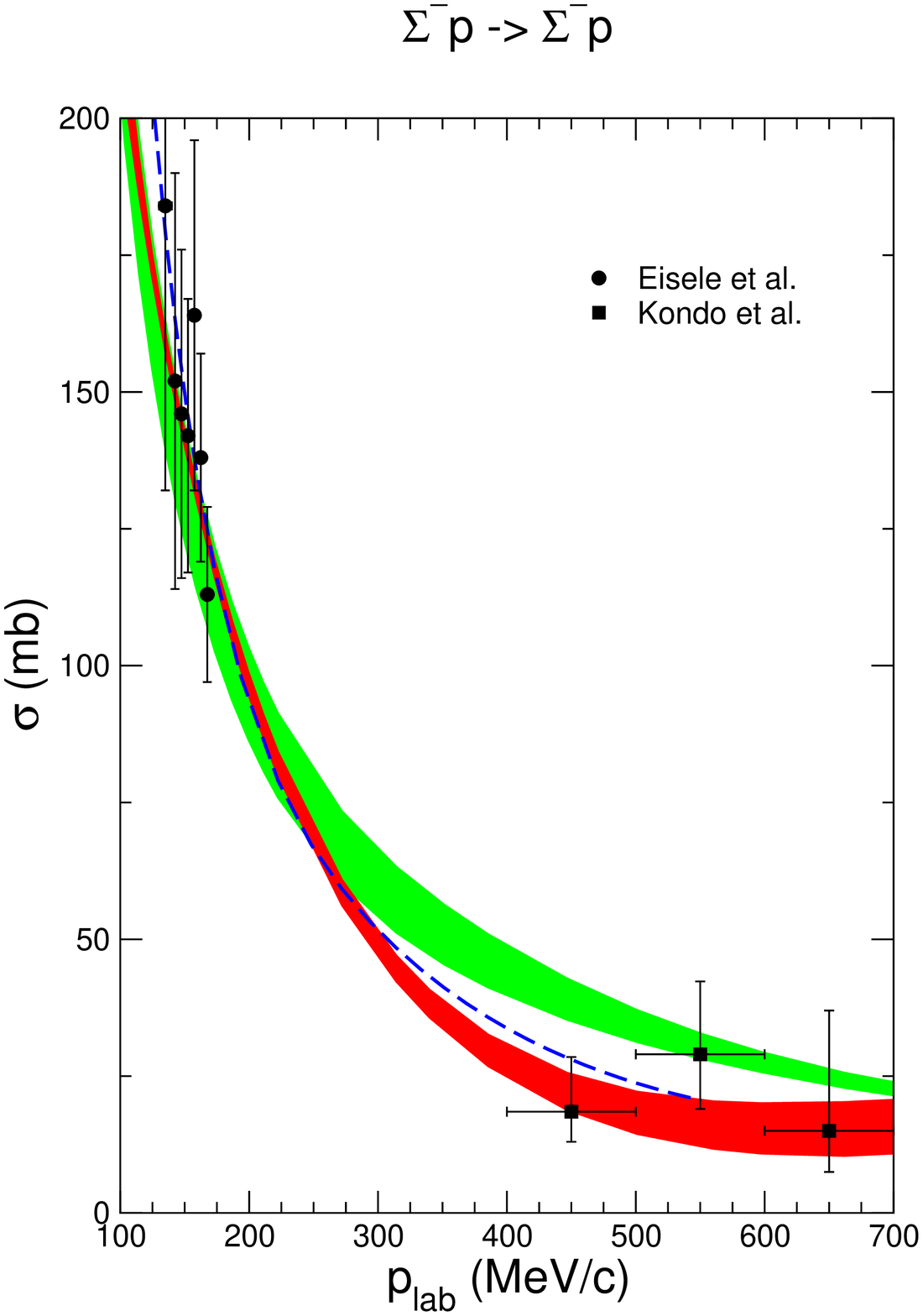}
\includegraphics[height=64mm]{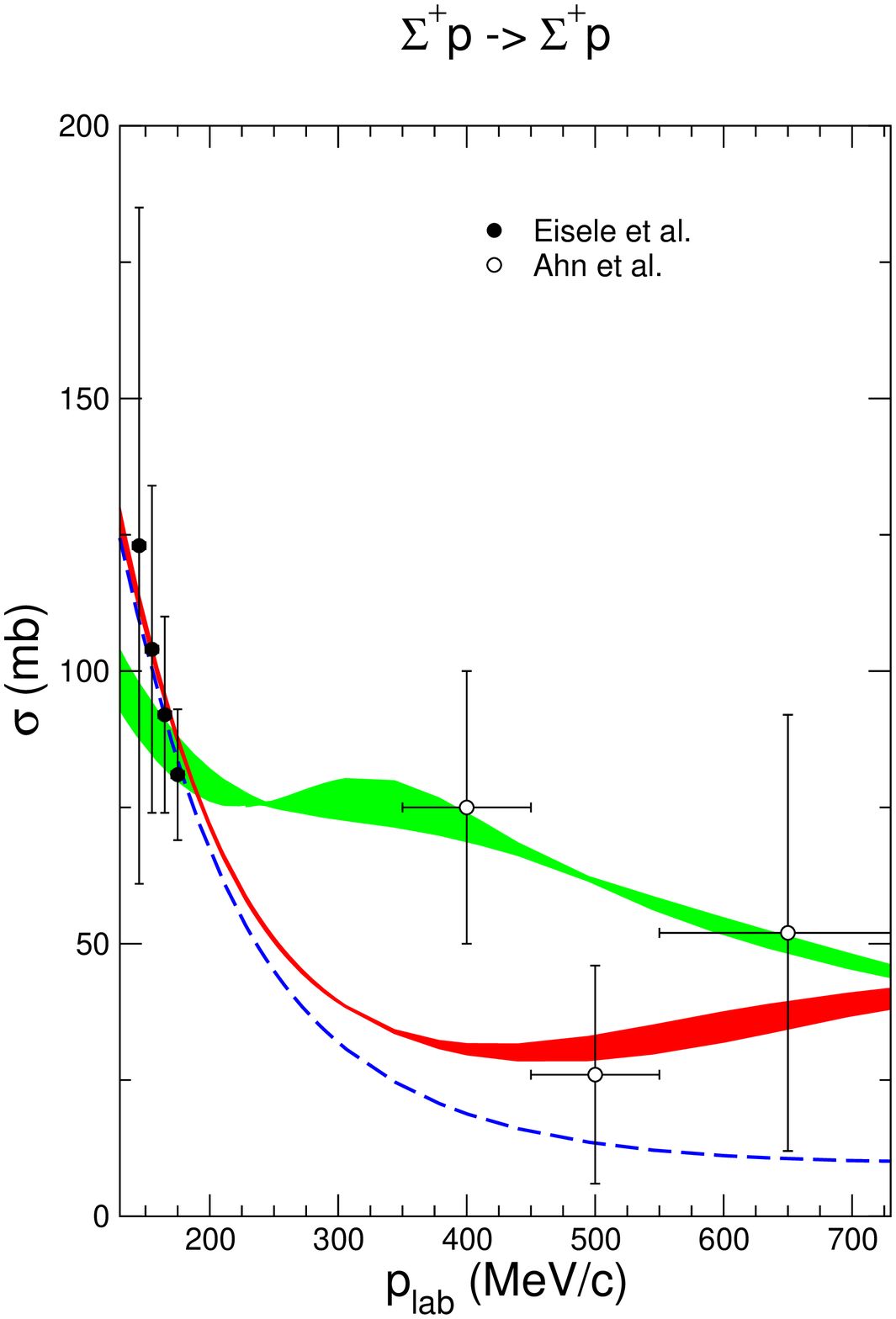}
\caption{As in Fig.~\ref{fig:R1}, but now the experimental cross sections 
are taken from Refs.~\cite{Ste70} ($\Si^- p \to \La n$, $\Si^- p \to \Si^0n$),
\cite{Kon00} ($\Si^- p \to \Si^-p$), and \cite{Ahn05} ($\Si^+ p \to \Si^+p$).
}
\label{fig:R2}
\end{center}
\end{figure}

\begin{table}
\caption{Comparison between the 36 $YN$ data and the theoretical results 
for the various cutoffs in terms of the achieved $\chi^2$. The last column
in the NLO section, denoted $600^*$, contains result for an interaction 
where all two-meson-exchange contributions
involving the $\eta$- and/or $K$ meson have been omitted, cf. text. 
}
\renewcommand{\arraystretch}{1.2}
\label{tab:R1}
\vspace{0.2cm}
\centering
\begin{tabular}{|c|c|rrrrrr|r|r|}
\hline
& data & \multicolumn{7}{|c|}{NLO} & LO \\
\hline
\multicolumn{2}{|l|}{$\Lambda$ (MeV)} &$450$ & $500$ & $550$& $600$& $650$& $700$ & $600^*$ & $600$ \\
\hline
\hline
$\La p \to \La p$          &Sechi-Zorn \cite{Sec68} &$2.8$   &$2.1$  &$2.0$ &$1.5$ &$1.6$ &$1.7$ &$1.6$ &$7.5$\\
                            &Alexander \cite{Ale68} &$4.2$   &$2.6$  &$1.6$ &$2.3$ &$2.4$ &$2.5$ &$2.2$ &$4.9$\\
$\Sigma^- p \to \La n$      &Engelmann \cite{Eng66} &$4.3$   &$3.7$  &$3.9$ &$4.1$ &$4.4$ &$4.5$ &$4.0$ &$5.5$\\
$\Sigma^- p \to \Sigma^0 n$ &Engelmann \cite{Eng66} &$5.6$   &$6.1$  &$5.8$ &$5.8$ &$5.7$ &$5.7$ &$5.9$ &$7.0$\\
$\Sigma^- p \to \Sigma^- p$    &Eisele \cite{Eis71} &$2.0$   &$2.0$  &$1.8$ &$1.9$ &$1.9$ &$2.0$ &$1.9$ &$2.4$\\
$\Sigma^+ p \to \Sigma^+ p$    &Eisele \cite{Eis71} &$0.5$   &$0.3$  &$0.4$ &$0.5$ &$0.3$ &$0.6$ &$0.9$ &$0.6$\\
\hline
$r_R$                          & \cite{Hep68,Ste70} &$0.3$   &$0.1$  &$0.2$ &$0.1$ &$0.2$ &$0.3$ &$0.1$ &$0.5$\\
\hline
total $\chi^2$ &               &                 $19.7$   &$16.8$  &$15.7$ &$16.2$ &$16.6$ &$17.3$ &$16.5$ &$28.3$\\
\hline
\end{tabular}
\renewcommand{\arraystretch}{1.0}
\end{table}

A total $\chi^2$ value of around $16$ is quite good. Indeed, the best values achieved
with phenomenological models, say the Nijmegen NSC97 meson-exchange potentials \cite{Rij99}, 
lie also in that region. We should add that our additional requirements that we want to 
produce a correctly bound hypertriton and that we want a repulsive $\Sigma N$ interaction
in the isospin $I=3/2$ channel leads to a slightly increased $\chi^2$. Without those
constraints we could achieve values which are around 5 \% smaller. 
In any case, one has to say that one should not overrate the $\chi^2$. 
Given that there are only 36 data points the $\chi^2$ per data point 
amounts to $\approx 0.5$ only -- which is somewhat low 
as compared to what one would expect from a set of statistically sound data.
As a matter of fact, the biggest single contribution to the
$\chi^2$ comes from the $\Sigma N$ charge-exchange reaction, see Tab.~\ref{tab:R1}, and 
specifically from a single data point near threshold that is far off all other data points, 
see Fig.~\ref{fig:R1}. When one omits this data point the total $\chi^2$ would be around
$10$ and the $\chi^2$ per data point would be $\approx 0.3$.
Improvements in the order of 5 \% on that level are certainly not significant. 
In this context let us mention the $\chi^2$ of preliminary results of our study that 
have been presented in \cite{Hai13,Hai13a} is not yet optimal. 
Specifically, there the description of the $\Si^+ p$ cross section was still inferior. 

In Table \ref{tab:R1} we include also results of an $YN$ interaction where from all
two-meson exchange contributions that arise to NLO according to SU(3) symmetry 
only the $\pi\pi$ exchange diagrams were kept.
All two-meson exchange diagrams involving the heavy mesons $\eta$ and/or $K$ were omitted. 
We performed an exemplary fit within this scenario for the cutoff $\Lambda = 600$~MeV
and the corresponding $\chi^2$ values can be found in the column labeled by $600^*$. 
It is obvious that a comparable fit to the data can be achieved within such a scenario,
too. 
Finally, in the last column of Table \ref{tab:R1}, $\chi^2$ results for our LO 
interaction from Ref.~\cite{Polinder:2006zh} (for $\Lambda = 600$ MeV) are
reproduced. Evidently, going to NLO allows to reduce the $\chi^2$ by roughly 50 \%! 

A comparison of our results with integrated cross sections at higher energies 
is presented in Fig.~\ref{fig:R2}. These data were not included in the fitting
procedure and, therefore, the shown results are genuine predictions of the
chiral EFT interaction. One can see that the cross sections achieved at NLO
are now closer to those obtained from the J\"ulich meson-exchange potential
than the ones at LO, and to some extent they are also more in line with the data. 
But given the large uncertainties in the experiments, even in the fairly
recent measurements of the $\Si^- p \to \Si^-p$ \cite{Kon00} and
$\Si^+ p \to \Si^+p$ \cite{Ahn05} reactions, precise conclusions are difficult
to draw. 

\begin{figure}[t]
\begin{center}
\includegraphics[height=64mm]{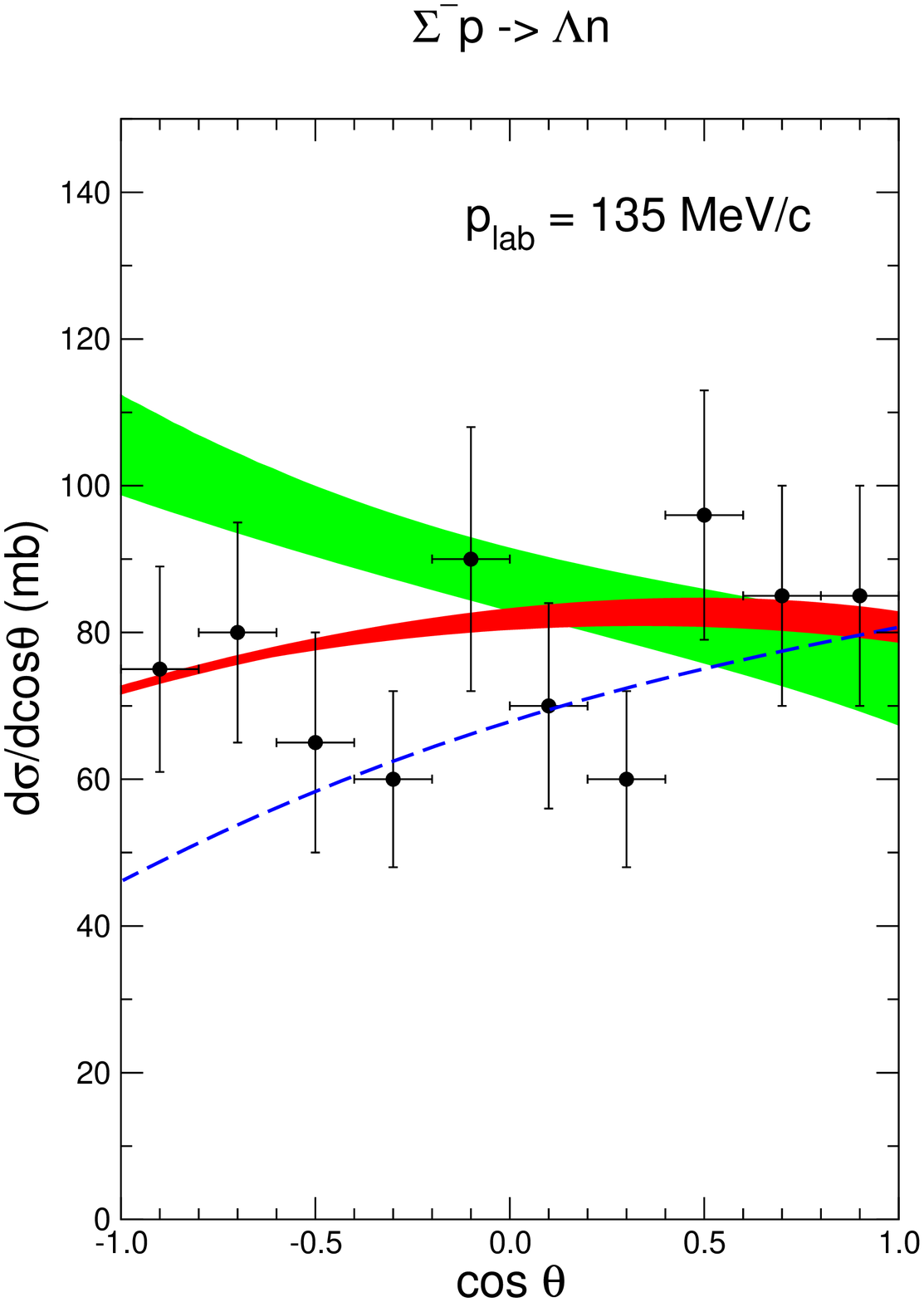}
\includegraphics[height=64mm]{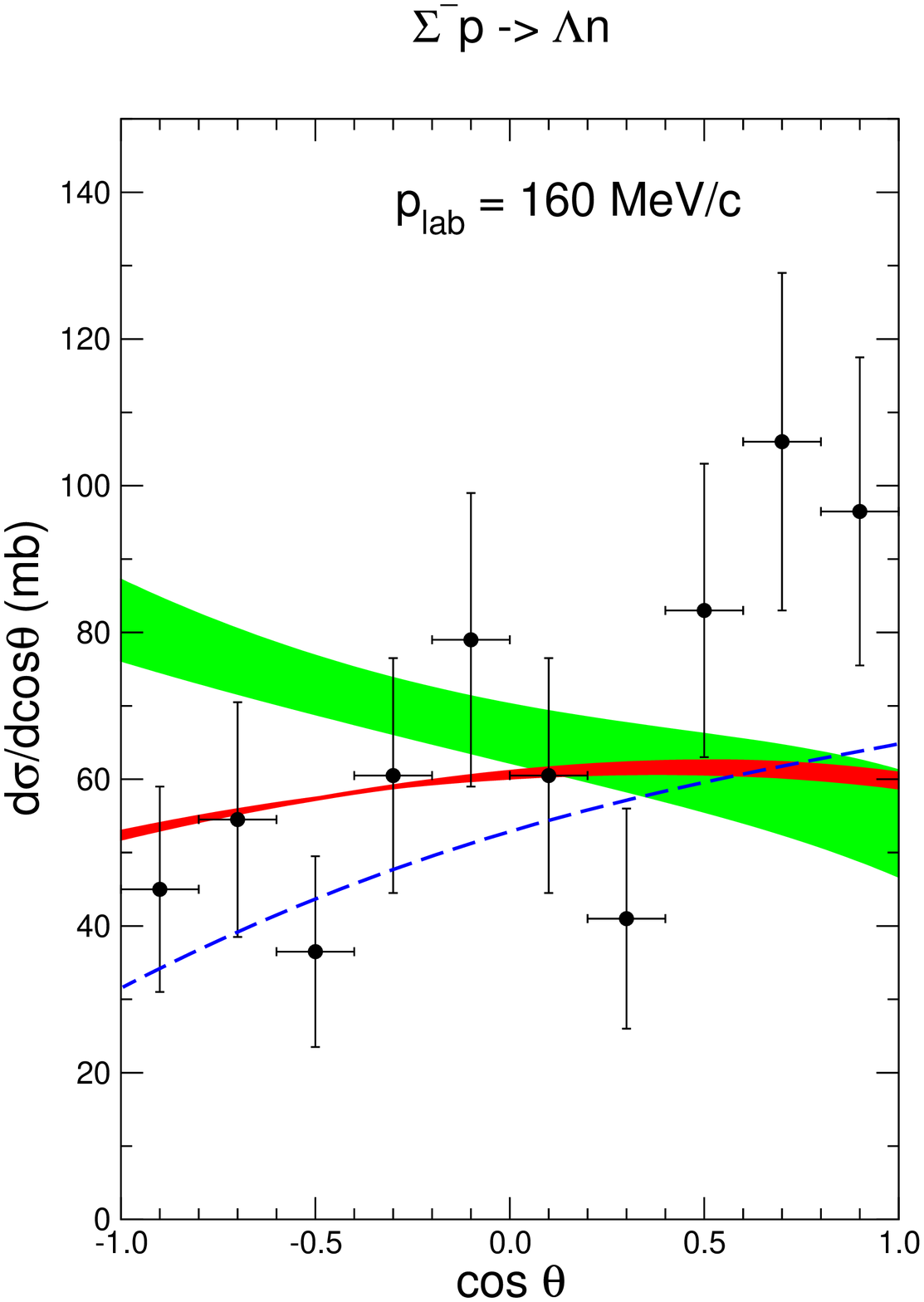}

\includegraphics[height=64mm]{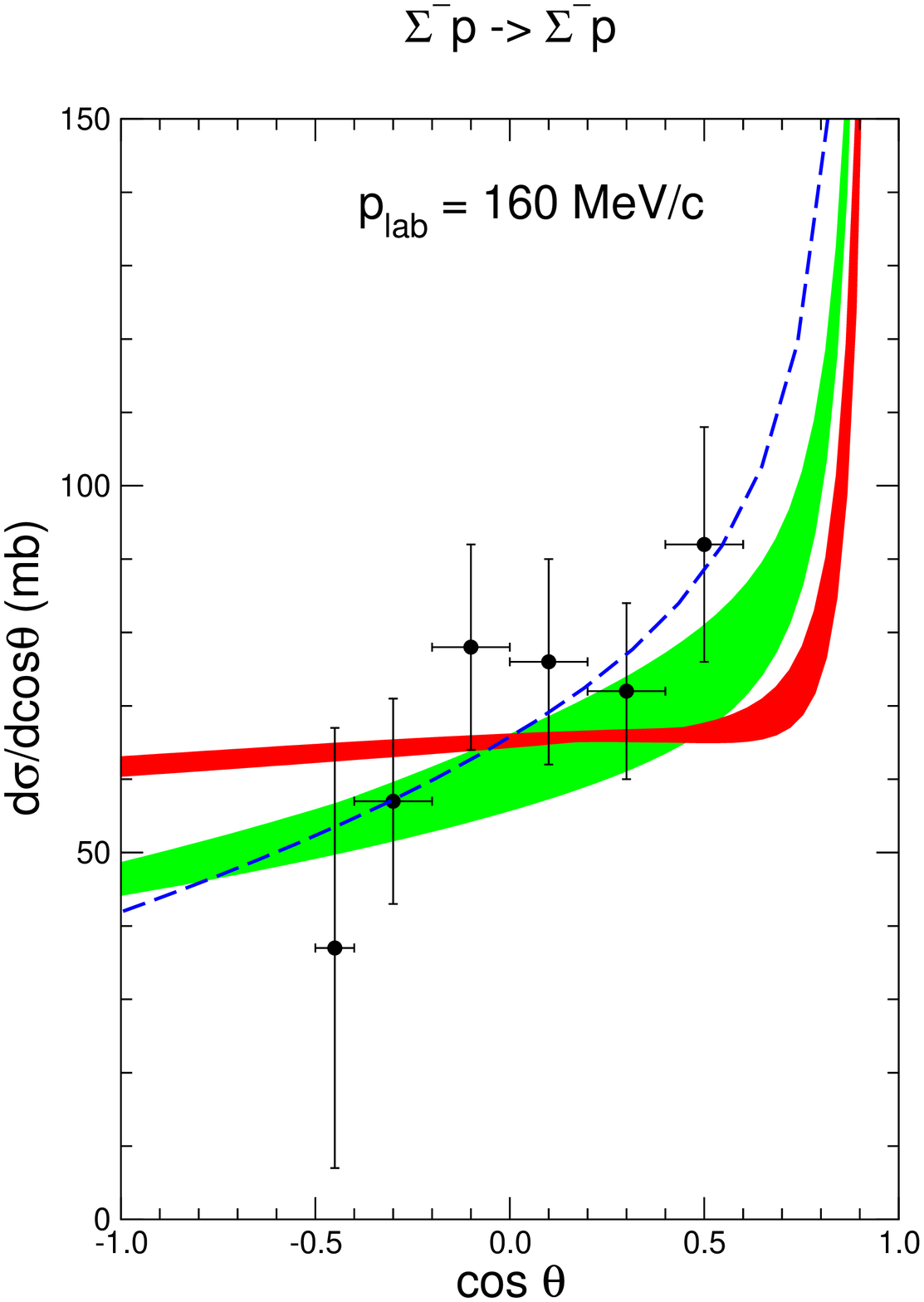}
\includegraphics[height=64mm]{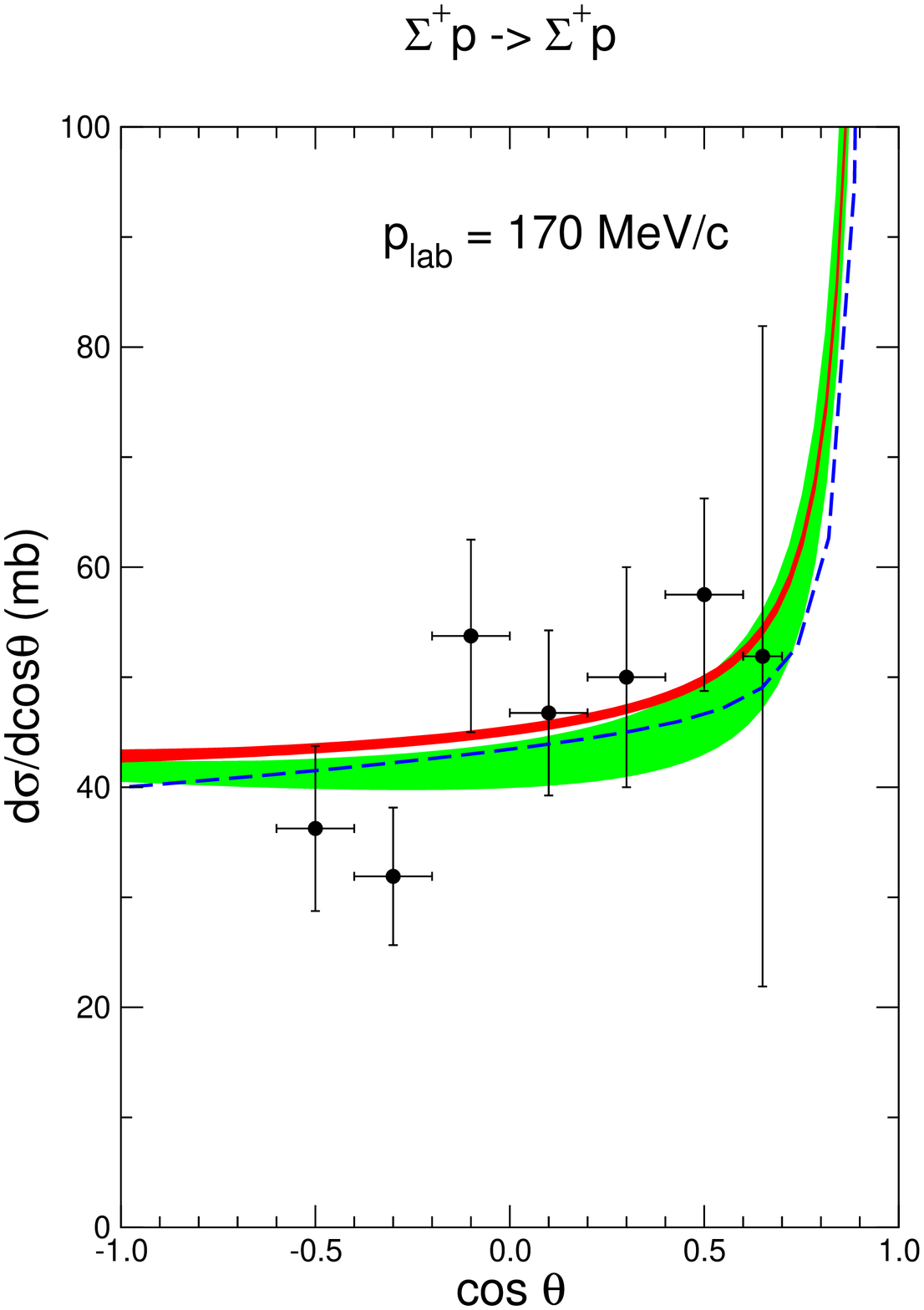}
\caption{
Differential cross section $d\sigma / d\cos \theta$ as a function of $\cos \theta$, 
where $\theta$ is the c.m. scattering angle, at various values of $p_{{\rm lab}}$.
The experimental differential cross sections are taken from \cite{Eng66} 
($\Si^-p \to \La n$, $\Si^-p \to \Si^0 n$) 
and from \cite{Eis71} ($\Si^-p \to \Si^- p$, $\Si^+p \to \Si^+ p$). 
Same description of curves as in Fig.~\ref{fig:R1}.
}
\label{fig:R3}
\end{center}
\end{figure}

\begin{figure}[ht]
\begin{center}
\includegraphics[height=64mm]{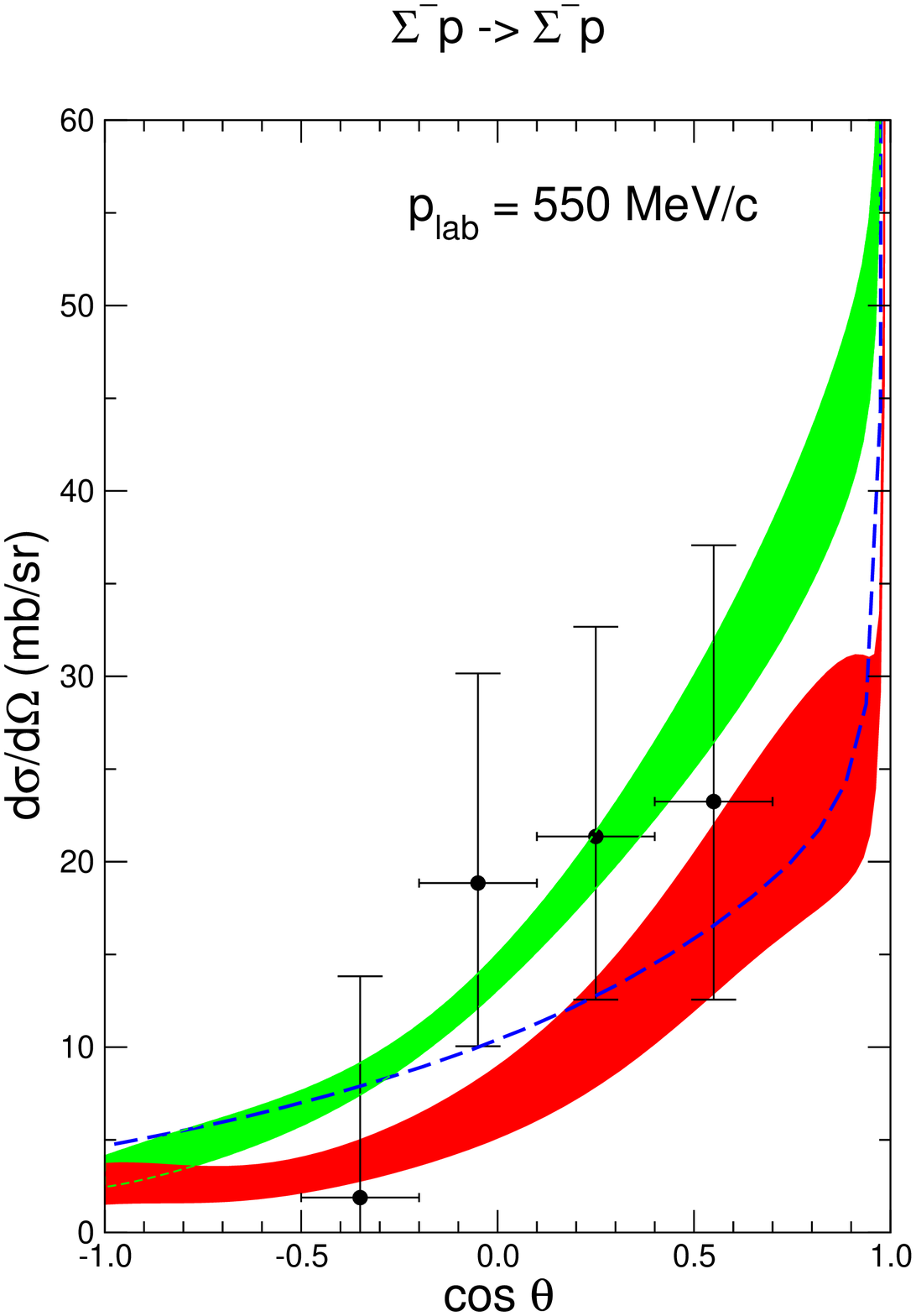}
\includegraphics[height=64mm]{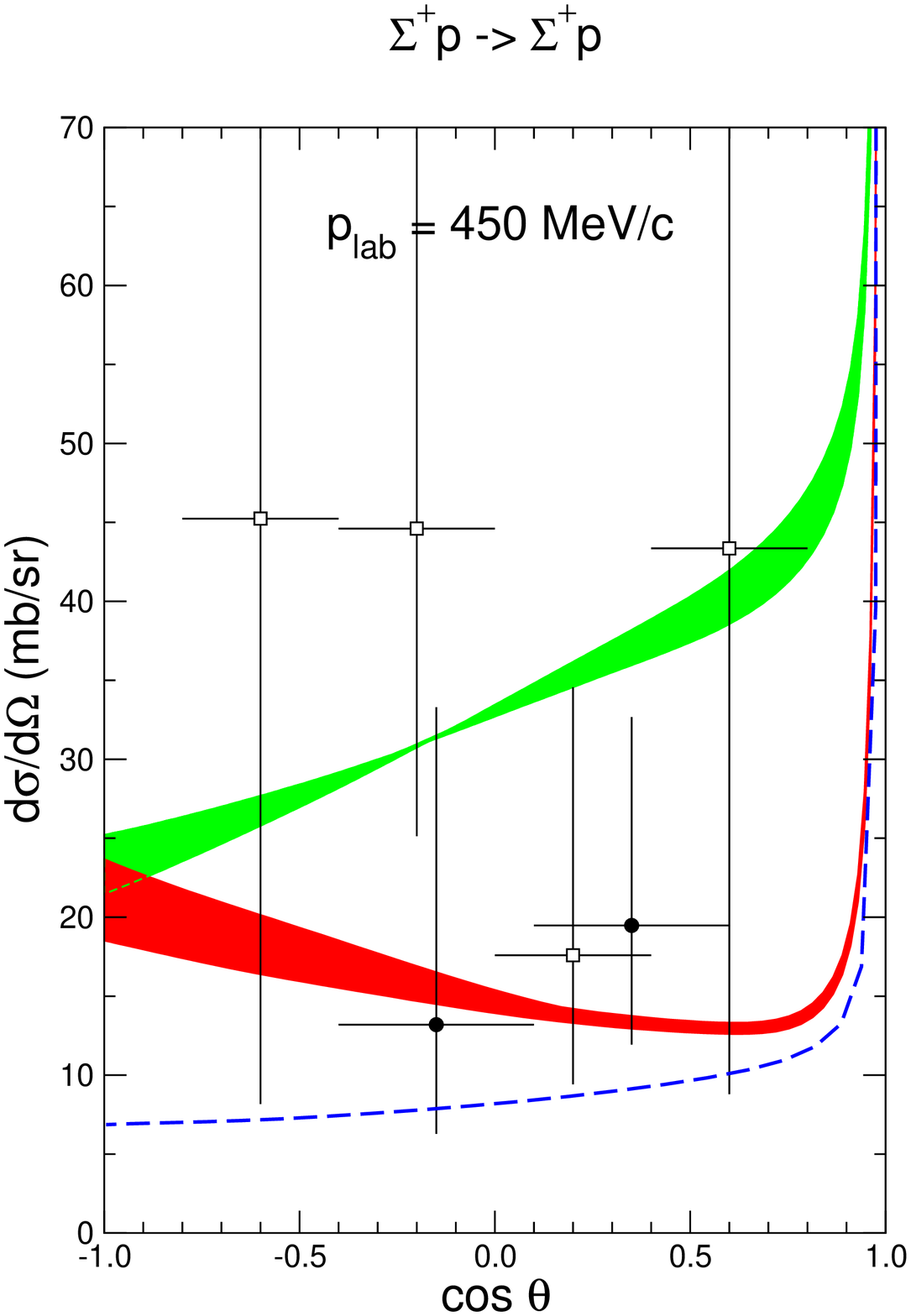}
\caption{
Differential cross section $d\sigma / d\cos \theta$ as a function of $\cos \theta$, 
where $\theta$ is the c.m. scattering angle, at various values of $p_{{\rm lab}}$.
The experimental differential cross sections are taken from 
\cite{Kon00} ($\Si^-p \to \Si^- p$),
and from \cite{Ahn99} (filled circles) and \cite{Ahn05} (open circles) 
($\Si^+p \to \Si^+ p$).
Same description of curves as in Fig.~\ref{fig:R1}.
}
\label{fig:R4}
\end{center}
\end{figure}

Differential cross sections are shown in Figs.~\ref{fig:R3} and \ref{fig:R4}
and compared with available measurements \cite{Eng66,Eis71,Kon00,Ahn05,Ahn99}.
Also these data were not included in the fitting procedure (as far as the LECs 
in the $S$-waves are concerned). However, as already mentioned in the preceding 
section, the differential cross sections for $\Si^-p \to \La n$ were considered,
together with the integrated $\La p$ cross section around
$p_{\rm lab} \approx 750-850$ MeV/c, in the ``by hand'' adjustment of the
LECs in the $P$-waves. 


As can be seen from Fig.~\ref{fig:R3}, the prediction of the NLO interaction 
for $\Si^-p \to \La n$ at $p_{\rm lab} = 135$~MeV/c agrees well with 
the trend of the data \cite{Eng66}. The amplitude is dominated by the 
$^3S_1\to\, ^3S_1$ and $^3D_1\to\, ^3S_1$ transitions so that the 
resulting angular distribution is rather flat. At LO there are significant 
$P$-wave transitions (see the phase shifts discussed below) that give
rise to an enhancement of the cross section at backward angles. 
Such large $P$-waves arise at LO solely from the one-pion exchange;
there are no contact terms yet in those partial waves that would allow one to
reduce the $P$-wave amplitudes as it was possible in the present NLO approach.
The resulting cross section at $p_{lab} = 160$~MeV/c is very similar. 
Here the data seem to indicate a clear enhancement in forward direction. 
It should be said, however, that the experimental values shown in
Fig.~\ref{fig:R3} are not the result of a measurement at the specified
momenta, but rather an average over different momentum intervals. 
Specifically, for $\Si^-p \to \La n$ the data \cite{Eng66} are
averages over the intervals $100 \leq p_{\Si^-} \leq 170$~MeV/c
and $150 \leq p_{\Si^-} \leq 170$~MeV/c, respectively. In view of the
large error bars we refrain here from averaging our theoretical
results and, following common practice, present predictions at the
average value of the momenta. 
The same is also true for the data from Ref.~\cite{Eis71} which
represent averages over
$150 \leq p_{\Si^-} \leq 170$~MeV/c (for $\Si^-p \to \Sigma^-p$)
and 
$160 \leq p_{\Si^+} \leq 180$~MeV/c (for $\Si^+p \to \Sigma^+p$),
respectively.

Fig.~\ref{fig:R3} suggests that a somewhat different cocktail of $P$-wave 
contributions, like the one predicted by the J\"ulich~'04 interaction 
(cf.\ the dashed lines), might be more in line with the experimental 
data.
However, we postpone a thorough determination of the LECs 
in the $P$-waves to future investigations. 
Here, as a first step, one would try to connect our interaction with
the effective $YN$ interactions in a nuclear medium as determined
from the analysis of $\gamma$-ray data for $\La$ hypernuclei
\cite{Millener} via a G-matrix calculation \cite{Reuber}.
The results of such an analysis could provide additional and valuable 
information on the spin--dependence of the $YN$ force and, specifically, on 
the spin-spin and spin-orbit interaction. Indeed, the spin-orbit splitting 
of the $\Lambda$ single-particle levels in nuclei is experimentally well 
established and very small \cite{Hashimoto06,Kaiser:2004fe} and, therefore, 
should constitute a stringent constraint on the interaction. 
We expect that then also the LEC for the $^1P_1$-$^3P_1$
transition potential can be fixed, which has been set to zero in the
present work. 

The data on differential cross sections at higher energies 
(cf. Fig.~\ref{fig:R4}) are averages over
$400 \leq p_{\Si^-} \leq 700$~MeV/c, for $\Si^-p \to \Sigma^-p$
and over 
$300 \leq p_{\Si^+} \leq 600$~MeV/c \cite{Ahn99} and
$350 \leq p_{\Si^+} \leq 750$~MeV/c \cite{Ahn05}, respectively, for 
$\Si^+p \to \Sigma^+p$. Also here the predictions of the $YN$ 
interactions are for the momenta as specified in Fig.~\ref{fig:R4}.

Scattering lengths for the $\Lambda p$ and $\Sigma^+ p$ interactions in
the $^1S_0$ and $^3S_1$ partial waves are summarized in Tab.~\ref{tab:R2}. 
Furthermore we provide results for the hypertriton binding energy. 
As already said, this binding energy had to be taken as additional 
constraint in the fitting procedure because otherwise it would have not 
been possible to fix the relative strength 
of the ($S$-wave) singlet- and triplet contributions to the $\Lambda p$ interaction. 
Tab.~\ref{tab:R2} lists also results for two meson-exchange potentials, namely
of the J\"ulich '04 model \cite{Hai05} and the Nijmegen NSC97f potential \cite{Rij99},
which both reproduce the hypertriton binding energy correctly. 
 
The $\Si^+ p$ scattering length in the $^3S_1$ partial wave is positive,
as it was already the case for our LO potential, indicating a repulsive 
interaction in this channel. 

\begin{table}[ht]
\caption{The $YN$ singlet (s) and triplet (t) scattering lengths (in fm) 
  and the hypertriton binding energy, $E_B$ (in MeV). 
  The binding energies for the hypertriton (last row) are calculated using
  the Idaho-N3LO $NN$ potential \cite{Entem:2003ft}. 
  The experimental value for the $^3_\La \rm H$ binding energy is -2.354(50) MeV.
  \label{tab:R2}
  }
\vskip 0.1cm 
\renewcommand{\arraystretch}{1.2}
\begin{center}
\begin{tabular}{|c|rrrrrr|r|rr|}
\hline
& \multicolumn{6}{|c|}{NLO} & \ LO \ & \ J\"ulich '04 \ & \ NSC97f \ \\
& \multicolumn{6}{|c|}{}    & \cite{Polinder:2006zh} & \cite{Hai05} & \cite{Rij99} \\
\hline
${\Lambda}$ [MeV] & 450 & 500 & 550 & 600 & 650 & 700 & 600 & & \\
\hline
$a^{\La p}_s$ & $-2.90$ & $-2.91$ & $-2.91$ & $-2.91$ & $-2.90$ & $-2.90$ & $-1.91$ & $-2.56$ & $-2.60$ \\
$r^{\La p}_s$ & $ 2.64$ & $ 2.86$ & $ 2.84$ & $ 2.78$ & $ 2.65$ & $ 2.56$ & $ 1.40$ & $ 2.74$ & $ 3.05$ \\
$a^{\La p}_t$ & $-1.70$ & $-1.61$ & $-1.52$ & $-1.54$ & $-1.51$ & $-1.48$ & $-1.23$ & $-1.67$ & $-1.72$ \\
$r^{\La p}_t$ & $ 3.44$ & $ 3.05$ & $ 2.83$ & $ 2.72$ & $ 2.64$ & $ 2.62$ & $ 2.13$ & $ 2.93$ & $ 3.32$ \\
\hline
$a^{\Si^+ p}_s$ & $-3.58$ & $-3.59$ & $-3.60$ & $-3.56$ & $-3.46$ & $-3.49$ & $-2.32$ & $-3.60$ & $-4.35$ \\
$r^{\Si^+ p}_s$ & $ 3.49$ & $ 3.59$ & $ 3.56$ & $ 3.54$ & $ 3.53$ & $ 3.45$ & $ 3.60$ & $ 3.24$ & $ 3.16$ \\
$a^{\Si^+ p}_t$ & $ 0.48$ & $ 0.49$ & $ 0.49$ & $ 0.49$ & $ 0.48$ & $ 0.49$ & $ 0.65$ & $ 0.31$ & $-0.25$ \\
$r^{\Si^+ p}_t$ & $-4.98$ & $-5.18$ & $-5.03$ & $-5.08$ & $-5.41$ & $-5.18$ & $-2.78$ & $-12.2$ & $-28.9$ \\
\hline
\hline
$(^3_\La \rm H)$ $E_B$ & $-2.39$ & $-2.33$ & $-2.30$ & $-2.30$ & $-2.30$ & $-2.32$ &$-2.34$ & $-2.27$ & $-2.30$ \\
\hline
\end{tabular}
\end{center}
\renewcommand{\arraystretch}{1.0}
\end{table}

The $\La p$ scattering lengths predicted at NLO turn out to be significantly 
larger than those obtained at LO -- as example for the latter we included the 
result for the cutoff $\Lambda= 600$ MeV in Tab.~\ref{tab:R2}. 
In case of the $^1S_0$ channel, they are even somewhat larger than the values 
of the meson-exchange potentials. 
We want to remind the reader that the hypertriton binding energy is much 
more sensitive to the $\La N$ $^1S_0$ strength than to that of the 
$^3S_1$-$^3D_1$ partial wave, as is known from studies in the 
past \cite{Gibson,Miyagawa95}. Thus, the value of the
$^1S_0$ scattering length is strongly influenced by our demand to 
reproduce a correctly bound hypertriton. Interestingly, in the incomplete 
NLO calculation (i.e.\ without two-meson--exchange contributions) 
presented in Ref.~\cite{Hai10} a bound hypertriton
was achieved with $\Lambda p$ $^1S_0$ scattering lengths around $-2.6$~fm, i.e.
close to the values of the two meson-exchange potentials, cf. Tab.~\ref{tab:R2}. 
 
The hypertriton results discussed above were all obtained without an 
explicit three-body force (3BF). It has been argued that the variation
of the three-baryon binding energy with the cutoff $\La$ could provide a
measure for the size of the 3BF \cite{Nog12}. If so one would expect 
its effect to be somewhere in the range of 10-90 keV, based on the 
values listed in Tab.~\ref{tab:R2}.
Formally the first non-vanishing contributions to the 3BF appear at 
next-to-next-to-leading order (NNLO) in the scheme that we
follow \cite{Epelbaum:2008ga}. 
But we want to point out that our present calculation includes already
some 3BF effects. These are generated automatically in the employed 
coupled-channel $\La N$-$\Si N$ formalism and occur in the form of the 
transition of the $\La$ to the $\Si$ in the intermediate ($YNN$) state.  
However, these contributions are two-body reducible and, therefore, do not
constitute a genuine (irreducible) 3BF. 
Note that discussions of 3BF effects in the strangeness sector in the 
literature \cite{Bhaduri,Gal71,Schaffner,Vidana} are often related 
to the case of an intermediate $\Si$, sometimes even exclusively. 
One should distinguish its role from that of an irreducible 3BF which 
would be generated, for example, by the excitation
of the $\Si$(1385) resonance in the intermediate state -- analogous
to the 3BF that arises in the standard three-nucleon problem due to 
the $\Delta$(1232) excitation.

Calculations for the four-body hypernuclei ${}^4_\Lambda {\rm H}$ and ${}^4_\Lambda {\rm He}$ 
based on the preliminary version of the NLO interaction presented in \cite{Hai13} 
can be found in Ref.~\cite{Nog12}. 
That interaction reproduces qualitatively the $\Lambda$ separation 
energies for ${}^4_\Lambda {\rm H}$ and, in particular, it yields the correct ordering 
of the $0^+$ and $1^+$ states. However, a quantitative agreement
with the experimental information is not achieved. 
Corresponding computations for the EFT interactions discussed in the
present paper are in progress \cite{Nog13}. 

\begin{figure}[t]
\begin{center}
\includegraphics[height=64mm]{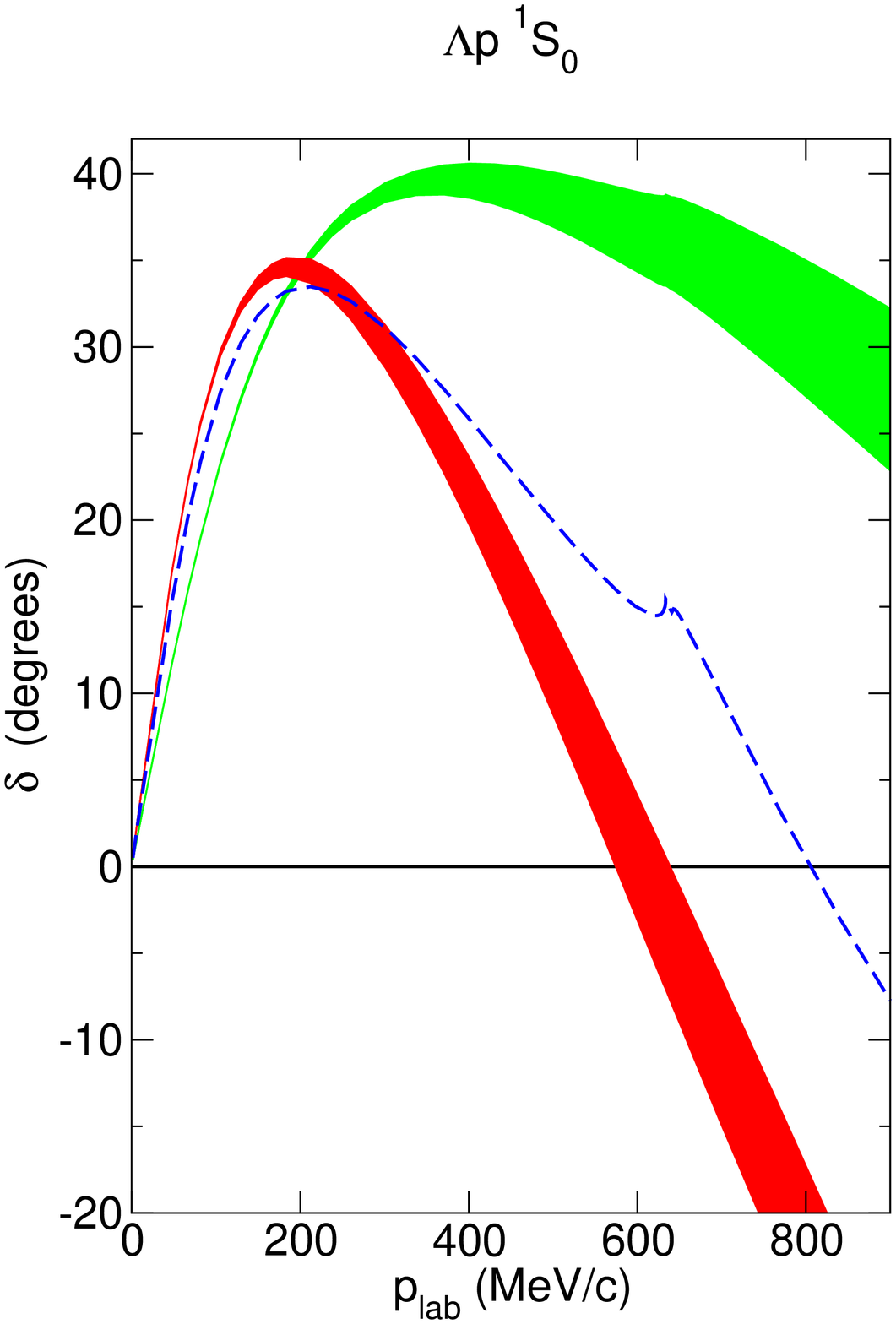}
\includegraphics[height=64mm]{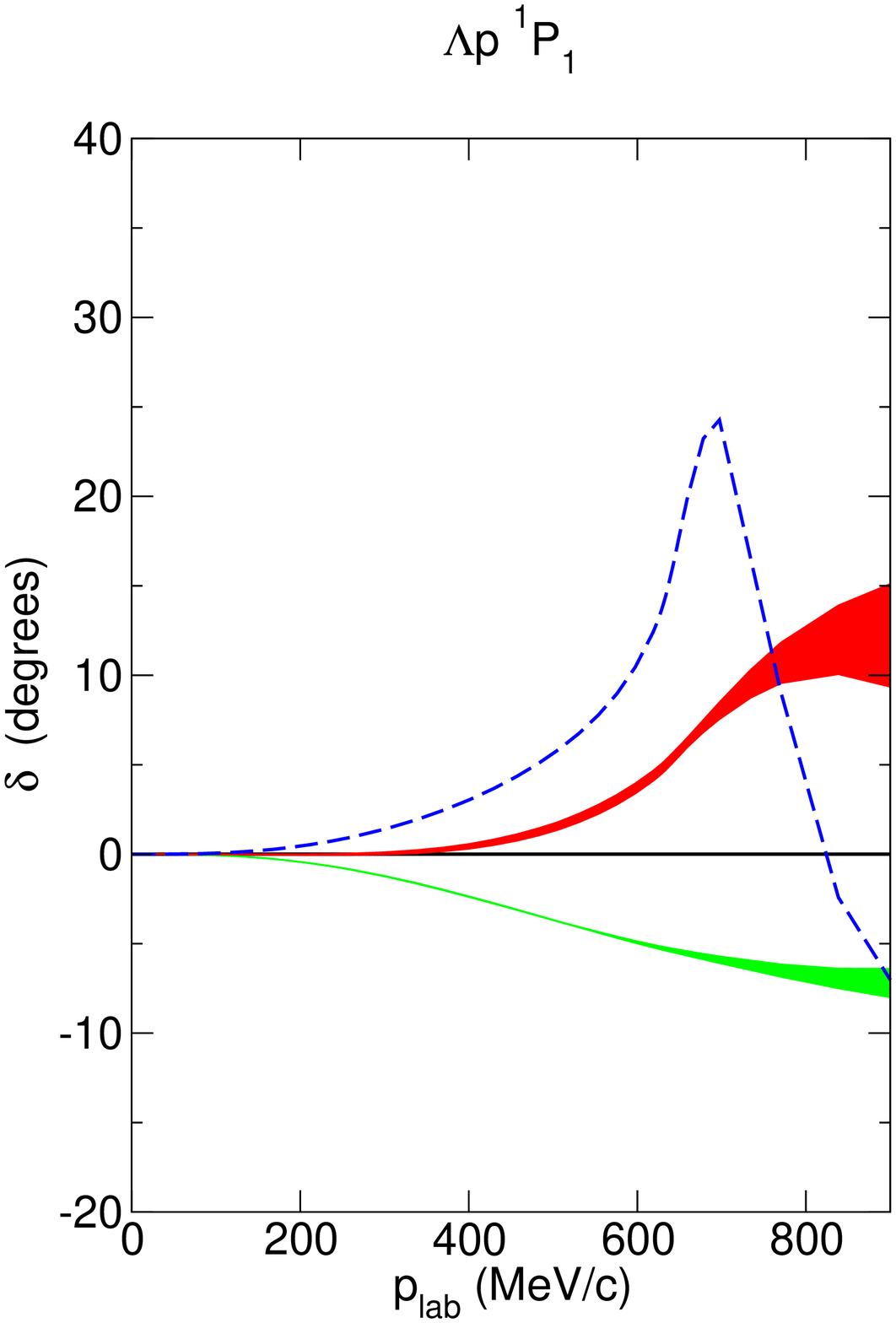}
\caption{
The $\Lambda p$ $^1S_0$ and $^1P_1$ phase shifts $\delta$ as a function of $p_{{\rm lab}}$. 
The red/dark band shows the chiral EFT results to NLO for variations of the cutoff
in the range $\Lambda =$ 500$,\ldots,$650~MeV, while the green/light band are
results to LO for $\Lambda =$ 550$,\ldots,$700~MeV. The dashed curve is the result of 
the J{\"u}lich '04 meson-exchange potential \cite{Hai05}.
}
\label{fig:R5}
\end{center}
\end{figure}

\begin{figure}[t]
\begin{center}
\includegraphics[height=64mm]{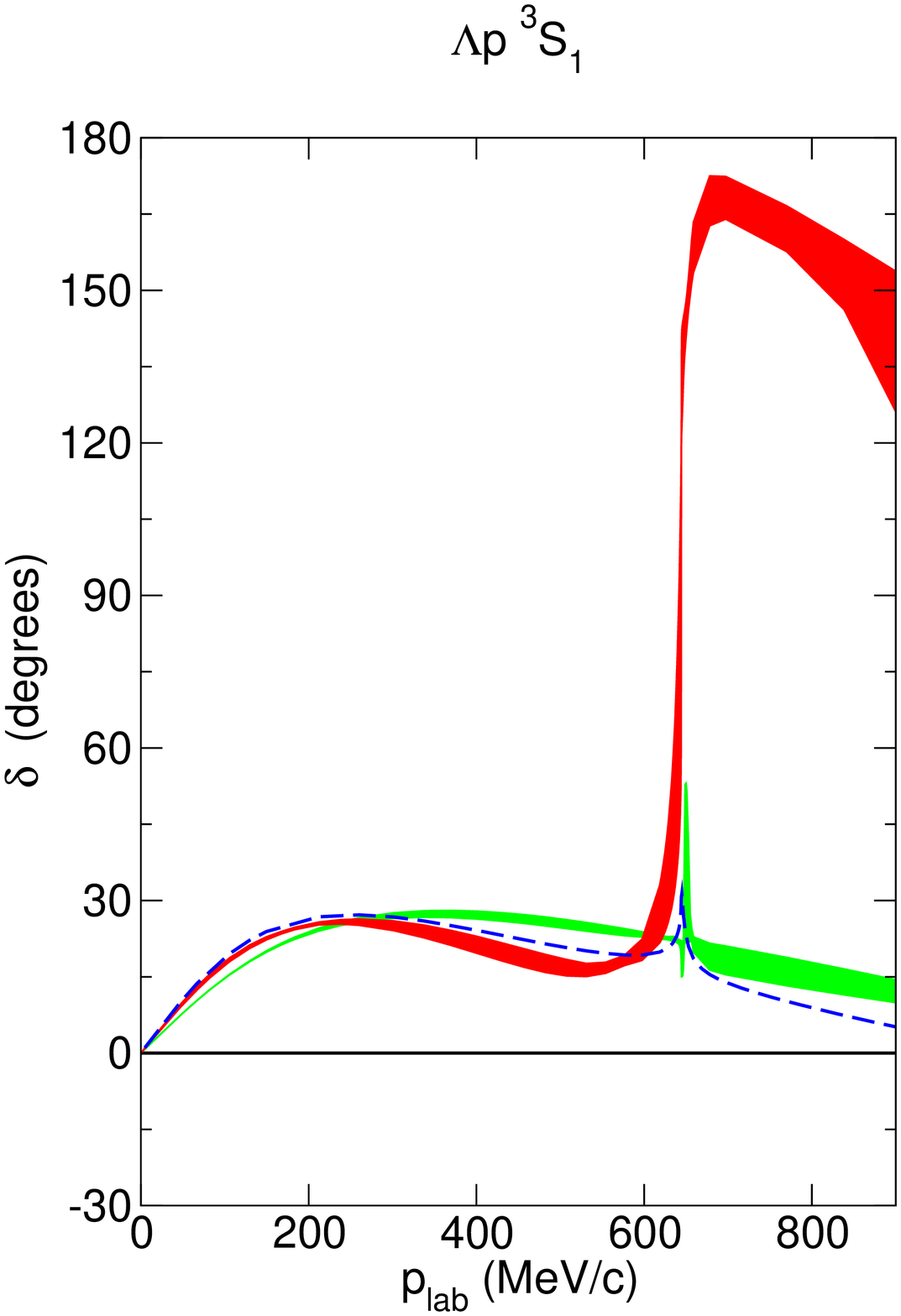}
\includegraphics[height=64mm]{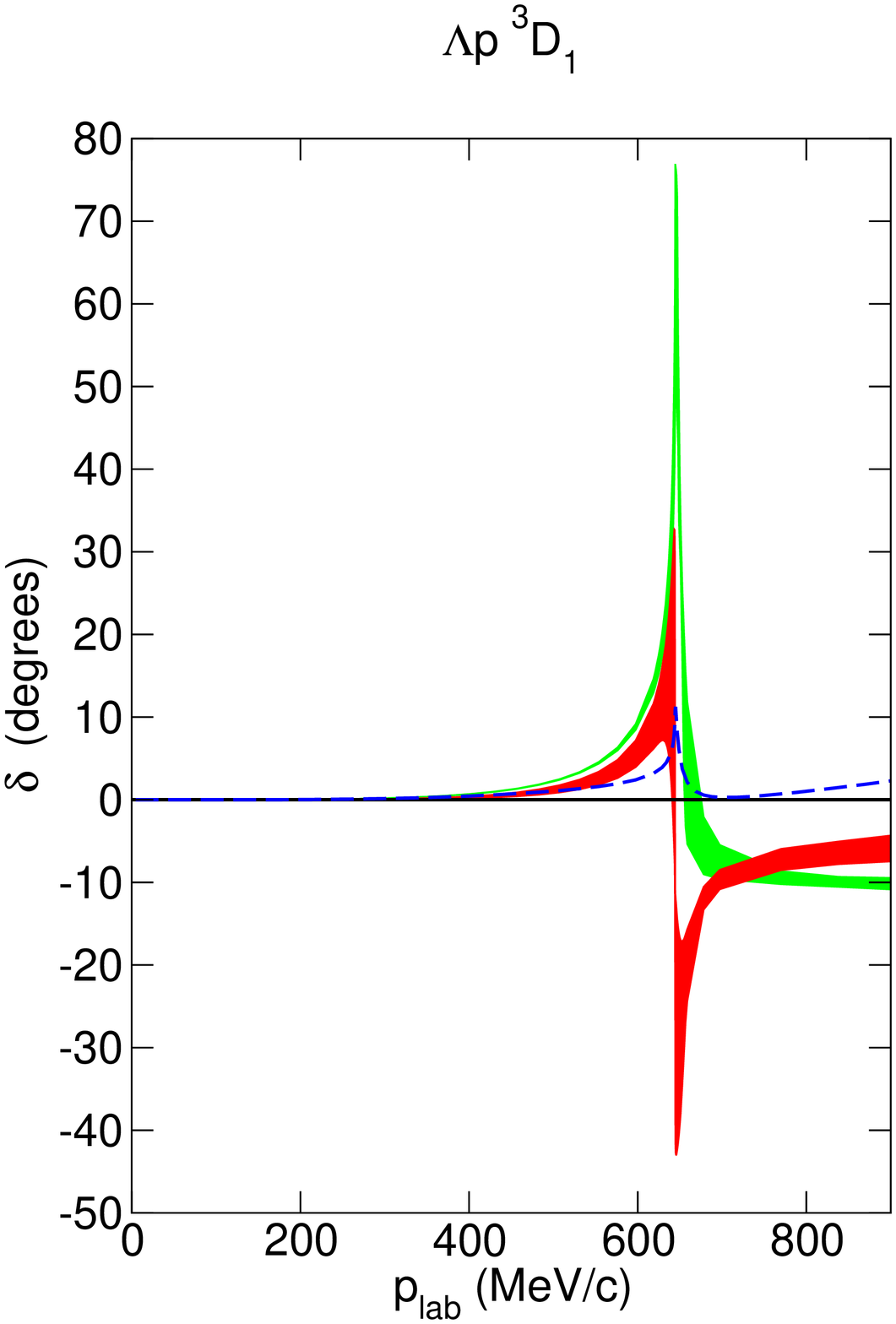}
\includegraphics[height=64mm]{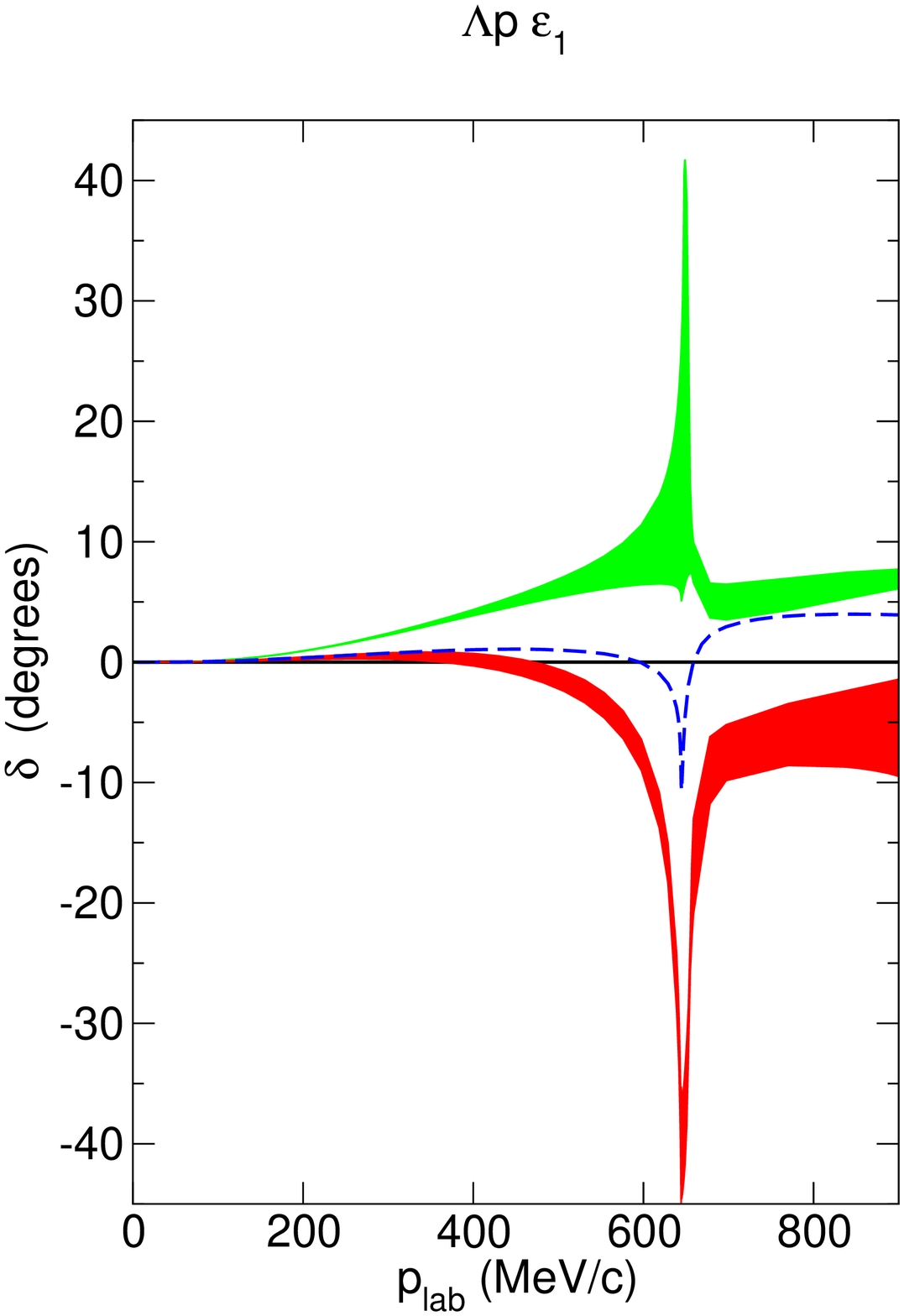}
\caption{
The $\Lambda p$ phase shifts for the coupled $^3S_1$-$^3D_1$ partial wave
as a function of $p_{{\rm lab}}$. 
Same description of curves as in Fig.~\ref{fig:R5}.
}
\label{fig:R6}
\end{center}
\end{figure}

\begin{figure}[t]
\begin{center}
\includegraphics[height=64mm]{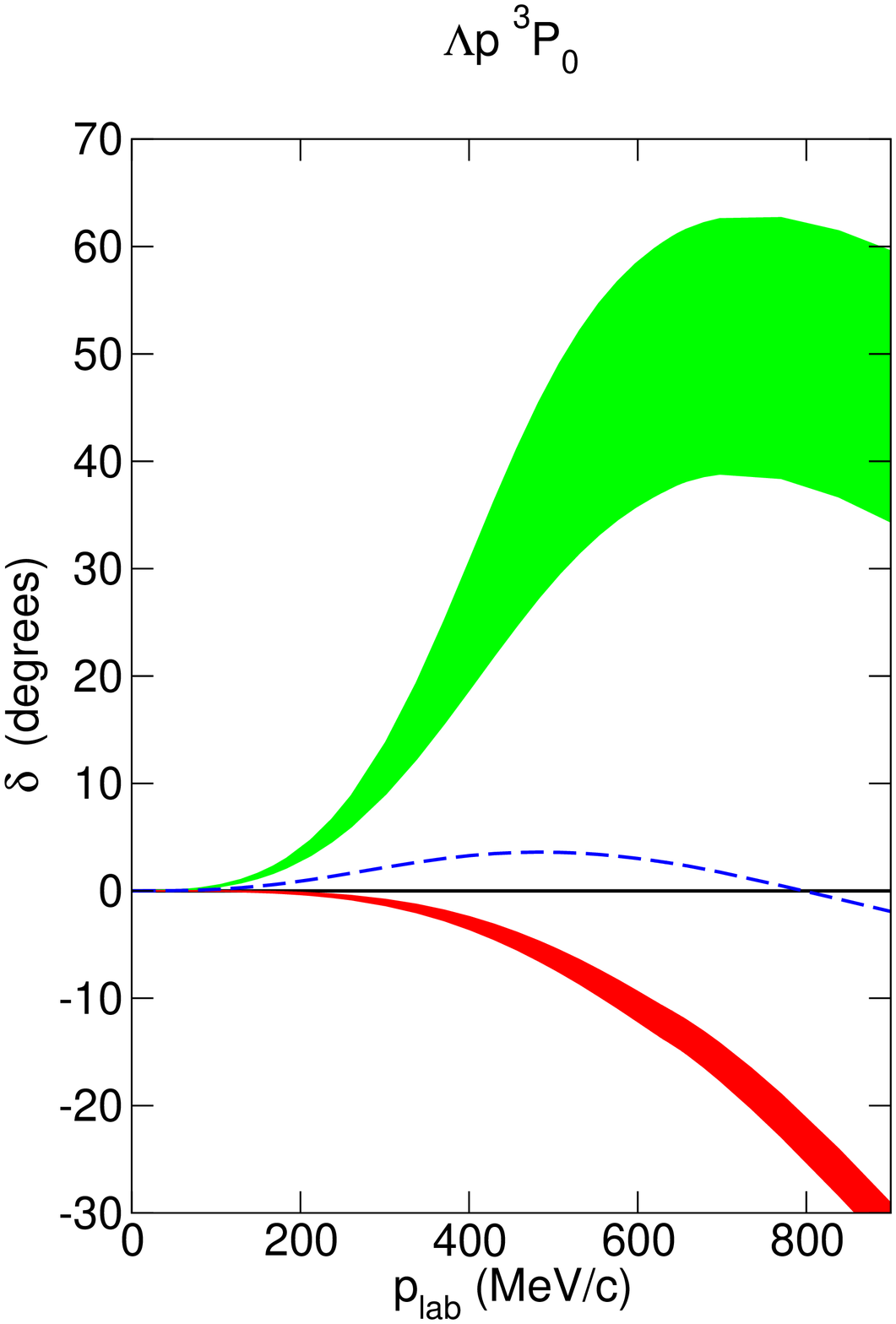}
\includegraphics[height=64mm]{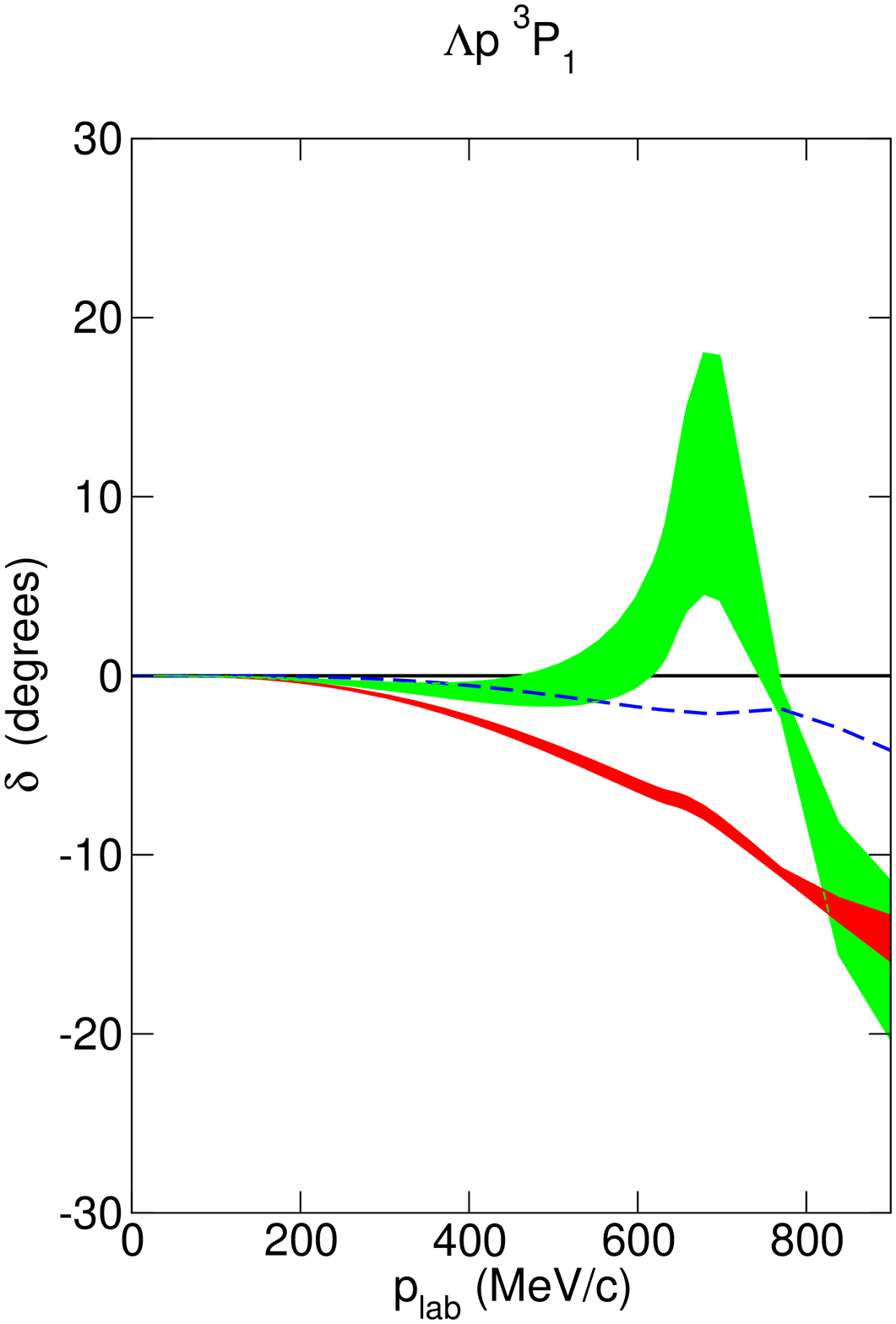}
\includegraphics[height=64mm]{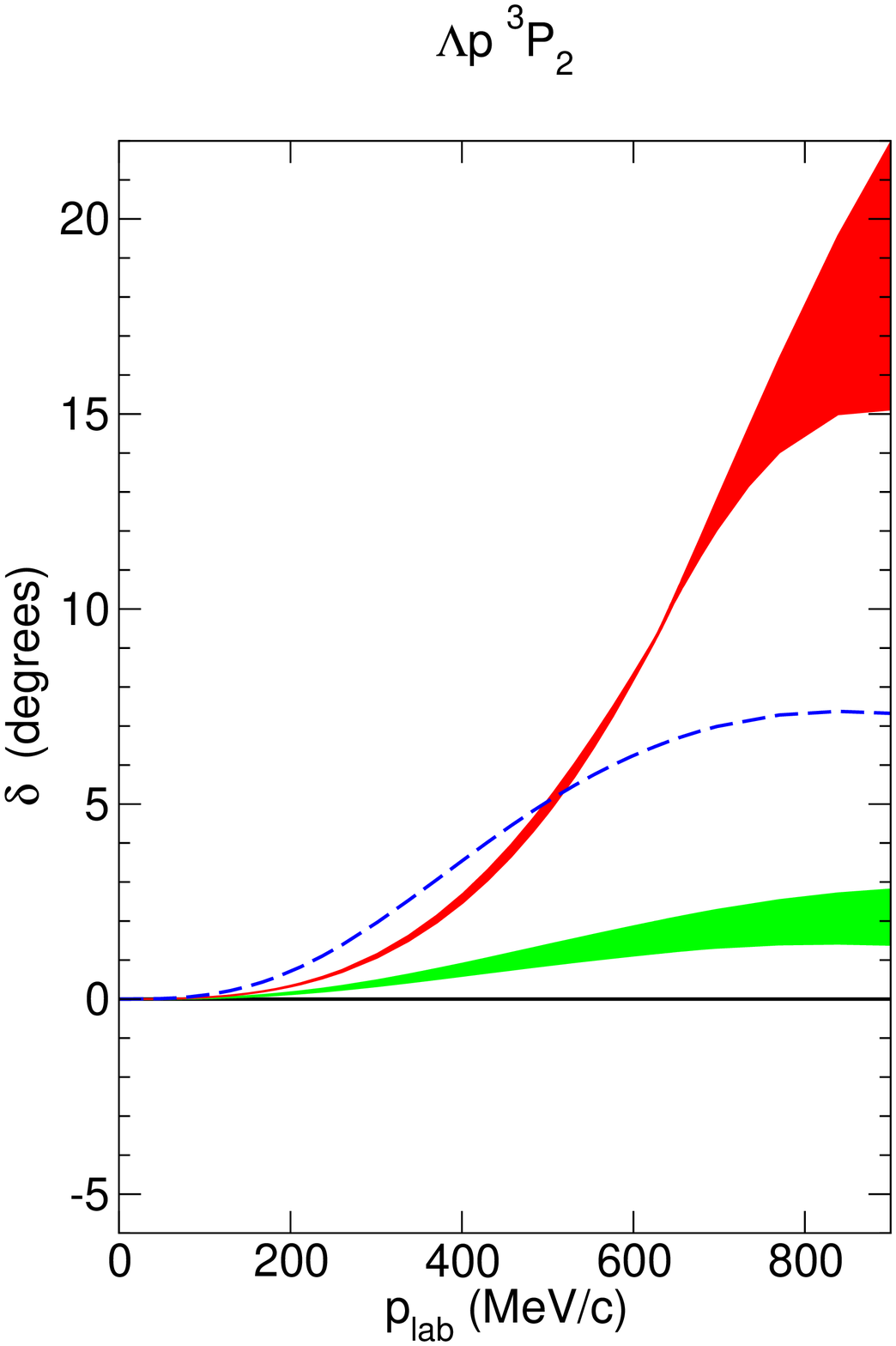}
\caption{
The $\Lambda p$ $^3P$-wave phase shifts $\delta$ as a function of $p_{{\rm lab}}$. 
Same description of curves as in Fig.~\ref{fig:R5}.
}
\label{fig:R7}
\end{center}
\end{figure}

\begin{figure}[t]
\begin{center}
\includegraphics[height=64mm]{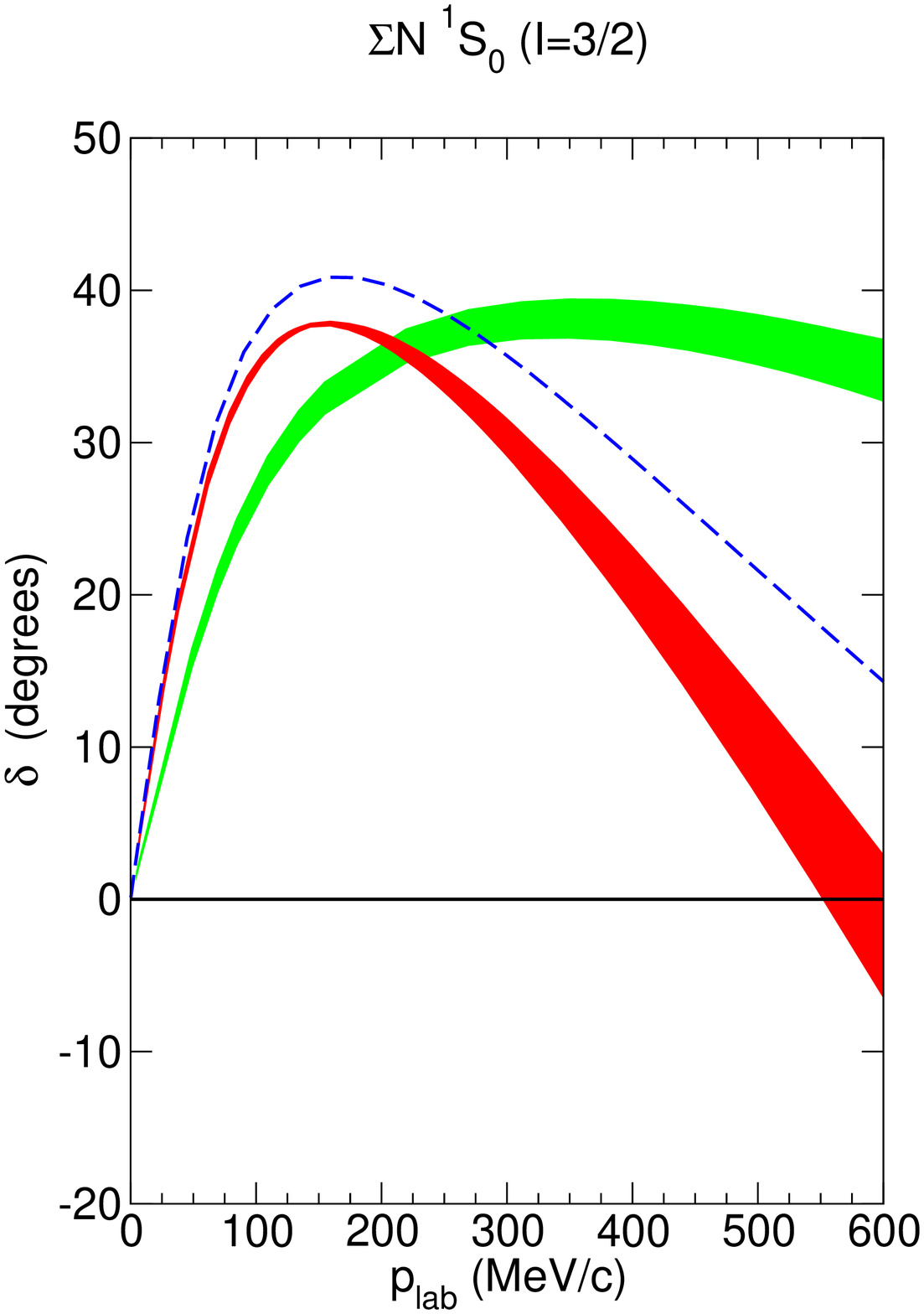}
\includegraphics[height=64mm]{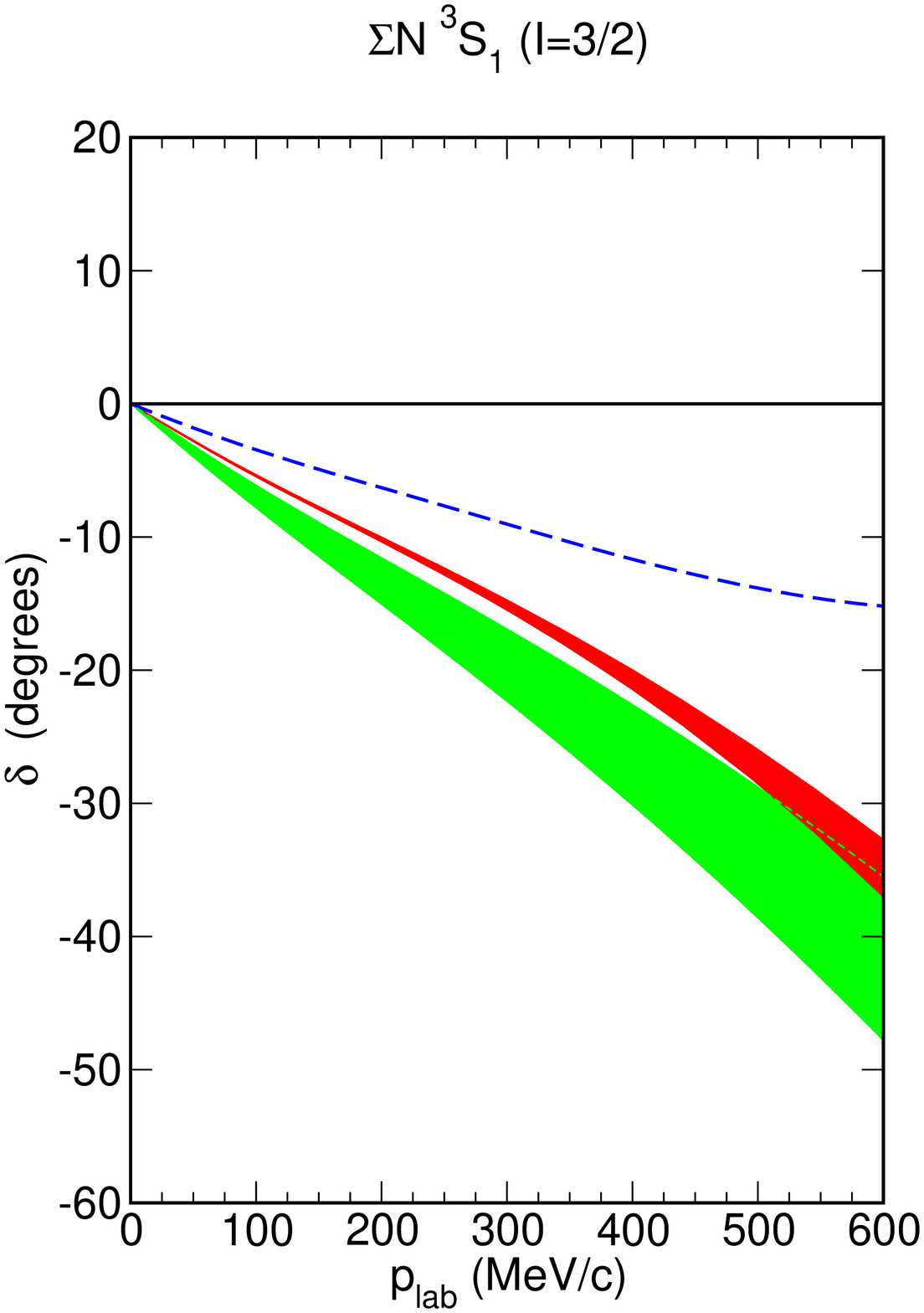}
\caption{
The $\Si^+ p$ $S$-wave phase shifts $\delta$ as a function of $p_{{\rm lab}}$. 
Same description of curves as in Fig.~\ref{fig:R5}.
}
\label{fig:R8}
\end{center}
\end{figure}

Finally, let us present predictions for a selection of $\La p$ and $\Si^+ p$ phase shifts,
evaluated in the isospin basis. They can be found in Figs.~\ref{fig:R5} - \ref{fig:R8}.
The behavior of the $^1S_0$ phase shift in the $\La p$ channel predicted at NLO is similar 
to the one of the J\"ulich '04 model and other meson-exchange potentials \cite{Rij99,Rij10}
though may be slightly more repulsive for higher momenta, cf. Fig.~\ref{fig:R5}.
The $^1P_1$ phase shift is also similar to the result of the J\"ulich model and has
opposite sign as compared to the LO result. 
Note that this partial wave is the only $P$-wave where we observed a noticeable cutoff 
dependence of the results and we counterbalanced this via a smooth variation of 
the LEC $C^{8_a}_{^1P_1}$, see Tab.~\ref{tab:F2}. In all other $P$-waves the value of
the additional LEC not determined from the $NN$ sector ($C^{8_s}_\xi$) 
is fixed independently of the cutoff $\Lambda$. 

Phase shifts for the coupled $^3S_1$-$^3D_1$ partial waves and the mixing
parameter $\epsilon_1$ can be found in Fig.~\ref{fig:R6}. Evidently, 
the $^3S_1$ phase shift based on the NLO interaction passes through 90$^\circ$ 
slightly below the $\Si N$ threshold. However, also in the $^3D_1$ phase shift 
and the mixing parameter we observe an appreciable increase near that threshold.
A rise of the $^3S_1$ (or $^3D_1$) phase shift beyond 90$^\circ$ is typical 
for the presence of an unstable bound state in the $\Si N$ system 
\cite{Badalyan82,Miyagawa99}, see also the discussion in \cite{Mac13}. 
In case of the $YN$ interaction at LO and 
the J\"ulich '04 model \cite{Hai05} none of the phases pass through 90$^\circ$ 
and an ordinary cusp is predicted. Such a behavior is caused by an inelastic 
virtual state in the $\Si N$ system. It should be said, however, that 
the majority of the meson-exchange potentials \cite{Hol89,Rij99,Rij10} 
produce an unstable bound state, similar to our NLO interaction. 
The only characteristic difference of the chiral EFT interactions to the 
meson-exchange potentials might be the mixing parameter $\epsilon_1$ 
which is fairly large in the former case and close to 45$^\circ$ at the
$\Si N$ threshold, see Fig.~\ref{fig:R6}. It is a reflection of the fact
that the pertinent $\Lambda p$ $T$-matrices (for the $^3S_1$$\to$$^3S_1$, 
$^3D_1$$\to$$^3D_1$, and $^3S_1$$\leftrightarrow$$^3D_1$ transitions)
are all of the same magnitude. Indeed, these amplitudes yield very similar
contributions to the $\Lambda p$ cross section in the vicinity of the 
$\Si N$ threshold.
 
In this context let us mention a recent experimental paper where the
energy region around the $\Sigma N$ threshold was investigated in 
the reaction $pp\to K^+ \La p$ via a measurement of the $\La p$ invariant 
mass \cite{Sam12} and where a pronounced structure was observed. 
For a discussion and summary of older measurements providing evidence 
for a strong enhancement of the $\La p$ amplitude near the $\Sigma N$ 
threshold see Ref.~\cite{Mac13}.

Predictions for the $^3P$ partial waves of the $\La p$ system 
are displayed in Fig.~\ref{fig:R7}. One can see that the $^3P_0$ and $^3P_1$ 
phase shifts are reduced at NLO as compared to those obtained at LO while 
they are larger in case of the $^3P_2$. Note that the behavior of the NLO
results is strongly influenced by the LECs as fixed from the 
corresponding $NN$ partial waves because, according to SU(3) symmetry,
the pertinent coefficient ($C^{27}$) dominates also the $\La N \to \La N$
interaction, see Tab.~\ref{tab:SU3}.  
Obviously, there are sizable quantitative differences between the results for 
the EFT interaction and the meson-exchange potential.

Results for the $\Si^+ p$ system are shown in Fig.~\ref{fig:R8},
where we restrict ourselves to the $S$-waves. 
We have switched off the Coulomb interaction for the computation
of the phase shifts so that the presented results are those for 
the $\Si N$ $I=3/2$ channel. There is no coupling to
the $\La N$ system and therefore the phase shifts are elastic in the
momentum region considered. 

Both partial waves are quite interesting. First, with regard to 
the $^1S_0$, strict SU(3) symmetry implies that $V_{NN} \equiv V_{\Si N}$, 
see Tab.~\ref{tab:SU3}, so that in an exactly SU(3) symmetric world
the corresponding $pp$ and $\Si^+ p$ phase shifts would be the same.   
In our calculation we break the symmetry already on the potential level
by using the physical masses of the exchanged pseudoscalar mesons
and in addition by using the physical baryon masses when solving the 
LS equation (\ref{LS}). As already mentioned in Sect.~\ref{sec:3} 
we had to introduce also an SU(3) breaking into the contact terms.
It turned out to be impossible to describe the $pp$ $^1S_0$ 
phase shifts and the $\Si^+ p$ cross sections with a consistent
set of LECs that fulfill SU(3) symmetry, at least on the level
of our NLO calculation. 
As a matter of fact, the $^1S_0$ amplitude of our NLO interaction 
alone saturates already more or less the experimental $\Si^+ p$ cross 
sections. Employing the LECs as fixed
from a fit to the $pp$ phase shifts yields a potential that is much more
attractive and that produces a near-threshold bound state in the $\Si^+ p$
system and, consequently, cross sections that are roughly four
times larger than experiment. 

Apparently meson-exchange interactions like the Nijmegen potentials are
able to describe the $NN$ and $YN$ systems simultaneously, without
any major obvious SU(3) breaking. 
However, usually a special (phenomenological) treatment of the short-ranged 
part of the potential is required, as discussed, for example, in Ref.~\cite{Rij10}. 
In the potentials of the J\"ulich group \cite{Hol89,Hai05}
SU(3) symmetry is broken via the employed vertex form factors. 

The coupled $^3S_1$-$^3D_1$ partial wave of the $\Si^+ p$ system has a strong 
influence on the properties of the $\Sigma$ in nuclear matter. Specifically, a
repulsive $\Sigma$ single-particle potential in nuclear matter \cite{Kohno06,Dab08}, 
as supported by present days experimental evidence \cite{SIG1,SIG2,SIG3},
can only be achieved with a repulsive $^3S_1$ interaction in the
$I=3/2$ channel. In the course of our investigation we found that we can
fit the available $YN$ data equally well with an attractive or a repulsive
$^3S_1$ interaction. The difference in the achieved $\chi^2$ is
marginal as already pointed out above. 
In view of the SU(3) structure given in Tab.~\ref{tab:SU3} this may be not
too surprising. The $^3S_1$ partial wave of the $\Si N$ $I=3/2$ channel
resides in the $10$ representation which does not contribute to any of
the other $NN$ and $YN$ systems. Of course, its contribution enters
indirectly because the measured (physical) reactions $\Si^- p \to \Si^- p$ 
and $\Si^- p \to \Si^0 n$ involve amplitudes that result from combinations 
of the $\Si N$ $I=3/2$ and $I=1/2$ interaction potentials. 

Our NLO interaction produces a moderately repulsive $^3S_1$ phase shift
as can be seen in Fig.~\ref{fig:R8}, comparable to the one predicted by
the LO potential. For the latter, calculations of the $\Sigma$ single-particle 
potential have been performed \cite{Kohno10} and indicate a values of
$U_\Si (k=0) \approx 12$ MeV at nuclear matter saturation density. 

Recent lattice QCD calculations \cite{Beane12} suggest a much more strongly 
repulsive $^3S_1$ phase shift in the $\Si N$ $I=3/2$ channel, when
extrapolating the lattice results obtained for $m_{\pi} \approx 389$~MeV to the 
physical pion mass. 
But within our framework we cannot accommodate a much more repulsive $^3S_1$ 
amplitude. Any sizable increase in the repulsion would yield a
$^3S_1$ amplitude which practically saturates the experimental $\Si^+ p$ 
cross section alone and, consequently, there would be no room anymore for the 
contribution from the spin-singlet amplitude --
which is likewise large as discussed above.
Thus, a more strongly repulsive $^3S_1$ phase shift would immediately result 
in a dramatic deterioration of the achieved $\chi^2$. 

\section{Summary and outlook}
\label{sec:5}

Chiral effective field theory, successfully applied in Refs.~\cite{Entem:2003ft,Epe05} 
to the $NN$ interaction, also works well for the baryon-baryon interactions 
in the strangeness $S=-1$ ($\Lambda N - \Sigma N$) and 
$S=-2$ ($\La \La -\Xi N - \Si\Si$) \cite{Polinder:2006zh,Polinder:2007mp} sectors. 
As shown in our earlier work \cite{Polinder:2006zh}, already at leading order 
the bulk properties of the $\Lambda N$ and $\Sigma N$ systems can be 
reasonably well accounted for. 
The new results for the $YN$ interaction presented here, obtained to 
next-to-leading order in the Weinberg counting, are very encouraging. 
First there is a visible improvement in the quantitative reproduction of
the available data on $\La N$ and $\Si N$ scattering and, secondly, the 
dependence on the regularization scheme is strongly reduced as compared to
the leading-order result. Indeed the 
description of the $YN$ system achieved at NLO is now on the same level of 
quality as the one by the most advanced meson-exchange $YN$ interactions. 

At the considered order there are contributions from one- and two-pseudoscalar-meson 
exchange diagrams and from four-baryon contact terms without and with two 
derivatives. SU(3) flavor symmetry is used as guiding principle in
the derivation of the interaction. This means that all the coupling constants at 
the various baryon-baryon-meson vertices are fixed from SU(3) symmetry and the 
symmetry is also exploited to derive relations between the contact terms. 
Furthermore, contributions from all mesons of the pseudoscalar octet 
($\pi$, $K$, $\eta$) are taken into account. 
The SU(3) symmetry is, however, broken by the masses of the pseudoscalar 
mesons and the baryons for which we take the known physical values.

Given the presently available data base with its still large uncertainties, 
we are able to achieve a combined description of the $\La N$ and $\Si N$ 
systems without any explicit SU(3) breaking in the contact interactions.
However, we found that a simultaneous description of the $NN$ interaction 
with contact terms fulfilling strict SU(3) symmetry is not possible. Here 
the strength of the contact interaction in the $27$-representation that is 
needed to reproduce the $pp$ (or $np$) 
$^1S_0$ phase shifts is simply not compatible with what is required 
for the description of the empirical $\Si^+ p$ cross section.
 
In any case, it is likely that future (and more precise) data will demand 
to depart from SU(3) symmetry in the contact terms even with regard to 
the $\La N$ and $\Si N$ interactions. 
Especially studies of few- and many-body systems involving hyperons, which
can be done in a consistent way in the framework followed in the present
work, could provide evidence for the need of an explicit SU(3) breaking.
So far reliable microscopic calculations that utilize directly the
elementary $YN$ interaction are only possible (and have been done) for 
systems with at most four baryons, namely
within the Faddeev-Yakubovsky approach \cite{Nog02}. 
However, it is expected that new approaches that have been developed
and refined over the past few years and that 
are successfully applied in studies of ordinary nuclei allow 
one to study also hypernuclei with a larger number of baryons 
with comparable accuracy. 
Thus, we consider the present investigation as a first and exploratory 
step towards a more thorough understanding of the baryon-baryon interaction 
in the strangeness sector. 

\section*{Acknowledgements}

We thank Evgeny Epelbaum for his collaboration in the early stages of this
investigation. This work is supported in part by the DFG and the NSFC through
funds provided to the Sino-German CRC 110 ``Symmetries and
the Emergence of Structure in QCD'' and by the EU Integrated 
Infrastructure Initiative HadronPhysics3. S.~Petschauer thanks the ``TUM Graduate School''.
Part of the numerical calculations have been performed on
the supercomputer cluster of the JSC, J\"ulich, Germany.

\appendix
\section{Two--pseudoscalar-meson exchange contributions}
\label{app:A}
\countzero 

In this section we present the next-to-leading order contributions from two-pseudoscalar-meson exchange shown in Fig.~\ref{fig:feynman}.
The calculation of these potentials was done according to the rules of SU(3) heavy-baryon chiral perturbation theory in the center-of-mass frame and in the isospin limit.
Ultraviolet divergences are treated by dimensional regularization, which introduces a scale \(\lambda\).
These divergences are parametrized in an \(R\)-term which is absorbed by contact terms.
In the used renormalization scheme it is defined as
\begin{equation}
 R = \frac 2 {d-4} + \gamma_E - 1 - \ln\left( 4\pi \right)\,,
\end{equation}
with the space-time dimension \(d\). 

As for the one-pseudoscalar-meson exchange, Eq.~\eqref{OBE}, 
the two-pseudoscalar-meson exchange potentials are given by a general expression, 
where the proper meson masses have to be inserted, and which has to be multiplied with appropriate SU(3) factors \(N\).
We display this factor next to the Feynman diagram and in the corresponding tables. The factors
contain coupling constants and isospin factors and are different for each combination of baryons and mesons.

In the following we will show the results for the five diagram types one after another.


\subsection{Planar box}

\begin{center}
\begin{minipage}{.4\textwidth} \centering
 \(\vcenter{\hbox{\includegraphics[width=\feynwidthbig]{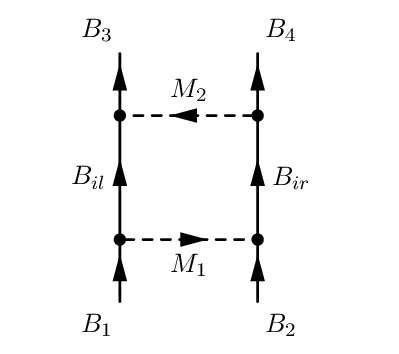}}}\)
 \vspace{-.5\baselineskip}
 \captionof{figure}{Planar box}\label{fig:pb}
\end{minipage}
\(N=f_{B_1B_{il}M_1}f_{B_{il}B_3M_2}f_{B_2B_{ir}M_1}f_{B_{ir}B_4M_2}{\mathcal I}_{B_1B_2\to B_3B_4}\)
\end{center}
The planar box, Fig.~\ref{fig:pb}, has an irreducible part and a reducible part coming from iterating the one-meson exchange to second order.
The reducible part is generated by solving the Lippmann-Schwinger equation and, therefore, is not part of the potential.
The irreducible part of this diagram can be obtained by regarding the residues of poles of the meson propagators, 
but leaving out poles of the baryon propagators.
One obtains a central potential (\(\mathbbm1 V_\mathrm C\)), 
a spin-spin potential
$(\mbox{\boldmath $\sigma$}_1\cdot\mbox{\boldmath $\sigma$}_2 V_S)$
and a tensor-type potential 
$(\mbox{\boldmath $\sigma$}_1\cdot{\bf q} \, \mbox{\boldmath $\sigma$}_2\cdot{\bf q}  V_T)$. 
With the momentum transfer \(q=\left|\bf{p^{\,\prime}} - \bf{p}\,\right|\) and the masses 
of the two exchanged mesons, \(m_1\) and \(m_2\), the irreducible potentials can be written in closed analytical form,
\begin{align}
\label{P1}
 V^\text{planar box}_\mathrm{irr,\,C}(q) =&\frac{N}{192 \pi ^2}\Bigg[\frac{5}{3}q^2+\frac{\left(m_1^2-m_2^2\right)^2}{q^2} + 16 \left(m_1^2+m_2^2\right) + \left(23 q^2+45 \left(m_1^2+m_2^2\right)\right)\left(R+2\ln \frac{\sqrt{m_1 m_2}}{\lambda }\right)\notag\\
 &+\frac{m_1^2-m_2^2}{q^4} \left(12 q^4+\left(m_1^2-m_2^2\right)^2-9 q^2 \left(m_1^2+m_2^2\right)\right) \ln \frac{m_1}{m_2}+\frac{2}{w^2\left(q\right)} \bigg(23 q^4-\frac{\left(m_1^2-m_2^2\right)^4}{q^4}\notag\\
 &+56 \left(m_1^2+m_2^2\right) q^2+8\frac{ m_1^2+m_2^2}{q^2}\left(m_1^2-m_2^2\right)^2+ 2 \left(21 m_1^4+22 m_1^2 m_2^2+21 m_2^4\right)\bigg) L\left(q\right)\Bigg]\,,
\end{align}
\begin{align}
\label{P2}
 V^\text{planar box}_\mathrm{irr,\,T}\left(q\right) =-\frac{N}{8 \pi ^2} \Bigg[L\left(q\right)-\frac{1}{2}-\frac{m_1^2-m_2^2}{2 q^2}\ln \frac{m_1}{m_2}+\frac{R}{2}+\ln \frac{\sqrt{m_1 m_2}}{\lambda }\Bigg] = - \frac1{q^2} V^\text{planar box}_\mathrm{irr,\,S}(q) \,,
\end{align}
where we have defined the functions
\begin{equation}
 w\left(q\right) = \frac1q\sqrt{\left(q^2+\left(m_1+m_2\right)^2\right)\left(q^2+\left(m_1-m_2\right)^2\right)}\,, \qquad
 L\left(q\right) = \frac{w\left(q\right)}{2q} \ln \frac{\left( qw\left(q\right) + q^2\right)^2 - \left(m_1^2-m_2^2\right)^2}{4m_1m_2q^2}\,.
\label{P3}
\end{equation}
The relation 
$(\mbox{\boldmath $\sigma$}_1\times {\bf q}) \cdot (\mbox{\boldmath $\sigma$}_2\times {\bf q}) = 
{\bf q}^2 \mbox{\boldmath $\sigma$}_1\cdot\mbox{\boldmath $\sigma$}_2 - 
(\mbox{\boldmath $\sigma$}_1\cdot{\bf q}) \, (\mbox{\boldmath $\sigma$}_2\cdot{\bf q})$ 
is exploited for the connection between the spin-spin and tensor-type potential.
The isospin factors \(\mathcal I\) can be found in Tab.~\ref{tab:isoP}. 
Two-meson exchange contributions that involve a single $K$ (or $\Kb$) lead to an interchange of
the nucleon and the hyperon in the final state. The recoupling of the corresponding isospin
states yields a factor $(-1)$ for some of the transitions that is already included in the values
given in Tab.~\ref{tab:isoP}. The same applies to the Tables given below. 
In this context let us mention that for diagrams with an interchange of the nucleon and the
hyperon in the final state, likewise an appropriate treatment of the spin-space part is required. In 
particular, the momentum transfer is then given by \(q=\left|\bf{p^{\,\prime}} + \bf{p}\,\right|\).

\begin{table}
\caption{Isospin factors ${\mathcal I}$ for planar box diagrams.
$B_{il}B_{ir}$ indicates the two baryons in the intermediate state and
$\pi\pi$ etc. the exchanged pair of mesons $M_1M_2$, cf. Fig.~\ref{fig:pb}. 
}
\begin{center}
\renewcommand{\arraystretch}{1.20}
\begin{tabular}[t]{|c||c|cccccc||c|ccc|}
\hline \hline
transition & \MMBB & $\pi\pi$ & $\pi\eta$ & $\eta\pi$ & $\eta\eta$ & $\pi K$ & $\eta K$ & \MMBB & 
$\Kb\pi$ & $\Kb \eta$ & $\Kb\,\Kb$ \\ 
(isospin)  &       & & & & & & & & & &\\
\hline \hline
$N N\to N N$ & & & & & & & & & & &\\
($I=0$) & $N N$         & $ 9$ & $-3$ & $-3$ & $1$ & $0$ & $0$ & $NN$ & $0$ & $0$ & $0$ \\
($I=1$) & $N N$         & $ 1$ & $ 1$ & $1$ & $1$ & $0$ & $0$ & $NN$ & $0$ & $0$ & $0$ \\
\hline 
$\Sigma N\to \Sigma N$  & & & & & & & & & & &\\
($I=1/2$) & $\Sigma N$  & $4$ & $-2$ & $ -2$ & $1$ & $2$ & $-1$ & $N\Sigma$ & $ 2$ & $-1$ & $1$ \\
          & $\Lambda N$ & $3$ & $ 0$ & $ 0$ & $ 0 $ & $3$ & $0$ & $N\Lambda$ & $ 3$ & $ 0$ & $3$ \\
($I=3/2$) & $\Sigma N$  & $1$ & $ 1$ & $ 1$ & $ 1 $ & $2$ & $ 2$ & $N\Sigma$ & $ 2$ & $ 2$ & $4$ \\
\hline 
$\Lambda N\to \Sigma N$ & & & & & & & & & & &\\
($I=1/2$) & $\Sigma N$  & $ 2\sqrt{3}$ & $-\sqrt3$ & $0$ & $0$ & $\sqrt3$ & $0$ & $N\Sigma$ & $2\sqrt{3}$ & $-\sqrt3$ & $\sqrt{3}$ \\
          & $\Lambda N$ & $0$ & $0$ & $-\sqrt{3}$ & $ 0$ & $0$ & $ -\sqrt{3}$ & $N\Lambda$ & $-\sqrt{3}$ & 0 & $-\sqrt{3}$ \\
\hline 
$\Lambda N\to \Lambda N$ & & & & & & & & & & &\\
($I=1/2$) & $\Sigma N$  & $3$ & $ 0$ & $ 0$ & $ 0$ & $3$ & $ 0$ & $N\Sigma$ & $ 3$ & $ 0$ & $3$ \\
          & $\Lambda N$ & $0$ & $ 0$ & $ 0$ & $ 1 $ & $0$ & $ 1$ & $N\Lambda$ & $ 0$ & $ 1$ & $1$ \\
\hline \hline
\end{tabular}
\label{tab:isoP}
\end{center}
\end{table}

Note that the potential given above and also the following potentials are finite for \(q\rightarrow0\).
Terms proportional to $1/q^2$ and/or $1/q^4$ in Eqs.~(\ref{P1}) and (\ref{P2}) are cancelled by
corresponding terms in the functions $L(q)$ and $w(q)$ of Eq.~(\ref{P3}) in the limit of small $q$. 
We perform an expansion of the potentials in a power series for small $q$ so that these 
cancellations can be taken into account analytically and we obtain stable results in the 
numerical calculations. 
For equal meson masses some terms in the potentials vanish and the expressions reduce to the results 
in Refs.~\cite{Epe00,Kaiser1997}. This is the case in the $NN$ interactions of 
Refs.~\cite{Epe00,Entem:2003ft,Epe05} based on chiral EFT, where only contributions from 
two-pion exchange are taken into account.

In the actual calculations only the non-polynomial part of Eqs.~(\ref{P1}) and (\ref{P2}) 
is taken into account, i.e.\ the pieces proportional to $L(q)$ and to $1/q^2$ and $1/q^4$. 
The polynomial part only renormalizes the LO and NLO contact terms and, therefore, is not 
considered. The contributions involving the regularization scheme (i.e.\ that depend on $R$) 
are likewise omitted. As already said, their effect is assumed to be also absorbed by the
contact terms and a renormalization of the coupling constants, see, e.g., the
corresponding discussion in Appendix A of \cite{Epe00} for the $NN$ case. 
We want to remark that the majority of those terms omitted involve the masses of
the pseudoscalar mesons and, therefore, generate an SU(3) symmetry breaking. Thus, 
the SU(3) symmetry imposed on our contact interaction (at least for $\La N$ and $\Si N$)
is understood as one that is fulfilled on the level of the renormalized coupling constants. 
 
All statements above apply also to the other contributions to the potential described below. 

\subsection{Crossed box}

\begin{center}
\begin{minipage}{.4\textwidth} \centering
 \(\vcenter{\hbox{\includegraphics[width=\feynwidthbig]{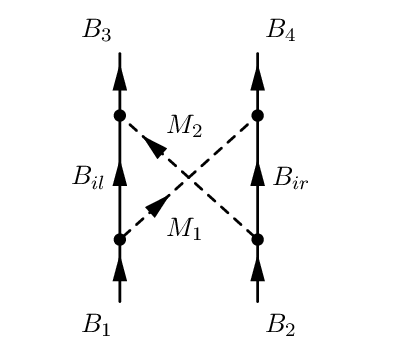}}}\)
 \vspace{-.5\baselineskip}
 \captionof{figure}{Crossed box}\label{fig:cb}
\end{minipage}
\(N=f_{B_1B_{il}M_1}f_{B_{il}B_3M_2}f_{B_2B_{ir}M_2}f_{B_{ir}B_4M_1}{\mathcal I}_{B_1B_2\to B_3B_4}\)
\end{center}
The crossed box diagrams, Fig.~\ref{fig:cb}, yield a central, a spin-spin, and a tensor-type potential.
Due to the similar structure but different kinematics, the potentials resulting from the crossed boxes
are the same as those of the planar box, up to a sign:
\begin{align}
 V^\text{crossed box}_\mathrm{C}(q) &= - V^\text{planar box}_\mathrm{C,\,irr}(q)\,,\\
 V^\text{crossed box}_\mathrm{T}(q) &= - \frac1{q^2} V^\text{crossed box}_\mathrm{S}(q) = \phantom{-} V^\text{planar box}_\mathrm{T,\,irr}(q)\,.
\end{align}
Note that there is no iterated part in case of the crossed boxes.
The corresponding isospin factors \(\mathcal I\) can be found in Tab.~\ref{tab:isoX}.


\begin{table}
\caption{Isospin factors ${\mathcal I}$ for crossed box diagrams. 
$B_{il}B_{ir}$ indicates the two baryons in the intermediate state and
$\pi\pi$ etc. the exchanged pair of mesons $M_1M_2$, cf.\ Fig.~\ref{fig:cb}.
}
\begin{center}
\renewcommand{\arraystretch}{1.20}
\begin{tabular}[t]{|c||c|cccccc||c|ccc|}
\hline \hline
transition & \MMBB & $\pi\pi$ & $\pi\eta$ & $\eta\pi$ & $\eta\eta$ & $\Kb\pi$ & $\Kb\eta$ 
& \MMBB & $\pi K$ & $\eta K$& $K K$ \\ 
(isospin)  & & & & & & & & & & & \\
\hline \hline
$N N\to N N$ & & & & & & & & & & &\\
($I=0$) & $N N$ & $-3$ & $-3$ & $-3 $ & $1$ & $0$ & $0$ &$\Si\Si $ & $0$ & $0$ & $3$ \\ 
        &       & $ $ & $  $ & $ $ & $ $ & $ $ & $ $ &$\La\Si,\Si\La $ & $0$ & $0$ & $3$ \\ 
        &       & $ $ & $  $ & $ $ & $ $ & $ $ & $ $ & $\La\La $ & $0$ & $0$ & $-1$ \\ 
($I=1$) & $N N$ & $ 5$ & $ 1$ & $ 1 $ & $1$ & $0$ & $0$ &$\Si\Si $ & $0$ & $0$ & $5$ \\ 
        &       & $ $ & $  $ & $ $ & $ $ & $ $ & $ $ &$\La\Si,\Si\La $ & $0$ & $0$ & $1$ \\ 
        &       & $ $ & $  $ & $ $ & $ $ & $ $ & $ $ & $\La\La $ & $0$ & $0$ & $ 1$ \\ 
\hline 
$\Sigma N\to \Sigma N$  & & & & & & & & & & &\\
($I=1/2$) & $\Sigma N$  & $0$ & $-2$ & $-2$ & $1$ & $0$ & $0$ & $\Si\Si$ & $ 0$ & $-1$ & $0$ \\ 
          & $N N$       & $0$ & $0$ & $0$ & $0 $ & $5$ & $-1$ & $\Xi \Si$ & $ 0$ & $0$ & $5$ \\ 
          & $\Lambda N$ & $-1$ & $ 0$ & $ 0$ & $ 0 $ & $0$ & $ 0$ & $\La \La$ & $ 1$ & $ 0$ & $0$ \\ 
          &             & $ $ & $  $ & $  $ & $   $ & $ $ & $  $ & $\Xi \La$ & $ 0$ & $ 0$ & $-1$ \\ 
          &             & $ $ & $ $ & $ $ & $ $ & $ $ & $  $ & $\La \Si$,$\Si \La$ & $2$ & $0$ & $0$ \\ 
($I=3/2$) & $\Sigma N$  & $3$ & $ 1$ & $ 1$ & $ 1 $ & $0$ & $0$ & $\Si\Si$ & $ 3$ & $ 2$ & $0$ \\ 
          & $N N$       & $0$ & $0$ & $0$ & $0 $ & $2$ & $2$ & $\Xi \Si$ & $ 0$ & $ 0$ & $2$ \\ 
          & $\Lambda N$ & $2$ & $ 0$ & $0$ & $ 0 $ & $0$ & $ 0$ & $\La\La$ & $ 1$ & $ 0$ & $0$ \\ 
          &             & $ $ & $  $ & $  $ & $   $ & $ $ & $  $ & $\Xi \La$ & $ 0$ & $ 0$ & $2$ \\ 
          &             & $ $ & $ $ & $  $ & $ $ & $ $ & $ $ & $\La\Si$,$\Si\La$ & $-1$ & $0$ & $0$ \\ 
\hline 
$\Lambda N\to \Sigma N$ & & & & & & & & & & &\\
($I=1/2$) & $\Sigma N$  & $-2\sqrt{3}$ & $-\sqrt{3} $ & $0$ & $0$  & $0$ & $0$& $\Si\Si$ & $2\sqrt{3}$ & $0$ & $0$ \\ 
          & $N N$       & $0$ & $0$ & $ 0$ & $0$ & $\sqrt{3} $ & $-\sqrt{3}$ & $\Xi\Si$ & $0$ & $0$ & $-\sqrt{3}$ \\ 
          & $\Lambda N$ & $0$ & $ 0$ & $ -\sqrt{3}$ & $0$ & $0$ & $0$& $\Si\La$ & $-\sqrt{3}$ & $0$ & $0$ \\ 
          &             & $ $ & $ $ & $  $ & $ $  & $ $ & $ $& $\La\Si$ & $0$ & $-\sqrt{3}$ & $0$ \\ 
          &             & $ $ & $ $ & $  $ & $ $  & $ $ & $ $& $\Xi\La$ & $0$ & $0$ & $\sqrt{3}$ \\ 
\hline 
$\Lambda N\to \Lambda N$ & & & & & & & & & & &\\
($I=1/2$) & $\Sigma N$  & $3$ & $ 0$ & $ 0$ & $ 0$ & $0$ & $ 0$& $\Si\Si$ & $ 3$ & $ 0$ & $0$ \\ 
          & $N N$       & $0$ & $ 0$ & $ 0$ & $ 0$ & $3$ & $ 1$& $\Xi\Si$ & $ 0$ & $ 0$ & $3$ \\ 
          & $\Lambda N$ & $0$ & $ 0$ & $ 0$ & $ 1$ & $0$ & $0$& $\Xi\La$  & $0$ & $0$ & $1$ \\ 
          &             & $ $ & $  $ & $  $ & $  $ & $ $ & $ $& $\La\La$ & $0$ & $1$ & $0$\\ 
\hline \hline
\end{tabular}
\label{tab:isoX}
\end{center}
\end{table}

\subsection{Triangles}

\begin{center}
\begin{minipage}{0.49\textwidth}
\centering
\includegraphics[width=\feynwidthbig]{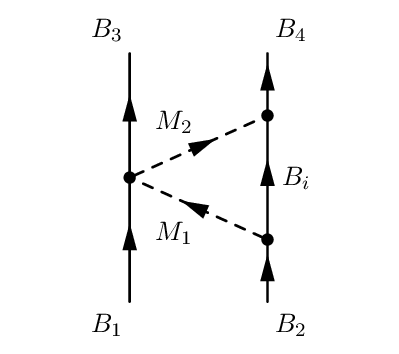}
 \vspace{-.5\baselineskip}
 \captionof{figure}{Left triangle}\label{fig:triL}
 \vspace{-\baselineskip}
\[
N=f_{B_2B_{i}M_1}f_{B_iB_4M_2}{\mathcal I}_{B_1B_2\to B_3B_4}
\]
\end{minipage}
\begin{minipage}{0.49\textwidth}
\centering
\includegraphics[width=\feynwidthbig]{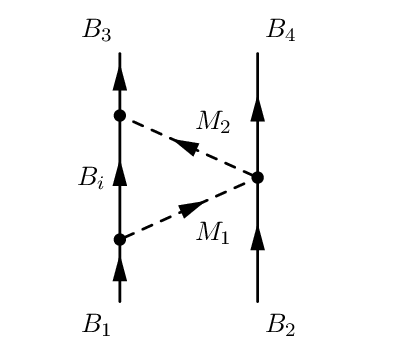}
 \vspace{-.5\baselineskip}
 \captionof{figure}{Right triangle}\label{fig:triR}
 \vspace{-\baselineskip}
\[
N=f_{B_1B_{i}M_1}f_{B_iB_3M_2}{\mathcal I}_{B_1B_2\to B_3B_4}
\]
\end{minipage}
\end{center}
The two triangle diagrams, Figs.~\ref{fig:triL} and \ref{fig:triR}, lead to equal potentials, but with different SU(3) factors.
They contribute only to the central potential and the corresponding expression is given by
\begin{align}
 V^\text{triangle}_\mathrm C (q) = &-\frac{N}{768 \pi ^2 f_0^2} \Bigg[-2 \left(m_1^2+m_2^2\right)+\frac{\left(m_1^2-m_2^2\right)^2}{q^2}-\frac{13}{3} q^2 + \left(8 \left(m_1^2+m_2^2\right)-\frac{2 \left(m_1^2-m_2^2\right)^2}{q^2}+10 q^2\right) L\left(q\right)\notag\\
 &+\frac{m_1^2-m_2^2}{q^4} \left(\left(m_1^2-m_2^2\right)^2-3 \left(m_1^2+m_2^2\right) q^2\right) \ln \frac{m_1}{m_2} + \left(9 \left(m_1^2+m_2^2\right)+5 q^2\right) \left(R+2\ln \frac{\sqrt{m_1 m_2}}{\lambda }\right)\,.
\end{align}
The isospin factors \(\mathcal I\) are stated in Tables \ref{tab:isoTl} and \ref{tab:isoTr}.


\begin{table}
\caption{Isospin factors ${\mathcal I}$ for triangle diagrams with the $BBMM$ vertex at the left baryon.
$B_i$ denotes the baryon in the intermediate state
and $\pi\pi$ etc. the exchanged pair of mesons $M_1M_2$, cf.\ Fig.~\ref{fig:triL}. 
}
\begin{center}
\renewcommand{\arraystretch}{1.20}
\begin{tabular}[t]{|c|c|cccc||c|ccc|}
\hline \hline
transition & \MMB & $\pi\pi$ & $\pi \Kb$& $\eta \Kb$ & $\Kb\, \Kb$ 
& \MMB & $K \pi$ & $K\eta$ & $K K$ \\ 
(isospin)  & & & & & & & & & \\
\hline \hline
$N N\to N N$ & & & & & & & & &\\
($I=0$) & $N $  & $12$ & $0$ & $0$&$0$ & $\Si$ & $0$ & $0$ & $-12$ \\
($I=1$) & $N $  & $-4$ & $0$ & $0$&$0$ & $\Si$ & $0$ & $0$ & $-8$  \\
        &         & & & & &$\La$ & $0$ & $0$ & $-4$   \\
\hline 
$\Sigma N\to \Sigma N$ & & & & & & & & &\\
($I=1/2$) & $N$  & $16$ & $ 1$ & $-\sqrt{3}$&$0$ & $\Si$ & $-2$ & $\sqrt{3}$ & $-4$ \\ 
          &      & $ 0$ & $0$& $0$& $0$ & $\La$ & $ 1$ & $0$ & $4$ \\ 
($I=3/2$) & $N$  & $-8$ & $-2$ & $2\sqrt{3}$&$0$ & $\Si$ & $ 4$ & $-2\sqrt{3}$ & $ 2$ \\ 
          &      & $0$ & $0$ & $0$ & $0$ & $\La$ & $-2$ & $0$ & $-2$ \\ 
\hline 
$\Lambda N\to \Sigma N$ & & & & & & & & &\\
($I=1/2$) & $N $   & $0$ & $3$ & $-3\sqrt{3}$&$0$ & $\Si$ & $-6$ & $3\sqrt{3}$ & $0$ \\ 
          &        & $0$ & $0$&$0$ &$0$ & $\La$ & $3$ & $ 0$ & $0$ \\ 
\hline 
$\Lambda N\to \Lambda N$ & & & & & & & & &\\
($I=1/2$) & $N$  & $0$ & $3\sqrt{3}$ & $3$&$0$ & $\Si$ & $-3\sqrt{3}$ & $0$ & $0$ \\ 
          &      & $0$ & $0$ & $0$ &$0$ & $\La$ & $0$ & $-3$  & $0$ \\ 
\hline\hline
\end{tabular}
\label{tab:isoTl}
\end{center}
\end{table}

\begin{table}
\caption{Isospin factors ${\mathcal I}$ for triangle diagrams with the $BBMM$ vertex at the right baryon.
$B_i$ denotes the baryon in the intermediate state
and $\pi\pi$ etc. the exchanged pair of mesons $M_1M_2$, cf.\ Fig.~\ref{fig:triR}.
}
\begin{center}
\renewcommand{\arraystretch}{1.20}
\begin{tabular}[t]{|c|c|cccc||c|ccc|}
\hline \hline
transition & \MMB & $\pi\pi$ & $\pi K$ & $\eta K$ & $KK$ 
& \MMB & $\Kb \pi$ & $\Kb\eta$ & $\Kb\,\Kb$ \\ 
(isospin)  & & & & & & & & & \\
\hline \hline
$N N\to N N$ & & & & & & & & &\\
($I=0$) & $N $  & $12$ & $0$ & $0$&$0$ & &  & &\\
 & $\Si$  & $0$ & $0$ & $0$&$-12$ & & & &\\
($I=1$) & $N $  & $-4$ & $0$ & $0$&$0$ & & & &\\
 & $\Si$  & $0$ & $0$ & $0$&$-8$ & & & &\\
 & $\La$  & $0$ & $0$ & $0$&$-4$ & & & &\\
\hline
$\Sigma N\to \Sigma N$ & & & & & & & & &\\
($I=1/2$) & $ \Si$ & $4$ & $-2$ & $\sqrt{3}$ & $0$& $N$ & $1$ & $-\sqrt{3}$ & $10$\\ 
          & $\La $ & $4$ & $ 1$ & $0$ & $0$ & $$& $$& $$& $$\\ 
          & $\Xi $ & $0$ & $0$ & $0$ & $-2$& $$& $$& $$& $$\\ 
($I=3/2$) & $ \Si$ & $-2$ & $ 4$ & $-2\sqrt{3}$ & $0$& $N$ & $-2$ & $2\sqrt{3}$ & $4$\\ 
          & $\La $ & $-2$ & $-2$ & $0$ & $0$ & $$& $$& $$& $$\\ 
          & $\Xi $ & $0$ &  $0$ & $0$ & $-8$& $$& $$& $$& $$\\ 
\hline 
$\Lambda N\to \Sigma N$ & & & & & & & & &\\
($I=1/2$) & $\Si$ & $4\sqrt{3}$ & $3\sqrt{3}$ & $0$ & $0$& $N$ & $-3\sqrt{3}$& $-3$ & $-2\sqrt{3}$\\ 
          & $\La$ & $0$ & $0$ & $3$ & $0$ & $$& $$& $$& $$\\ 
          & $\Xi$ & $0$ & $0$ & $0$ & $-2\sqrt{3}$& $$& $$& $$& $$\\ 
\hline 
$\Lambda N\to \Lambda N$ & & & & & & & & &\\
($I=1/2$) & $ \La$ & $0$ & $ 0$ & $-3$ & $0$& $N$ & $ 3\sqrt{3}$ & $3$ & $6$\\ 
          & $ \Si$ & $0$ & $ -3\sqrt{3}$ & $0$ & $0$ & $$& $$& $$& $$\\ 
          & $\Xi $ & $0$ & $0$ & $0$ & $-6$& $$& $$& $$& $$\\ 
\hline \hline
\end{tabular}
\label{tab:isoTr}
\end{center}
\end{table}

\subsection{Football}

\begin{center}
\begin{minipage}{.25\textwidth} \centering
 \(\vcenter{\hbox{\includegraphics[width=\feynwidthbig]{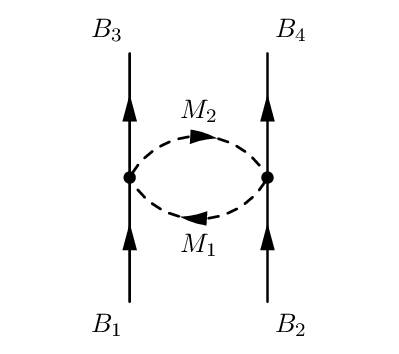}}}\)
 \vspace{-.5\baselineskip}
 \captionof{figure}{Football diagram}\label{fig:foot}
 \vspace{\baselineskip}
\end{minipage}
\qquad\qquad\qquad
\(N={\mathcal I}_{B_1B_2\to B_3B_4}\)
\end{center}
The football diagrams, Fig.~\ref{fig:foot}, give contributions to the central potential only,
\begin{align}
 V^\text{football}_\mathrm C (q) = &\frac{N}{3072 \pi ^2 f_0^4}\Bigg[-2 \left(m_1^2+m_2^2\right)-\frac{\left(m_1^2-m_2^2\right)^2}{2 q^2}-\frac{5}{6} q^2   +   \left(3 \left(m_1^2+m_2^2\right)+q^2\right) \left(\frac R2 + \ln \frac{\sqrt{m_1 m_2}}{\lambda }\right) \notag\\
 &-\frac{m_1^2-m_2^2}{2 q^4} \left(\left(m_1^2-m_2^2\right)^2+3 \left(m_1^2+m_2^2\right) q^2\right) \ln \frac{m_1}{m_2} +w^2\left(q\right) L\left(q\right) \Bigg]\,.
\end{align}
The isospin factors \(\mathcal I\) can be found in Table \ref{tab:isoF}. 


\begin{table}
\caption{Isospin factors ${\mathcal I}$ for football diagrams.
$\pi\pi$ etc. indicates the exchanged pair of mesons $M_1M_2$, cf. Fig.~\ref{fig:foot}.
}
\begin{center}
\renewcommand{\arraystretch}{1.20}
\begin{tabular}[t]{|c|ccccccc|}
\hline \hline
transition & $\pi\pi$ & $K\pi$ & $K \eta$ & 
$\pi \Kb$ & $\eta \Kb$ & $K K$ & $\Kb\,\Kb$ \\ 
(isospin)  & & & & & & & \\
\hline \hline
$N N\to N N$            & & & & & & & \\
($I=0$) & $-24$ & 0 & 0 & 0 & 0 & $12$ & $12$ \\ 
($I=1$) & $ 8$  & 0 & 0 & 0 & 0 & $20$ & $20$ \\ 
\hline 
$\Sigma N\to \Sigma N$   & & & & & & & \\
($I=1/2$) & $-32$ & $-3$ & $-3$ & $-3$ & $-3$ & $-8$ & $-8$ \\ 
($I=3/2$) & $16$  & $ 6$ & $6$ & $ 6$ & $6$ & $4$  & $4$ \\ 
\hline 
$\Lambda N\to \Sigma N$  & & & & & & & \\
($I=1/2$) & $0$ & $-9$ & $-9$ & $-9$ & $-9$ & $0$ & $0$  \\ 
\hline 
$\Lambda N\to \Lambda N$ & & & & & & & \\
($I=1/2$) & $0$ & $ 9$ & $ 9$ & $ 9$ & $ 9$ & $0$ & $0$  \\ 
\hline \hline
\end{tabular}
\label{tab:isoF}
\end{center}
\end{table}


\section{SU(3) breaking}

\subsection{SU(3) breaking in the contact terms}

In addition to the SU(3) symmetric contact terms given in Sect.~\ref{sec:2} 
that arise at NLO, there are further contact terms at this order that lead 
to an explicit SU(3) symmetry breaking. 
These terms contain new, i.e.\ additional, low-energy constants.
As already mentioned in Sect.~\ref{sec:3}, the lack of experimental data makes it
practically impossible to fix those contact terms and, therefore, we decided 
to set all the corresponding constants to zero. 
However, for completeness and for future reference, we summarize here the 
structure of the pertinent contributions. 

First there would be, in principle, relativistic corrections ($1/M_B$)
to the leading order contact terms \cite{Polinder:2006zh},
\begin{align}
 \mathcal L_1=C_1^i\trace{{(\bar B^\alpha}{(\Gamma^iB)_\alpha}{\bar B^\beta}{(\Gamma^iB)_\beta}}\,,\quad
 \mathcal L_2=C_2^i\trace{{\bar B^\alpha}{\bar B^\beta}{(\Gamma^iB)_\beta}{(\Gamma^iB)_\alpha}}\,,\quad
 \mathcal L_3=C_3^i\trace{{\bar B^\alpha}{(\Gamma^iB)_\alpha}}\trace{{\bar B^\beta}{(\Gamma^iB)_\beta}}\,,
\end{align}
which break SU(3) symmetry because of different baryon masses. Here a sum over the different elements of 
the Dirac algebra, \(\Gamma^i\in\{\mathbbm1,\gamma_\mu,\gamma_5,\gamma_5\gamma_\mu,\sigma_{\mu\nu}\}\), 
is implied. The indices \(\alpha\) and \(\beta\) are Dirac indices.
However, since the corrections to the baryon mass in the chiral limit are of order 
\(\mathcal O(q^2)\), explicit symmetry breaking due to these corrections does not appear 
up to NLO.

However, NLO contact terms with an insertion of the external field \(\chi\), which is of 
order \(\mathcal O(q^2)\), are possible. In the case of baryon-baryon scattering that field amounts to 
\begin{equation}
 \chi = 2B_0 \begin{pmatrix} m_u & 0 & 0 \\ 0 & m_d & 0 \\ 0 & 0 & m_s \end{pmatrix} \approx \begin{pmatrix} m_\pi^2 & 0 & 0 \\ 0 & m_\pi^2 & 0 \\ 0 & 0 & 2m_K^2-m_\pi^2 \end{pmatrix} \,.
\end{equation}
The following baryon-baryon contact terms with insertions of \(\chi\) are possible \cite{Pet13}:
\begin{align}
 \mathcal L_1 &= C_1^i \trace{{\bar B^\alpha} \,\chi\,{(\Gamma^iB)_\alpha}{\bar B^\beta}{(\Gamma^iB)_\beta}}\,, \notag\\
 \mathcal L_2 &= C_2^i \trace{{\bar B^\alpha} {(\Gamma^iB)_\alpha} \,\chi\,{\bar B^\beta}{(\Gamma^iB)_\beta}}\,, \notag\\\
 \mathcal L_3 &= C_3^i  \trace{{\bar B^\alpha}\,\chi\,{\bar
     B^\beta}{(\Gamma^iB)_\beta}{(\Gamma^iB)_\alpha}} + \trace{{\bar
     B^\alpha}{\bar B^\beta}{(\Gamma^iB)_\beta}\,\chi\,
   {(\Gamma^iB_2})_\alpha}\,, \notag 
\end{align}
\begin{align}
 \mathcal L_4 &= C_4^i  \trace{{\bar B^\alpha}{\bar B^\beta}\,\chi\,{(\Gamma^iB)_\beta}{(\Gamma^iB)_\alpha}}\,, \notag\\
 \mathcal L_5 &= C_5^i  \trace{{\bar B^\alpha}{\bar B^\beta}{(\Gamma^iB)_\beta}{(\Gamma^iB)_\alpha}\,\chi\,}\,, \notag\\
 \mathcal L_6 &= C_6^i  \trace{{\bar B^\alpha}{(\Gamma^iB)_\alpha}\,\chi\,}\trace{{\bar B^\beta}{(\Gamma^iB)_\beta}}\,, \notag\\
 \mathcal L_7 &= C_7^i  \trace{{\bar B^\alpha}\,\chi\,}\trace{{(\Gamma^iB)_\alpha\bar B^\beta}{(\Gamma^iB)_\beta}} + \trace{{\bar B^\alpha}{(\Gamma^iB)_\alpha\bar B^\beta}}\trace{{(\Gamma^iB)_\beta\,\chi\,}}\,,
\end{align}
and lead to an explicit SU(3) symmetry breaking linear in the quark masses, since we use \(m_u=m_d\neq m_s\).
Using these Lagrangians one obtains terms that contribute to the \(^1S_0\) and \(^3S_1\) partial waves (only). 
In the following we list the results for different transitions (with isospin \(I\)).

\(NN\rightarrow NN,\ I=0\):
\begin{align}
 V(^1S_0) &=\textstyle 0\,,  \\
 V(^3S_1) &=\textstyle 4\pi\Big[ 4 C^1_{^3S_1} m_\pi^2+C^2_{^3S_1} \left(8 m_K^2-4 m_\pi^2\right)+C^6_{^3S_1} \left(8 m_K^2-4 m_\pi^2\right) \Big] \notag\\
               &=\textstyle m_\pi^2 \hat C^{10^*} \,,
\end{align}

\(NN\rightarrow NN,\ I=1\):
\begin{align}
 V(^1S_0) &=\textstyle 4\pi\Big[ 4 C^1_{^1S_0} m_\pi^2+C^2_{^1S_0} \left(8 m_K^2-4 m_\pi^2\right)+C^6_{^1S_0} \left(8 m_K^2-4 m_\pi^2\right) \Big] \notag\\
               &=\textstyle m_\pi^2 \hat C^{27} \,,\\
 V(^3S_1) &=\textstyle 0\,,
\end{align}

 \(\Lambda N\rightarrow\Lambda N,\ I=1/2\):
 \begin{align}
  V(^1S_0) =&\textstyle 4\pi\Big[ \frac{1}{3} C^1_{^1S_0} \left(4 
 m_K^2+m_\pi^2\right)+C^2_{^1S_0} \left(3 m_K^2-\frac{4 
 }{3}m_\pi^2\right)+C^3_{^1S_0} \left(\frac{4 
 }{3}m_K^2-m_\pi^2\right)+\frac{1}{6}C^4_{^1S_0}m_\pi^2 \notag\\
                       &\textstyle+\frac{1}{6} C^5_{^1S_0} \left(2 
 m_K^2-m_\pi^2\right) +\frac{2}{3} C^6_{^1S_0} \left(5 m_K^2-2 
 m_\pi^2\right)+\frac43 C^7_{^1S_0} \left(m_K^2-m_\pi^2\right) \Big] 
 \notag\\
                      =&\textstyle \frac1{10} m_\pi^2 \left(9\hat 
 C^{27}+\hat C^{8s}\right) \,,\\
  V(^3S_1) =&\textstyle 4\pi\Big[ \frac{1}{3} C^1_{^3S_1} \left(4 
 m_K^2-m_\pi^2\right)+\frac{1}{3} C^2_{^3S_1} \left(7 m_K^2-4 
 m_\pi^2\right)+C^3_{^3S_1} \left(4 m_K^2-m_\pi^2\right)+\frac{3}{2} 
 C^4_{^3S_1} m_\pi^2 \notag\\
                       &\textstyle+C^5_{^3S_1} \left(3 m_K^2-\frac{3 
 }{2}m_\pi^2\right)+\frac{2}{3} C^6_{^3S_1} \left(5 m_K^2-2 
 m_\pi^2\right)+4 C^7_{^3S_1} \left(m_K^2-m_\pi^2\right) \Big] \notag\\
                      =&\textstyle \frac12 m_\pi^2 \left(\hat 
 C^{10^*}+\hat C^{8a}\right)\,,
 \end{align}

 \(\Lambda N\rightarrow\Sigma N,\ I=1/2\):
 \begin{align}
  V(^1S_0) &=\textstyle 4\pi\Big[ C^1_{^1S_0} m_\pi^2+C^2_{^1S_0} 
 m_K^2+C^3_{^1S_0} \left(m_\pi^2-2 m_K^2\right)-\frac{1}{2}C^4_{^1S_0} 
 m_\pi^2+\frac{1}{2} C^5_{^1S_0} \left(m_\pi^2-2 m_K^2\right)-2 
 C^7_{^1S_0} \left(m_\pi^2+ m_K^2\right) \Big] \notag\\
                              &=\textstyle \frac3{10} m_\pi^2 
 \left(\hat C^{8s}-\hat C^{27}\right) + \left(m_K^2-m_\pi^2\right) \hat 
 C_1 \,,\\
  V(^3S_1) &=\textstyle 4\pi\Big[ -C^1_{^3S_1} m_\pi^2-C^2_{^3S_1} 
 m_K^2+C^3_{^3S_1} \left(2 m_K^2+m_\pi^2\right)+\frac{3 }{2}C^4_{^3S_1} 
 m_\pi^2+C^5_{^3S_1} \left(3 m_K^2-\frac{3 }{2}m_\pi^2\right)+2 
 C^7_{^3S_1} \left( m_K^2-m_\pi^2\right) \Big] \notag\\
                              &=\textstyle \frac12 m_\pi^2 \left(\hat 
 C^{10^*}-\hat C^{8a}\right) + \left(m_K^2-m_\pi^2\right) \hat C_2\,,
 \end{align} 

\(\Sigma N\rightarrow\Sigma N,\ I=1/2\):
\begin{align}
 V(^1S_0) &=\textstyle 4\pi\Big[ -C^1_{^1S_0} m_\pi^2-C^2_{^1S_0} m_K^2+3 C^3_{^1S_0} m_\pi^2+\frac{3 }{2}C^4_{^1S_0} m_\pi^2+\frac32 C^5_{^1S_0} \left(2 m_K^2-m_\pi^2\right)+2 C^6_{^1S_0} m_K^2 \Big] \notag\\
                      &=\textstyle  \frac1{10} m_\pi^2 \left(\hat C^{27}+9\hat C^{8s}\right) + \left(m_K^2-m_\pi^2\right) \hat C_3 \,,\\
 V(^3S_1) &=\textstyle 4\pi\Big[ C^1_{^3S_1} m_\pi^2+C^2_{^3S_1} m_K^2+3 C^3_{^3S_1} m_\pi^2+\frac{3}{2}C^4_{^3S_1}m_\pi^2+\frac32 C^5_{^3S_1} \left(2 m_K^2-m_\pi^2\right)+2 C^6_{^3S_1} m_K^2 \Big] \notag\\
                     &=\textstyle \frac12 m_\pi^2 \left(\hat C^{10^*}+\hat C^{8a}\right) + \left(m_K^2-m_\pi^2\right) \hat C_4\,,
\end{align}

\(\Sigma N\rightarrow\Sigma N,\ I=3/2\):
\begin{align}
 V(^1S_0) &=\textstyle 4\pi\Big[ 2 C^1_{^1S_0} m_\pi^2+2 C^2_{^1S_0} m_K^2+2 C^6_{^1S_0} m_K^2 \Big] \notag\\
                      &=\textstyle m_\pi^2 \hat C^{27} + \left(m_K^2-m_\pi^2\right) \hat C_5 \,,\\
 V(^3S_1) &=\textstyle 4\pi\Big[ -2 C^1_{^3S_1} m_\pi^2-2 C^2_{^3S_1} m_K^2+2 C^6_{^3S_1} m_K^2 \Big] \notag\\
                      &=\textstyle m_\pi^2 \hat C^{10}\,.
\end{align}

We introduced here appropriately redefined constants $\hat C$ so that the SU(3) breaking is clearly visible.
In case of flavor symmetry where \(m_\pi=m_K\) the leading order SU(3) relations, cf.\ Tab.~\ref{tab:SU3}, are 
obtained and the constants can be absorbed in the leading order contact terms.
If \(m_\pi\neq m_K\) one obtains additional (though suppressed) constants that are proportional 
to \(m^2_K-m^2_\pi\).

\subsection{SU(3) breaking in the OBE contribution}

At NLO and NNLO there are corrections to the one-meson exchange potential 
due to differences in the baryon masses. Energy conservation leads to
\begin{equation} \label{eq:treeencons}
\sqrt{{\bf p}^{\,2}+M_{B_1}^2} + \sqrt{{\bf p}^{\,2}+M_{B_2}^2} = 
\sqrt{{\bf p}^{\,\prime2}+M_{B_3}^2} + \sqrt{{\bf p}^{\,\prime2}+M_{B_4}^2}\,,
\end{equation}
and, therefore, in some cases ${\bf p}^2 \neq {\bf p}^{\prime2}$ 
and/or \(q_0\neq0\), where 
\begin{equation}
  q^0 = \Delta E = E^3_{p^\prime}-E^1_p = E^2_p - E^4_{p^\prime}\,.
\end{equation}

Using \(M_{B_i}=M_\mathrm 0+\mathcal{O}\left(p^2\right)\) \cite{Bernard1995} and performing
an expansion in \(1/M_\mathrm 0\) one obtains 
\begin{align} \label{eq:treeNLO}
V^{OBE}_{B_1B_2\to B_3B_4} =&- 
f_{B_1B_3P}f_{B_2B_4P}\mathcal I_{B_1B_2\rightarrow B_3B_4}\,\frac{1}{{\bf q}^{\,2} -q_0^2 + m_P^2 } 
\Bigg[\boldsymbol\sigma_1 \cdot {\bf q}\, 
\boldsymbol\sigma_2 \cdot {\bf q} \notag\\
  &\quad + \frac{{\bf p}^{\,\prime2}-{\bf p}^{\,2}}{4M_\mathrm 
0^2}\left(
\boldsymbol{\sigma}_1 \cdot {\bf p}'\,
\boldsymbol{\sigma}_2 \cdot {\bf p}' - 
\boldsymbol{\sigma}_1 \cdot {\bf p}\,
\boldsymbol{\sigma}_2 \cdot {\bf p}  
\right) 
+ \frac{q_0}{M_0}\left(\boldsymbol\sigma_1\cdot\mathbf p \,\boldsymbol\sigma_2\cdot\mathbf p^{\,\prime} - \boldsymbol\sigma_1\cdot\mathbf p^{\,\prime} \boldsymbol\sigma_2\cdot\mathbf p\, \right) \Bigg]\,.
\\&\notag
\end{align}
The first term gives rise to the leading order tensor potential, see 
Eq.~\eqref{OBE}, but with a shift ${\bf q}^{\,2}\rightarrow {\bf 
q}^{\,2}-q_0^2$ caused by the mass differences
of the baryons, i.e.\ \(q^0\approx\Delta M\)  
where $\Delta M = (M_{B_1} + M_{B_4} - M_{B_3} - M_{B_2})/2 $
\cite{Rij10}. 
The last two terms in Eq.~\eqref{eq:treeNLO} give a formal contribution beyond LO.
The term proportional to $\left({\bf p}^{\,\prime2}-{\bf p}^{\,2}\right)$
contributes, in general, only off-shell. 
An exception are transitions where the baryon masses in the initial 
state are not equal to those of the final state, cf.\ Eq.~\eqref{eq:treeencons}. 
For the $YN$ interaction considered here this is only the case for
the $V_{\La N \to \Si N}$ transition potential.
In the present study we have neglected all these corrections. 

There are also deviations of the meson-baryon coupling constants from
the SU(3) values which, in principle, should be taken into account in 
a NLO calculation. Specifically, there is an explicit SU(3) symmetry breaking 
in the empirical values of the decay constants \cite{PDG},
\begin{equation}
f_\pi =  92.4 \ {\rm MeV}, \ \ f_\eta=(1.19\pm 0.01) f_\pi, \ \ f_K=(1.30\pm 0.05) f_\pi \ .
\end{equation}
A somewhat smaller SU(3) breaking occurs also in the axial coupling constants,
see \cite{Ratcliffe,Yamanishi,Donoghue} but also \cite{Ber01,General}. 
All these effects are likewise not taken into account in the present study. Rather
we use the standard SU(3) relations for the baryon-baryon-meson coupling constants
Eq.~\eqref{su3} with the values $g_A = 1.26$ and $f_0\approx f_\pi = 93~\mathrm{MeV}$.


\begin{thebibliography}{99}

\bibitem{JPARC} see list of proposals at 
{\it http://j-parc.jp/researcher/Hadron/en/Proposal\_e.html}

\bibitem{Panda} W. Erni {\it et al.} [${\rm \bar P}$anda Collaboration], arXiv:0903.3905 [hep-ex].

\bibitem{Hiruma} F. Hiruma {\it et al.}, 
{\it http://j-parc.jp/researcher/Hadron/en/pac\_1101/pdf/KEK\_J-PARC-PAC2010-12.pdf}
  
\bibitem{COSY} M. R\"oder, PhD thesis, Bochum (2011); 
M. R\"oder et al., in preparation. 

\bibitem{Hicks} K. Hicks {\it et al.}, 
{\it Measurement of the $\La N$ interaction with a deuterium target}, \hfill \break
{\rm http://www.rcnp.osaka-u.ac.jp/Divisions/plan/q-pac/ex-appro/prop/Q026.pdf}

\bibitem{Pieper}
  S.~C.~Pieper and R.~B.~Wiringa,
  Ann.\ Rev.\ Nucl.\ Part.\ Sci.\  {\bf 51} (2001) 53. 
\bibitem{Kievsky}
  A.~Kievsky, S.~Rosati, M.~Viviani, L.~E.~Marcucci and L.~Girlanda,
  J.\ Phys.\ G {\bf 35} (2008) 063101.
\bibitem{NCSM} 
  P.~Navr\'atil, S.~Quaglioni, I.~Stetcu and B.~R.~Barrett,
  J.\ Phys.\ G {\bf 36} (2009) 083101.

\bibitem{Roth:2011vt}
  R.~Roth, S.~Binder, K.~Vobig, A.~Calci, J.~Langhammer and P.~Navratil,
  Phys.\ Rev.\ Lett.\  {\bf 109} (2012) 052501
  [arXiv:1112.0287 [nucl-th]].

\bibitem{Borasoy:2006qn}
  B.~Borasoy, E.~Epelbaum, H.~Krebs, D.~Lee and U.-G.~Mei{\ss}ner,
  Eur.\ Phys.\ J.\ A {\bf 31} (2007) 105
  [nucl-th/0611087].
\bibitem{Lee} 
  D. Lee, Prog. Part. Nucl. Phys. {\bf 63} (2009) 117. 

\bibitem{Hagen:2007ew}
  G. Hagen, T. Papenbrock, D.J. Dean, A. Schwenk, A. Nogga, M. Wloch and P. Piecuch,
  Phys.\ Rev.\ C {\bf 76} (2007) 034302.

\bibitem{Hagen:2007hi}
  G. Hagen, D.J. Dean, M. Hjorth-Jenson, T. Papenbrock and A. Schwenk,
  Phys.\ Rev.\ C {\bf 76} (2007) 044305.

\bibitem{Epelbaum:2011md}
  E.~Epelbaum, H.~Krebs, D.~Lee and U.-G.~Mei{\ss}ner,
  Phys.\ Rev.\ Lett.\  {\bf 106} (2011) 192501
  [arXiv:1101.2547 [nucl-th]].

\bibitem{Nemura} 
  H.~Nemura, Y.~Akaishi and Y.~Suzuki,
  Phys.\ Rev.\ Lett.\  {\bf 89} (2002) 142504.
\bibitem{Hiyama} 
  E.~Hiyama, M.~Kamimura, Y.~Yamamoto, T.~Motoba and T.~A.~Rijken,
  Prog.\ Theor.\ Phys.\ Suppl.\  {\bf 185} (2010) 106.

\bibitem{Beane11}
  S.~R.~Beane, W.~Detmold, K.~Orginos and M.~J.~Savage,
  Prog.\ Part.\ Nucl.\ Phys.\  {\bf 66} (2011) 1.
\bibitem{Aoki12}
  S.~Aoki {\it et al.}  [HAL QCD Collaboration],
  Prog. Theor. Exp. Phys. {\bf 2012} (2012) 01A105.

\bibitem{Beane12}
  S.~R.~Beane {\it et al.},
  Phys. Rev. Lett. {\bf 109} (2012) 172001.

\bibitem{Polinder:2006zh}
  H.~Polinder, J.~Haidenbauer and U.-G.~Mei{\ss}ner,
  Nucl.\ Phys.\ A {\bf 779} (2006) 244.

\bibitem{Wei90}
S.~Weinberg, Phys. Lett. B {\bf 251} (1990) 288.

\bibitem{Wei91}
S.~Weinberg, Nucl. Phys. B {\bf 363} (1991) 3.

\bibitem{Bhaduri}
R. K. Bhaduri, B. Loiseau, and Y. Nogami, Ann. Phys. (N. Y.) {\bf} 44 (1967) 57.

\bibitem{Gal71} A. Gal, J.M. Soper, and R.H. Dalitz, 
Ann. Phys. (N. Y.) {\bf} 63 (1971) 53.

\bibitem{Schaffner}
J. Schaffner-Bielich, Nucl. Phys. A {\bf 835} (2010) 279. 
\bibitem{Vidana}
  I.~Vidana, D.~Logoteta, C.~Providencia, A.~Polls and I.~Bombaci,
  Europhys.\ Lett.\  {\bf 94} (2011) 11002. 

\bibitem{Entem:2003ft}
D.~R. Entem, R.~Machleidt, Phys. Rev. C {\bf 68} (2003) 041001.

\bibitem{Epe05}
  E.~Epelbaum, W.~Gl\"ockle, U.-G.~Mei{\ss}ner,
  Nucl.\ Phys.\ A {\bf 747} (2005) 362.

%

\bibitem{Epelbaum:2008ga} 
  E.~Epelbaum, H.~-W.~Hammer and U.-G.~Mei{\ss}ner,
  Rev.\ Mod.\ Phys.\  {\bf 81} (2009) 1773.

\bibitem{Machleidt:2011zz}
  R.~Machleidt and D.~R.~Entem,
  Phys.\ Rept.\  {\bf 503} (2011) 1
  [arXiv:1105.2919 [nucl-th]].

\bibitem{Epe98}
  E.~Epelbaum, W.~Gl\"ockle, U.-G.~Mei{\ss}ner,
  Nucl.\ Phys.\ A {\bf 637} (1998) 107.

\bibitem{Epe00}
  E.~Epelbaum, W.~Gl\"ockle, U.-G.~Mei{\ss}ner,
  Nucl.\ Phys.\ A {\bf 671} (2000) 295.

\bibitem{Kor01}
C.~L. Korpa, A.~E.~L. Dieperink, R.~G.~E. Timmermans, Phys. Rev. {\bf C} 65 (2001)
  015208.

\bibitem{Beane:2003yx}
S.~R. Beane, P.~F. Bedaque, A.~Parre\~no, M.~J. Savage, Nucl. Phys. A {\bf 747} (2005)
  55.

\bibitem{Hol89}
B.~Holzenkamp, K.~Holinde, J.~Speth, Nucl. Phys. A {\bf 500} (1989) 485.

\bibitem{Hai05}
  J.~Haidenbauer, U.-G.~Mei{\ss}ner,
  Phys.\ Rev.\ C {\bf 72} (2005) 044005.

\bibitem{Rij99}
T.~A. Rijken, V.~G.~J. Stoks, Y.~Yamamoto, Phys. Rev. C {\bf 59} (1999) 21.

\bibitem{Rij10}
T.~A. Rijken, M.~M.~Nagels, Y.~Yamamoto, Prog. Theor. Phys. Suppl. {\bf 185} (2010) 14.

\bibitem{Nog05}
A. Nogga, R.~G.~E. Timmermans and U. van Kolck, Phys. Rev. C {\bf 72} (2005) 054006.

\bibitem{Pav06}
M. Pavon Valderama, E. Ruiz Arriola, Phys. Rev. C {\bf 74} (2006) 054001.

\bibitem{Mach12}
  R.~Machleidt, Q.~MacPherson, E.~Marji, R.~Winzer, C.~.Zeoli and D.~R.~Entem,
  arXiv:1210.0992 [nucl-th].

\bibitem{Phillips13}
  D.~R.~Phillips,
  arXiv:1302.5959 [nucl-th].

\bibitem{Hai10a}
  J.~Haidenbauer, U.-G.~Mei{\ss}ner,
  Phys.\ Lett.\ B {\bf 684} (2010) 275.

\bibitem{Haidenbauer:2007ra}
  J.~Haidenbauer, U.-G.~Mei{\ss}ner, A.~Nogga and H.~Polinder,
  Lect.\ Notes Phys.\ {\bf 724} (2007) 113.

\bibitem{Pet11} S. Petschauer, diploma thesis, TU Munich, 2011. 

\bibitem{Swa63}
J.~J. de~Swart, Rev. Mod. Phys. {\bf 35} (1963) 916.

\bibitem{Dover1991}
  C.~B.~Dover and H.~Feshbach,
  Annals Phys.\  {\bf 217} (1992) 51.

\bibitem{Pet13} S. Petschauer and N. Kaiser, in preparation. 

\bibitem{Polinder:2007mp}
  H.~Polinder, J.~Haidenbauer and U.-G.~Mei{\ss}ner,
  Phys.\ Lett.\ B {\bf 653} (2007) 29.

\bibitem{Bernard1995} V.~Bernard, N.~Kaiser, U.-G.~Mei\ss{}ner, Int.\ J.\ Mod.\ Phys.\ E {\bf 4} (1995) 193.

\bibitem{VP}
  C.M. Vincent and S.C. Phatak,          
  Phys.\ Rev.\ C {\bf 10} (1974) 391.

\bibitem{Sec68}
B.~Sechi-Zorn, B.~Kehoe, J.~Twitty, R.~A. Burnstein, Phys. Rev. {\bf 175} (1968)
  1735.

\bibitem{Ale68}
G.~Alexander, U.~Karshon, A.~Shapira, G.~Yekutieli, R.~Engelmann, H.~Filthuth,
  W.~Lughofer, Phys. Rev. {\bf 173} (1968) 1452.

\bibitem{Eng66}
R.~Engelmann, H.~Filthuth, V.~Hepp, E.~Kluge, Phys. Lett. {\bf 21} (1966) 587.

\bibitem{Eis71}
F.~Eisele, H.~Filthuth, W.~F{\"o}lisch, V.~Hepp, G.~Zech, Phys. Lett. {\bf 37B} (1971)
  204.

\bibitem{Hep68}
V. Hepp and H. Schleich, Z. Phys. {\bf 214} (1968) 71.


\bibitem{Ste70}
D.~Stephen, Ph.D. thesis, University of Massachusetts, unpublished (1970).

\bibitem{Swa62}
J.~J. de~Swart, C.~Dullemond, Ann. Phys. {\bf 19} (1962) 485.

\bibitem{SIG1}
S.~Bart et al., Phys. Rev. Lett. {\bf 83} (1999) 5238.
\bibitem{SIG2}
H.~Noumi et al., Phys. Rev. Lett. {\bf 89} (2002) 072301; {\bf 90} (2003) 049902 (E).
\bibitem{SIG3}
P.K.~Saha et al., Phys. Rev. C {\bf 70} (2004) 044613.

\bibitem{Kohno06}
  M.~Kohno, Y.~Fujiwara, Y.~Watanabe, K.~Ogata and M.~Kawai,
  Phys.\ Rev.\ C {\bf 74} (2006) 064613.
\bibitem{Dab08}
  J.~Dabrowski and J.~Rozynek,
  Phys.\ Rev.\ C {\bf 78} (2008) 037601.

\bibitem{Stoks}
  V.~G.~J.~Stoks, R.~A.~M.~Klomp, C.~P.~F.~Terheggen and J.~J.~de Swart,
  Phys.\ Rev.\ C {\bf 49} (1994) 2950.

\bibitem{Kad71}
J.~A. Kadyk, G.~Alexander, J.~H. Chan, P.~Gaposchkin, G.~H. Trilling, Nucl.
  Phys. B {\bf 27} (1971) 13.

\bibitem{Hau77}
J.~M. Hauptman, J.~A. Kadyk, G.~H. Trilling, Nucl. Phys. B {\bf  125} (1977) 29.

\bibitem{Ratcliffe} P.~G.~Ratcliffe, Phys. Lett. B {\bf  365} (1996) 383.

\bibitem{Yamanishi} T.~Yamanishi, Phys. Rev. D {\bf 76} (2007) 014006.

\bibitem{PDG} J. Beringer {\it et al.} [Particle Data Group], 
  Phys. Rev. D {\bf 86} (2012) 010001.

\bibitem{Hai10} J. Haidenbauer, EPJ Web of Conferences {\bf 3} (2010) 01009.

\bibitem{Hai13} J.~Haidenbauer,
  arXiv:1301.1141 [nucl-th], Nucl.\ Phys. A, in press. 

\bibitem{Hai13a} J.~Haidenbauer, in {\it Proceedings of 
The 7th International Workshop on Chiral Dynamics},
6-10 August, 2012, Newport News, USA, in press. 

\bibitem{Kon00}
Y.~Kondo et~al., Nucl. Phys. A {\bf 676} (2000) 371.

\bibitem{Ahn05}
J.~K. Ahn et~al., Nucl. Phys. A {\bf 761} (2005) 41.
\bibitem{Ahn99}
J.~K. Ahn et~al., Nucl. Phys. A {\bf 648} (1999) 263.


\bibitem{Millener} D.~J. Millener,
  Nucl.\ Phys. A {\bf 881} (2012) 298.
\bibitem{Reuber}
  A.~Reuber, K.~Holinde and J.~Speth,
  Nucl.\ Phys.\ A {\bf 570} (1994) 543.

\bibitem{Hashimoto06}
  O.~Hashimoto and H.~Tamura,
  Prog.\ Part.\ Nucl.\ Phys.\  {\bf 57} (2006) 564.

\bibitem{Kaiser:2004fe}
  N.~Kaiser and W.~Weise,
  Phys.\ Rev.\ C {\bf 71} (2005) 015203

\bibitem{Gibson} B.~F. Gibson and D.~R. Lehman, Phys. Rev. C {\bf 22} (1980) 2024.

\bibitem{Miyagawa95}
  K.~Miyagawa, H.~Kamada, W.~Gl\"ockle, V.~G.~J.~Stoks,
  Phys.\ Rev.\ C {\bf 51} (1995) 2905.

\bibitem{Nog12} A.~Nogga, Proceedings of the {\it XI International Conference on Hypernuclear and 
Strange Particle Physics}, 
Barcelona, October 1 - 5, 2012, Nucl. Phys. A, in press. 

\bibitem{Nog13} A.~Nogga, J. Haidenbauer, and U.-G. Mei{\ss}ner, in preparation. 

\bibitem{Badalyan82}
A.~M. Badalyan, L.~P. Kok, M.~I. Polikarpov, Y.~A. Simonov, Phys. Rep. {\bf 82}
  (1982) 31.

\bibitem{Miyagawa99}
K.~Miyagawa, H.~Yamamura, Phys. Rev. C {\bf 60} (1999) 024003.

\bibitem{Mac13} 
  H.~Machner, J.~Haidenbauer, F.~Hinterberger, A.~Magiera, J.~A.~Niskanen, J.~Ritman and R.~Siudak,
  Nucl. Phys. A {\bf 901} (2013) 65. 
%
\bibitem{Sam12} 
  S.~A.~El-Samad {\it et al.}  [COSY TOF Collaboration],
  arXiv:1206.0426 [nucl-ex].

\bibitem{Kohno10} M.~Kohno, Phys.\ Rev.\ C {\bf 81} (2010) 014003.

\bibitem{Nog02}
  A.~Nogga, H.~Kamada and W.~Gloeckle,
  Phys.\ Rev.\ Lett.\  {\bf 88} (2002) 172501.

\bibitem{Kaiser1997} N.\ Kaiser, R.\ Brockmann, W.\ Weise, 
  Nucl.\ Phys. A {\bf 625} (1997) 758.

\bibitem{Donoghue} J.~F. Donoghue and B.~R. Holstein, Phys. Rev. D {\bf 25} (1982) 2015.

\bibitem{Ber01}
  V.~Bernard, L.~Elouadrhiri and U.-G.~Mei{\ss}ner,
  J.\ Phys.\ G {\bf 28} (2002) R1.

\bibitem{General} I.~J. General and S.~R. Cotanch, Phys. Rev. C {\bf 69} (2004) 035202.

\end{thebibliography}
\end{document}